\documentclass[11pt]{article}
\usepackage[usenames,dvipsnames,svgnames,table]{xcolor} 
\usepackage[obeyspaces,hyphens,spaces]{url}
\usepackage{jcapmod}
\usepackage[english]{babel}
\usepackage{amsmath, amssymb, amsbsy, amstext, amsthm,mathtools}
\usepackage{hyperref}
\usepackage{graphicx}
\usepackage{amsfonts}
\usepackage{fouridx}
\usepackage{cases}
\usepackage{bm}
\usepackage{bbm}
\usepackage{cleveref}
\usepackage{slashed}
\usepackage{mdframed}
\usepackage{tensor}
\usepackage{setspace}
\usepackage{braket}
\usepackage{subcaption}

\usepackage{tikz}
\usetikzlibrary{calc,shadings,patterns,tikzmark,fadings, decorations.markings, decorations.pathmorphing}
\usetikzlibrary{arrows}

\usepackage[export]{adjustbox}

\definecolor{Blue}{rgb}{0.25, 0.41, 0.88}
\definecolor{Red}{rgb}{0.92,0.,0.}
\definecolor{darkorange}{rgb}{1.0,0.549,0.}
\definecolor{cobalt}{RGB}{44, 98, 120}
\definecolor{Mathematica1}{rgb}{0.368417, 0.506779, 0.709798}
\definecolor{Mathematica2}{rgb}{0.880722, 0.611041, 0.142051}
\definecolor{Mathematica3}{rgb}{0.560181, 0.691569, 0.194885}
\definecolor{Mathematica4}{rgb}{0.922526, 0.385626, 0.209179}
\definecolor{Mathematica5}{rgb}{0.528488, 0.470624, 0.701351}
\definecolor{Mathematica6}{rgb}{0.772079, 0.431554, 0.102387}
\definecolor{Mathematica7}{rgb}{0.363898, 0.618501, 0.782349}
\definecolor{Mathematica8}{rgb}{1, 0.75, 0}
\definecolor{Mathematica9}{rgb}{0.647624, 0.37816, 0.614037}
\definecolor{plotBlue}{RGB}{94, 130, 181}
\definecolor{plotRed}{RGB}{233, 85, 54}
\definecolor{plotGreen}{RGB}{142, 176, 50}
\definecolor{plotPurple}{RGB}{135, 120, 178}
\definecolor{cornellRed}{HTML}{B31B1B}
\definecolor{cornellBlue}{HTML}{0068AC}
\definecolor{cornellGreen}{HTML}{6EB43F}

\newcolumntype{C}[1]{>{\centering\let\newline\\\arraybackslash\hspace{0pt}}m{#1}}

\newcommand{\confsc}{\Phi}

\newcommand{\tFo}[4]{{}_2 F_1\!\left[\genfrac..{-1pt}{0}{\raisebox{-1pt}{$#1, \,\, #2$}}{\raisebox{1pt}{$#3$}} \, \bigg| \, #4\, \right]}

\newcommand{\tFt}[5]{{}_2 F_2\!\left[\genfrac..{-1pt}{0}{\raisebox{-1pt}{$#1, \,\, #2$}}{\raisebox{1pt}{$#3, \,\, #4$}} \, \bigg| \, #5\, \right]}

\makeatletter
\newlength{\apb@width}
\newcommand{\autoparbox}[2][c]{\settowidth{\apb@width}{#2}\parbox[#1]{\apb@width}{#2}}

\makeatother

\makeatletter
\newsavebox\myboxA
\newsavebox\myboxB
\newlength\mylenA

\newcommand*\xoverline[2][0.75]{
    \sbox{\myboxA}{$\m@th#2$}%
    \setbox\myboxB\null
    \ht\myboxB=\ht\myboxA%
    \dp\myboxB=\dp\myboxA%
    \wd\myboxB=#1\wd\myboxA
    \sbox\myboxB{$\m@th\overline{\copy\myboxB}$}
    \setlength\mylenA{\the\wd\myboxA}
    \addtolength\mylenA{-\the\wd\myboxB}%
    \ifdim\wd\myboxB<\wd\myboxA%
       \rlap{\hskip 0.5\mylenA\usebox\myboxB}{\usebox\myboxA}%
    \else
        \hskip -0.5\mylenA\rlap{\usebox\myboxA}{\hskip 0.5\mylenA\usebox\myboxB}%
    \fi}
\makeatother

\numberwithin{equation}{section}

\def\beq{\begin{equation}}
\def\eeq{\end{equation}}

\def\bea{\begin{eqnarray}}
\def\eea{\end{eqnarray}}

\newcommand{\ud}{\mathrm{d}}

\newcommand{\mb}[1]{{\mathbf{#1}}}
\newcommand{\minus}{{\scalebox {0.8}[1.0]{$-$}}}

\theoremstyle{definition}

\DeclareRobustCommand{\SkipTocEntry}[4]{}

\definecolor{blue2}{cmyk}{1, 0.1, 0.1, 0.1}
\definecolor{byzantium}{rgb}{0.44, 0.16, 0.39}

\definecolor{pyBlue}{RGB}{31, 119, 180}
\definecolor{pyRed}{RGB}{214, 39, 40}
\definecolor{pyGreen}{RGB}{44, 160, 44}
\definecolor{pyBlue2}{RGB}{0, 111, 237}
\definecolor{pyRed2}{RGB}{224, 52, 36}

\DeclareMathOperator{\sech}{sech}
\DeclareMathOperator{\arctanh}{arctanh}

\tikzstyle{intSty}=[draw=white, thick, line width=0.24mm]
\tikzstyle{vert}=[draw=white, thick, line width=0.24mm]
\tikzstyle{vertC}=[draw=white, thick, line width=0.24mm]
\tikzstyle{inflSty}=[cornellRed]
\tikzstyle{vertexProp}=[densely dashed, line width=0.3mm, dash phase=1pt]
\tikzstyle{sigmaProp}=[line width=0.3mm]
\tikzstyle{intVertSty}=[draw=white, line width=0.2mm]

\usepackage[parfill]{parskip}

\begin{document}

\pagenumbering{roman}
\begin{titlepage}
\baselineskip=15.5pt \thispagestyle{empty}

\bigskip\

\vspace{1cm}
\begin{center}
{\fontsize{18}{24}{\bfseries Primordial Non-Gaussianity from Light Compact Scalars}}
\end{center}
\vspace{0.1cm}
\begin{center}
{\fontsize{12}{18}\selectfont Priyesh Chakraborty} 
\end{center}

\begin{center}
\vskip8pt
\textit{Department of Physics, Harvard University, Cambridge, MA 02138, USA}

\end{center}

\vspace{1.2cm}
\hrule \vspace{0.3cm}
\noindent {\bf Abstract}\\[0.1cm]
We study the non-Gaussianities generated by light axions, or compact scalar fields, during inflation. To correctly calculate their impact on primordial statistics, we argue that it is necessary to account for the periodicity, or gauge symmetry, of the compact scalars. We illustrate this point by comparing the predictions for the squeezed kinematic limit of the primordial bispectrum generated by two cases---a non-compact scalar $\sigma$ and a compact scalar $\varphi$. We demonstrate that while a light non-compact scalar predicts a bispectrum of the so-called local shape, the light compact scalar predicts a qualitatively different shape characterised by the ratio of the Hubble scale to its field-space circumference. In doing so, we show that ignoring the gauge symmetry of the compact scalar during inflation leads to spurious infrared enhancements, which are softened by working with appropriate gauge-invariant operators. In addition, we connect our results for the primordial bispectrum with late-time cosmological observables and show that it is possible to measure the decay constant of the compact scalar using galaxy clustering measurements.
	
\vskip10pt
\hrule
\vskip10pt

\end{titlepage}

\newcommand\emd{\xi}
\newcommand\imd{\zeta}

\thispagestyle{empty}
\setcounter{page}{2}
\begin{spacing}{1.03}
\tableofcontents
\end{spacing}

\clearpage
\pagenumbering{arabic}
\setcounter{page}{1}

\newpage

\section{Introduction}
With new cosmological data growing by the year, there is also growing interest in utilizing this data to sharpen our understanding of particle physics. One particular direction which has seen significant theoretical and observational effort is to precisely measure the probability distribution of primordial density fluctuations. Current cosmic microwave background (CMB) and large-scale structure (LSS) measurements have found that the initial conditions of the universe were largely Gaussian. Deviations from this Gaussianity, if detected, could be generated by interactions in the inflationary epoch and therefore such a detection channel is a way to observe signatures of new particles, i.e.\ to realise a cosmological collider \cite{Chen:2009we,Chen:2009zp,Baumann:2011nk,Assassi:2012zq,Chen:2012ge,Pi:2012gf,Noumi:2012vr,Baumann:2012bc,Flauger:2013hra,Assassi:2013gxa,Gong:2013sma,Green:2013rd,Arkani-Hamed:2015bza,Lee:2016vti,Flauger:2016idt,Chen:2016uwp,Chen:2016hrz,Kumar:2017ecc,Hook:2019zxa,Kumar:2019ebj,Pajer:2020wnj,Pajer:2020wxk,Lu:2021wxu,Wang:2021qez,Reece:2022soh,Pimentel:2022fsc,Jazayeri:2022kjy,Werth:2023pfl,Pinol:2023oux}. Already, CMB and large-scale structure data has started to be utilized to search for such cosmological collider signals \cite{Green:2023uyz,Sohn:2024xzd,Cabass:2024wob}. In this spirit, we study the cosmological collider phenomenology of a class of beyond the standard model particles---axions. 

Axions are historically motivated as a solution to the strong CP problem in the standard model of particle physics \cite{PhysRevD.16.1791,PhysRevLett.38.1440,PhysRevLett.40.223,PhysRevLett.40.279}, but nevertheless find utility in capturing other BSM phenomena as well. For instance, axions could populate the dark sector \cite{Preskill:1982cy,Dine:1982ah,Abbott:1982af}. Separately, they are also ubiquitous in theories of quantum gravity \cite{Svrcek:2006yi,Arvanitaki:2009fg,Mehta:2021pwf,Demirtas:2018akl,Demirtas:2021gsq,Gendler:2023kjt}. A defining feature of these particles, which we denote throughout as $\varphi$, is that they are compact scalar fields i.e.\ their physics is constrained by the gauge symmetry $\varphi \sim \varphi + 2\pi f$, where $f$ is known as the decay constant of the compact scalar. This decay constant is sensitive to ultraviolet physics. For example, in extra-dimensional models the axion decay constant is related to the Kaluza-Klein scale $M_{\rm KK}$ and therefore measuring $f$ provides us with information about potential extra-dimensional physics \cite{Reece:2023czb}. Such axions can be very light, with masses well below the eV scale \cite{Mehta:2021pwf}, and is much smaller than Hubble constant during inflation \cite{Takahashi:2018tdu}. As such, for the purposes of this paper we will assume that they are massless.

We will study the inflationary non-Gaussianities of such a compact scalar field, working within the framework of the effective field theory of single-field slow roll inflation \cite{Cheung:2007st}. Crucially, this observational channel does not rely on these compact scalars being a part of the dark sector in the late universe, and so they need not have a non-trivial present day abundance. Therefore primordial non-Gaussian signatures are distinct from other cosmological probes such as dark matter isocurvature which minimally assumes that such scalars populate the dark sector.\footnote{Our minimal assumption will be that the compact scalar is in its Bunch-Davies state during inflation and therefore weakly breaks the de Sitter isometries. This does not have to be true. Indeed, the de Sitter isometries may be violently broken on short-scales and still be consistent with CMB and LSS data \cite{Planck:2018jri}.} Our main goal will be to highlight the importance of the gauge symmetry of the compact scalar in identifying the leading contribution to the primordial bispectrum. To be consistent with the gauge symmetry, we must couple the comoving curvature with operators such as $\cos(\varphi/f)$. One might expect that we can perturbatively expand this operator $\cos(\varphi/f)=1+\frac{1}{2 f^2}\varphi^2 + \cdots$ and capture the leading order effects with the first non-trivial term $\varphi^2$. Consistent with the findings of \cite{Chakraborty:2023eoq}, we will show that this is incorrect.

In order to highlight this, we will study two scenarios. On one hand we will couple the comoving curvature $\zeta$ to a compact scalar $\varphi$, keeping track of the periodicity of the potential by working with the appropriate gauge invariant operator $\cos(\varphi/f)$. Then, to capture the effects of keeping only the leading order term of the compact scalar's potential, we will couple the inflaton to a scalar $\sigma$ via the operator $\sigma^2$, giving it a small mass $m_\sigma$ which we take to zero at the end of the calculation. We will refer to the latter case as the non-compact scenario.

Particularly when $\sigma$ is light, we recover the standard local shape bispectrum. This shape is characterised by its scaling in the so-called squeezed limit ($k_3 \to 0$)
\begin{equation}
    B_{\zeta}(k_1, k_2, k_3) \sim P_\zeta(k_1) P_\zeta(k_3) \left(\frac{k_3}{k_1}\right)^{0}\,,
\end{equation}
where $\zeta$ is the comoving curvature perturbation. We will show that the compact theory, where we work with the full gauge-invariant operator, instead produces the scaling
\begin{equation}
    B_{\zeta}(k_1, k_2, k_3) \sim P_\zeta(k_1) P_\zeta(k_3)\left[ \left(\frac{k_3}{k_1}\right)^{\beta} +  \left(\frac{k_3}{k_1}\right)^{2} + \cdots \right]
\end{equation}
where $\beta \equiv \frac{1}{2}(\frac{H}{2\pi f})^2$ measures the ratio of the Hubble constant to the circumference of $\varphi$'s field space. Therefore the ratio $\beta$ sets the asymptotic scaling of the bispectrum in the squeezed limit
\begin{equation}
    \begin{aligned}
        B_{\zeta}(k_1, k_2, k_3) \sim &\, P_\zeta(k_1) P_\zeta(k_3) \left(\frac{k_3}{k_1}\right)^{\beta}\, \quad (\beta<2) \\
         \sim & \, P_\zeta(k_1) P_\zeta(k_3) \left(\frac{k_3}{k_1}\right)^{2}\, \quad (\beta>2)\,.
    \end{aligned}
\end{equation}
The reason is that for light scalar fields in de Sitter, infrared fluctuations are large and force the scalar to disperse further in its field space. This is commonly understood in the language of stochastic inflation, in which the field executes a random walk and explores the ends of its potential \cite{Starobinsky:1986fx,Starobinsky:1994bd}. For a compact scalar, it therefore becomes necessary to work with the full periodic potential, or in other words with gauge invariant vertex operators $e^{i\varphi/f}$. This phenomenon in de Sitter is evocative of the dynamics of massless fields in two-dimensional flat space, as was explored in \cite{Ford:1985qh} and in more recent work \cite{DiPietro:2023inn}. 

\noindent \textbf{Outline} \, We begin with a discussion of free and composite fields in pure de Sitter in Section~\ref{sec:pure_ds}. We discuss the K\"all\'en-Lehmann representation which will be especially fruitful for our inflationary calculation, and review the free field theory of the compact scalar. In Section~\ref{sec:eft_of_inf} we calculate the primordial bispectrum for the exchange of a non-compact scalar and of the compact scalar, and show the results for the squeezed limit for both cases. Our strategy will be to resolve the exchanged operator into a sum over free field states using its spectral representation, as was done in \cite{Marolf:2010zp,DiPietro:2021sjt,Xianyu:2022jwk,Qin:2023bjk,Chakraborty:2023eoq,Chakraborty:2023qbp,Qin:2024gtr}, and leverage the bispectrum for the exchange of a single massive scalar to determine our answer.\footnote{This technique has also been applied to AdS Witten diagrams, see e.g.~\cite{PhysRevD.33.389,Penedones:2010ue,Fitzpatrick:2011hu}} Finally, in Section~\ref{sec:conclusions} we will conclude by connecting our results with expectations for late-universe cosmological measurements such as galaxy clustering data and comment on the implications of our results for more realistic models of compact scalars.

In Appendix~\ref{app:inin} we show the calculation of the primordial bispectrum using the in-in formalism in detail. While in the main text we are primarily concerned with the squeezed bispectrum, which is sensitive to the on-shell production of particles in the inflationary bulk spacetime, there is also a background contribution which is degenerate with local self-interactions of $\zeta$. We discuss this part of the bispectrum, commonly called the EFT piece, in Appendix~\ref{app:eft_seed}. In Appendix~\ref{app:bulk_cft} we apply the spectral method to the exchange of a CFT operator in the inflationary bulk. We do this for two reasons---firstly as a cross-check of the spectral method and secondly since the CFT exchange bispectrum displays some useful similarities with the compact scalar exchange. In Appendix~\ref{app:btree_largeDlim} we derive the large $\Delta$ asymptotic behavior of the tree-level seed function $\hat{b}_{\rm NA}(\Delta;u)$. In Appendix~\ref{app:vertex_spec_asymp} we do the same for the spectral density of the vertex operator and connect our result with the same in flat-space. Finally, in Appendix~\ref{app:uv_cutoff} we discuss an explicit string theory UV completion for the compact scalar and discuss the associated UV cutoff.

\section{Free and Composite Fields in de Sitter Space}\label{sec:pure_ds}
Throughout this paper, we will be working in the flat-slicing of the (quasi) de Sitter (dS) spacetime. In $(d+1)$-dimensions the de Sitter metric in these coordinates takes the form 
\begin{equation}
    \ud s^2 = -\ud t^2 + a^2(t) \ud \mb{x}^2 = a^2(\eta)\left[-\ud \eta^2 + \ud \mb{x}^2\right]\,,
\end{equation}
where $H$ is the Hubble constant and $a(t)=e^{H t}$ and $a(\eta)=-1/(H \eta)$ is the scale factor. The second equality shows the metric written in conformal coordinates, and $\eta$ is known as conformal time which takes values in $(-\infty,0)$, where the lower and upper bounds correspond to the infinite past and infinite future respectively. Where necessary, we will use $\alpha\equiv d/2$ to keep track of the number of spacetime dimensions. 

Let us first recall some aspects of two-point functions in dS. While typical (quasi) de Sitter calculations are done by Fourier transforming the spatial coordinates, we will find it useful to discuss physics in configuration space first and subsequently connect these statements to Fourier space.

The two-point function $G_\mathcal{O}(x,y)$ of a local operator $\mathcal{O}(x)$ is constrained by the dS isometries to be functions of the dS invariant distance $\xi$. In the flat-slicing it takes the form,
\begin{equation}
    \xi = 1- \frac{|\mb{x}-\mb{x}'|^2-(\eta-\eta')^2}{2\eta \eta'} \equiv 1-\frac{2}{\lambda}
\end{equation}
where we define the variable $\lambda$ since it will be useful to help keep track of distances.\footnote{$\lambda$ here is related to the $\zeta$ from \cite{Chakraborty:2023eoq} by a minus sign. We have changed the notation here to avoid confusion with the inflationary comoving curvature fluctuations.} Particularly $\lambda\in (0,1)$ represents the infrared, i.e.\ long spatial distances ($x \to \infty$) or late-times ($\eta, \eta' \to 0$).

The two-point function of a free scalar field with mass $m$ is determined by the scaling dimension $\Delta$. For principal series scalars ($m/H > \alpha$) the scaling dimension $\Delta=\alpha+i \sqrt{m^2/H^2-\alpha^2}$ is complex valued. As is standard, we will occasionally use $\nu \equiv \sqrt{\frac{m^2}{H^2}-\alpha^2}$ in lieu of the mass. For complementary series scalars ($m/H < \alpha$) the scaling dimension $\Delta = \alpha- \sqrt{\alpha^2-m^2/H^2}$ is real valued and thus $\Delta \in (0,\alpha)$. The free field propagator is
\begin{equation}
    G(\Delta;\xi) = \frac{H^{2\alpha-1}}{(4\pi)^{\alpha+\tfrac{1}{2}}}\frac{\Gamma(\Delta)\Gamma(\bar{\Delta})}{\Gamma(\alpha+\tfrac{1}{2})} \tFo{\Delta}{\bar{\Delta}}{\alpha+\tfrac{1}{2}}{\frac{1+\xi}{2}}\,,
\end{equation}
where ${}_{2}F_{1}\left[a,b;c|z\right]$ is the Gauss hypergeometric function and $\bar{\Delta}\equiv 2\alpha-\Delta$ is the shadow scaling dimension. Note that the free propagator can be written in an explicitly shadow symmetric form by utilizing a Kummer relation
\begin{equation}\label{eq:kummer}
    G(\Delta;\lambda) = \mathcal{A}(\Delta) \lambda^{\Delta} \tFo{\Delta}{\Delta-\alpha+\tfrac{1}{2}}{2\Delta-2\alpha+1}{\lambda} + (\Delta \leftrightarrow \bar{\Delta})\,,
\end{equation}
where
\begin{equation}
    \mathcal{A}(\Delta) \equiv \frac{H^{2\alpha-1}}{(4\pi)^{\alpha+\tfrac{1}{2}}}\frac{\Gamma(\Delta)\Gamma(2\alpha-2\Delta)}{\Gamma(\alpha+\tfrac{1}{2}-\Delta)}\,.
\end{equation}
The two-point correlator of a massive free field decays as a power law in the infrared,
\begin{equation}
    G(\Delta;\lambda) \xrightarrow{\lambda \to 0} \mathcal{A}(\Delta)\lambda^{\Delta}\left[1+\mathcal{O}(\lambda)\right] + (\Delta \leftrightarrow \bar{\Delta})\,.
\end{equation}
Importantly, a massless scalar corresponding to $\Delta=0$ admits a propagator which \textit{grows} with distance. In four dimensions,
\begin{equation}
    G(0;\lambda) = \tfrac{1}{2}H^2\lambda+ H^2\log\left(\lambda\right).
\end{equation}
This infrared growth is a key feature of physics in (quasi) de Sitter spacetimes and is ultimately responsible for a wealth of interesting phenomena such as the absence of spontaneous symmetry breaking \cite{Ford:1985qh,DiPietro:2023inn} and strong-coupling dynamics for light scalars \cite{Starobinsky:1986fx,Starobinsky:1994bd}. This large infrared growth will also be the main driving force of the results discussed in this paper.

Another important free-field state which will be relevant for our discussion is the conformally coupled scalar $\confsc$ with scaling dimension $\Delta_\confsc=\alpha-\frac{1}{2}$ which admits a power law propagator,
\begin{equation}    
    G_\confsc(\lambda) = H^{2\alpha-1}\frac{\Gamma(\alpha-\tfrac{1}{2})}{(4\pi)^{\alpha+\tfrac{1}{2}}} \lambda^{\alpha-\tfrac{1}{2}} \xrightarrow{\alpha \to \tfrac{3}{2}} H^2\frac{\lambda}{16\pi^2}\,.
\end{equation}

Apart from free scalar fields, it will also be useful for us to refer to a bulk CFT operator $\mathcal{O}_\delta$ with scaling dimension $\delta$, which also admits a simple power law two-point correlation function,\footnote{This is a standard flat-space CFT operator conformally mapped to the de Sitter bulk, and should not be confused with the late-time boundary CFT operators discussed in the dS/CFT context. \cite{Green:2013rd}}
\begin{equation}
    G_\delta(\xi) = \frac{H^{2\delta}}{(1-\xi)^{\delta}} = H^{2\delta}\left(\frac{\lambda}{2}\right)^{\delta}
\end{equation}
where $\delta \geq \alpha-\tfrac{1}{2}$ in accordance with unitarity. This is of course not a free-field for general $\delta$, but becomes a (rescaled) conformally coupled scalar when $\delta = \alpha- \tfrac{1}{2}$. Indeed for any integer multiple $\delta=n\left(\alpha-\tfrac{1}{2}\right)$, the bulk CFT operator can be understood, up to the normalization, as the composite operator $\confsc^n$.

\subsection{K\"all\'en-Lehmann Representation in de Sitter}
A general two-point function $G_\mathcal{O}(\xi)$ admits two useful integral representations. The first is the Watson-Sommerfeld representation \cite{Marolf:2010zp},
\begin{equation}
    G_\mathcal{O}(\xi) =  -\int_{\gamma}\frac{\ud J}{2\pi i} (2J+2\alpha) [G_\mathcal{O}]_{J} G(-J;\xi)
\end{equation}
where $G(\Delta;\xi)$ is the free propagator with scaling dimension $\Delta=-J$ and the contour $\gamma$ runs parallel to the ${\rm Im}(J)$ axis slightly to the left of zero. The function $[G_\mathcal{O}]_J$ is called the Euclidean momentum coefficients, because they define the angular momentum expansion of the position space propagator on the Euclidean sphere. This representation can be obtained by analytically continuing the position space propagator on the sphere \cite{Marolf:2010zp,Hogervorst:2021uvp,Chakraborty:2023qbp}.

\begin{figure}
    \centering
    \includegraphics[width=0.45\textwidth]{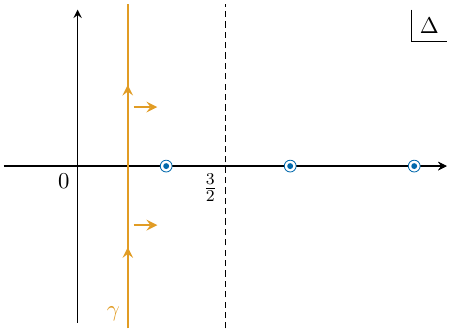}
    \includegraphics[width=0.45\textwidth]{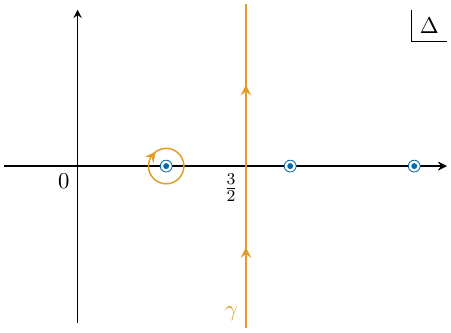}
    \caption{We show a contour plot illustrating the Watson-Sommerfeld integral (\ref{eq:watson_somm}) over the contour $\gamma$ on the left. On the right we show the result of pushing this contour onto the principal series line and thereby obtaining the K\"all\'en-Lehmann representation. The blue points represent potential simple poles of $[G_\mathcal{O}]_{\minus \Delta}$ that we may encounter as we deform the contour, the residues of which can be interpreted as complementary state contributions to the spectral representation.}
    \label{fig:watson_to_kallen}
\end{figure}

Given a propagator $G_\mathcal{O}(\xi)$, the momentum coefficients can be defined on the entire complex $J$-plane using the Lorentzian inversion formula \cite{Hogervorst:2021uvp,Chakraborty:2023qbp},
\begin{equation}\label{eq:mom_coeff_lorentz}
    [G_\mathcal{O}]_J = \mathcal{N}_J \int_{-\infty}^{0} \ud\lambda\, \lambda^{J-1} \tFo{J+\alpha+\tfrac{1}{2}}{J+1}{2J+2\alpha+1}{\lambda} {\rm disc}\, G_\mathcal{O}(\lambda)\,,
\end{equation}
where the normalization is,
\begin{equation}
    \mathcal{N}_J = \frac{2 \pi ^{\alpha +1} \Gamma (J+1)(-1)^{J-1}}{4^J \Gamma (J+\alpha +1)}\,.
\end{equation}
$G_\mathcal{O}(\xi)$ also admits a K\"all\'en-Lehmann representation, which is intimately related to the Watson-Sommerfeld representation \cite{Hogervorst:2021uvp,Loparco:2023rug}. To see this note that the Watson-Sommerfeld integral can equivalently be written in terms of $\Delta = -J$ after which it takes the form,
\begin{equation}\label{eq:watson_somm}
    G_\mathcal{O}(\xi) =  \int_{\gamma}\frac{\ud \Delta}{2\pi i} (2\alpha-2\Delta) [G_\mathcal{O}]_{\minus \Delta} G(\Delta;\xi)
\end{equation}
where the contour $\gamma$ runs parallel to the ${\rm Im}(\Delta)$ axis and slightly to the right of the origin. We can push the contour towards the principal series line, picking up any poles along the way, on which the shadow symmetry of the free propagator forces us to extract the shadow symmetric part of the momentum coefficient $[G_\mathcal{O}]_{\minus \Delta}$. The result is,
\begin{equation}\label{eq:gen_spec_rep}
    G_\mathcal{O}(\xi) = \int_{\alpha-i\infty}^{\alpha+i\infty} \frac{\ud \Delta}{2\pi i}\rho_{\mathcal{O}}(\Delta) G(\Delta;\xi) + (\text{complementary states})
\end{equation}
where $\rho_\mathcal{O}(\Delta)$ is called the spectral density and it is related to the momentum coefficients by $\rho_{\mathcal{O}}(\Delta)=\tfrac{1}{2}(2\alpha-2\Delta)[G_\mathcal{O}]_{-\Delta} + (\Delta \leftrightarrow \bar{\Delta})$, i.e. it is shadow symmetric. Here the integral sums up all the contributions to the correlator from principal series states and the contributions from complementary series states appear as residues due to pole crossings through the integration contour. We show a figure illustrating this procedure in Figure~\ref{fig:watson_to_kallen}. This representation can also be understood, as usual, as a simple consequence of the resolution of the identity. Furthermore, unitarity demands that $\rho_{\mathcal{O}}(\Delta) \geq 0$ on the principal series line.

The spectral density can be computed for a given dS propagator using the so-called Euclidean AdS inversion formula \cite{Loparco:2023rug},
\begin{equation}\label{eq:Eads_inv_formula}
    \rho_{\mathcal{O}}(\Delta) = \tilde{\mathcal{N}} \int_{-\infty}^{-1} \ud \xi (\xi^2-1)^{\alpha-1/2}\tFo{\Delta}{2\alpha-\Delta}{\alpha+\tfrac{1}{2}}{\frac{1+\xi}{2}} G_\mathcal{O}(\xi)\,,
\end{equation}
where the EAdS normalisation is\footnote{Note that we have an extra factor of $2\pi$ compared to \cite{Loparco:2023rug} which is due to our normalisation convention for the spectral density.}
\begin{equation}
    \tilde{\mathcal{N}} = \frac{2\pi^{\alpha+\tfrac{3}{2}}}{\Gamma(\Delta-\alpha)\Gamma(\alpha-\Delta) \Gamma(\alpha+\tfrac{1}{2})}\,.
\end{equation}
Importantly, the spectral densities we will encounter will all be meromorphic functions. As explained in \cite{Hogervorst:2021uvp}, the singularities of the spectral density in the ${\rm Re}(\Delta)>\alpha$ region of the complex $\Delta$-plane encode the infrared behavior of the propagator. This can be seen by using the free propagator as written in (\ref{eq:kummer}) in the spectral integral (\ref{eq:gen_spec_rep}) and closing the contour towards the right of the principal series line. Denoting the locations of the poles of $\rho_\mathcal{O}(\Delta)$ as $\Delta_*$ we have\footnote{Importantly, all spectral densities we will encounter contain a factor of $\nu \sinh(\pi \nu)$, due to which the spectral density $\rho_\mathcal{O}(\Delta)$ has zeros at $\Delta = \alpha + n$, for non-negative integer $n$. This eliminates a class of spurious poles of the free scalar propagator at $\Delta=\alpha+n$.}
\begin{equation}\label{eq:ir_expansion_spec}
    G_\mathcal{O}(\lambda) = -2\sum_{\Delta_*} {\rm Res}\, \rho_{\mathcal{O}}(\Delta) \mathcal{A}(\Delta) \lambda^{\Delta}\, \tFo{\Delta}{\Delta-\alpha+\tfrac{1}{2}}{2\Delta-2\alpha+1}{\lambda}\,.
\end{equation}
As an aside, we note that the momentum coefficients can also be computed from the spectral density. To see this note that we can resolve $G_{\mathcal{O}}(\lambda)$ in (\ref{eq:mom_coeff_lorentz}) into its K\"all\'en-Lehmann representation to obtain
\begin{equation}\label{eq:spec_density_to_mom_coeff}
    \begin{aligned}
        [G_\mathcal{O}]_J =&\, \int_\gamma\frac{\ud \Delta}{2\pi i} \rho_{\mathcal{O}}(\Delta)\left[\mathcal{N}_J \int_{-\infty}^{0} \ud\lambda\, \lambda^{J-1} \tFo{J+\alpha+\tfrac{1}{2}}{J+1}{2J+2\alpha+1}{\lambda} {\rm disc}\, G(\Delta;\lambda)\right] \\
        =&\, \int_{\gamma}\frac{\ud \Delta}{2\pi i}\frac{\rho_\mathcal{O}(\Delta)}{(J+\Delta)(J+\bar{\Delta})}\,,
    \end{aligned}
\end{equation}
where we have used the fact that the momentum coefficients of the free propagator are $[G(\Delta)]_J^{-1} = (J+\Delta)(J+\bar{\Delta})$. Note that the contour $\gamma$ lies along the principal series axis while separating the two poles at $\Delta=-J$ and at $\Delta=2\alpha+J$.

Let us briefly discuss some known examples of spectral densities of operators we will encounter in our inflationary calculations. For a free principal series scalar $\sigma$ with scaling dimension $\Delta_\sigma=\alpha+i \nu_\sigma$ the spectral density is
\begin{equation}
    \rho_\sigma(\alpha+i\nu) = \tfrac{1}{2}\delta_{\rm D}(\nu-\nu_\sigma)+\tfrac{1}{2}\delta_{\rm D}(\nu+\nu_\sigma)\,,
\end{equation}
where $\delta_{\rm D}(\nu)$ is the Dirac delta function. The spectral density for the composite operator $\sigma^2$ is also known \cite{Hogervorst:2021uvp},
\begin{equation}\label{eq:sig2_spec_func}
    \begin{aligned}
        \rho_{\sigma^2}(\Delta) = \,&H^{2\alpha-1}\frac{\nu  \sinh (\pi  \nu )}{2^3 \pi ^{\alpha +2} \Gamma (\alpha )} \frac{\Gamma \left(\frac{\Delta }{2}\right)^2 \Gamma \left(\alpha -\frac{\Delta }{2}\right)^2}{\Gamma (\Delta ) \Gamma (2 \alpha -\Delta )}  \\
        &\times \Gamma \left(\Delta_\sigma -\tfrac{1}{2}\Delta \right) \Gamma \left(\Delta_\sigma -\alpha +\tfrac{1}{2}\Delta \right) \Gamma \left( \alpha -\Delta_\sigma + \tfrac{1}{2}\Delta\right) \Gamma \left(2 \alpha -\Delta_\sigma -\tfrac{1}{2}\Delta\right)\,.
    \end{aligned}
\end{equation}
This spectral function admits three families of poles
\begin{equation}\label{eq:bulk_cft_spect}
    \Delta_*=2\Delta_\sigma + 2n, \, 2\bar{\Delta}_\sigma + 2n,\, \alpha+2n \quad (n \in \mathbb{Z}_{\geq 0})\,,
\end{equation}
which are known to correspond to the so-called double trace operators on the late-time dS boundary \cite{Hogervorst:2021uvp}. It will also be useful to recall the spectral density for a bulk CFT operator \cite{Hogervorst:2021uvp},
\begin{equation}
    \rho_\delta(\alpha+i \nu) = H^{2\delta-2\alpha+1}\frac{\pi ^{\alpha +\frac{1}{2}} 2^{2 \alpha -\delta +2}}{\Gamma (\delta ) \Gamma \left(\delta -\alpha +\frac{1}{2}\right)} \nu \sinh(\pi \nu) |\Gamma\left(\delta -\alpha- i \nu\right)|^2  \,.
\end{equation}
Indeed we can see that $\delta \geq \alpha-\tfrac{1}{2}$ ensures the positivity of the spectral density and is therefore required by unitarity. The singularities on the right side of the contour for the spectral integral are
\begin{equation}
    \Delta_* = \delta + n\,, \quad n \in \mathbb{Z}_{\geq 0}\,.
\end{equation}
Now let us move on to the main theory of interest---the compact scalar field.

\subsection{Compact Scalars in de Sitter}
A compact scalar $\varphi$ is defined by the gauge symmetry $\varphi \sim \varphi + 2\pi$.\footnote{We will be working with a dimensionless axion so that the decay constant only appears in the kinetic term.} In pure de Sitter the massless axion is described by the acion
\begin{equation}
    S_\varphi = \int \ud t \,\ud ^{2\alpha} \mb{x}\,\sqrt{-g} \left[\tfrac{1}{2}f^2 (\partial \varphi)^2 \right].
\end{equation}
It will often be convenient to keep track of the decay constant (or periodicity) using the parameter
\begin{equation}
    \beta \equiv \frac{1}{2}\left(\frac{H}{2\pi f}\right)^2\,.
\end{equation}
We will only be interested in $\beta \lesssim 10$, but it is worth noting that $\beta$ usually cannot be made arbitrarily large without exponentially suppressing the couplings in the theory \cite{Hook:2023pba}. Despite this suppression, we will leave open the possibility for larger $\beta$ in our main results. We will defer a more complete discussion of the UV cutoff, upper bounds on $\beta$ and the couplings to Appendix~\ref{app:uv_cutoff}.\footnote{We thank an anonymous referee for pointing out the EFT argument, which establishes that the cutoff must be logarithmically sensitive to the coupling, with model dependent prefactors. It would be interesting to carefully assess the detectability of such scalars when $\beta$ is large. The scenario in which the cutoff is larger than $4\pi f$ has been studied recently in the context of axion monodromy inflation (\cite{Creminelli:2025tae}) in which, unlike here, the compact scalar is responsible for driving inflation.}

Due to the periodicity of $\varphi$, the only healthy operators in the theory are those that are built using gauge invariant operators $\partial_\mu \varphi$ and $e^{\pm i \varphi}$. We will primarily be interested in the propagation of vertex operators in the bulk, i.e.\ $\mathcal{V}=:e^{i \varphi}:$, which have been defined with respect to a normal ordering procedure in dimensional regularisation \cite{Chakraborty:2023eoq}
\begin{equation}
    \mathcal{V}(x) \equiv \exp\left[\tfrac{1}{2}\beta\tilde{G}(1)+\tfrac{1}{2}\beta  \left(H_{2 \alpha -1}-H_{\alpha -\frac{1}{2}}+H_{2 \alpha }\right)\right] e^{i\varphi(x)}\,,
\end{equation}
where $H_{n}$ are the harmonic numbers (analytically continued to the complex plane) and
\begin{equation}
    \begin{aligned}
        \tilde{G}(1) &=\, -2\Gamma(\alpha)\left(\frac{H}{\mu}\right)^{2\alpha-3}\left[\psi ^{(0)}(2 \alpha )+\gamma_{\rm E}-\pi  \tan (\pi  \alpha )\right] \\
        &\to \,  \left(\frac{2}{\epsilon }-\frac{7}{2}\right) + \log \left(\frac{ \tilde{\mu}^2}{H^2}\right)
    \end{aligned}
\end{equation}
is the coincident limit of the massless two-point function in dimensional regularization. We have introduced the mass scale $\mu$ to ensure $[f]=1$ in arbitrary spacetime dimensions. In the last line we have specialised to $\alpha = \frac{3-\epsilon}{2}$ dimensions taking $\epsilon \to 0$ and defined $\tilde{\mu}^2 = 4\pi e^{\gamma_{\rm E}}\mu^2$. As with the Sine-Gordon model in two-dimensional flat-space \cite{PhysRevD.11.2088}, we can interpret the scale $\tilde{\mu}$ as the cutoff of the theory. Our choice of normal-ordering keeps $\mathcal{V}$ dimensionless, which we will find convenient. However, we note that we can always recover the canonical mass dimensions by rescaling $\mathcal{V} \to H^{\beta}\mathcal{V}$ at any point.

In the free-theory, correlation functions of the vertex operators must satisfy a ``charge-neutrality'' condition, i.e.\ correlators such as $\langle \mathcal{V}(x) \mathcal{V}(y) \rangle$ vanish. This is most easily seen in Euclidean signature, i.e.\ the sphere, where integration over the Euclidean zero mode enforces this constraint. In general, the vertex operators do not behave like standard free fields, clearly, since they are highly composite operators. Specifically, they do not Wick factorize in perturbation theory the way free field correlators do, which complicates perturbation theory at higher order \cite{Chakraborty:2023eoq}. However, since we are merely interested in tree-level exchange, this feature will not be relevant for us.

We denote the vertex propagator as
\begin{equation}
    \langle \mathcal{V}(x) \mathcal{V}^\dagger(y) \rangle \equiv G_{\mathcal{V}}(\lambda) = \lambda^{\beta} \exp\left(\frac{\beta}{2}\lambda\right),
\end{equation}
where we have specialised to four dimensions, and will do so hereafter. Note that this propagator has a qualitatively different singularity structure--- unlike composite operators such as $\sigma^2$ or $\mathcal{O}_\delta$, which exhibit a power law singularity, the vertex propagator has an essential singularity on the null cone ($\xi \to 1$ or $\lambda \to \infty$). However from the long-distance or boundary perspective, this propagator contains a hierarchy of states of definite scaling dimension \cite{Chakraborty:2023eoq}. 

Instead of working with the propagator directly, we will instead find it useful to rely on its spectral representation. To calculate the spectral density, we note that the vertex two-point function admits a simple expansion in power laws,
\begin{equation}\label{eq:vertex_to_bulkcft}
    G_{\mathcal{V}}(\lambda) = 2^\beta \sum_{n=0}^{\infty} \frac{\beta^n}{n!} \left(\frac{\lambda}{2}\right)^{\beta+n} 
\end{equation}
which importantly converges everywhere except at $\lambda = \infty$ (or $\xi = 1$), i.e.\ in the deep UV. This is useful because each power law can be interpreted as a bulk CFT state. In other words, we can interpret the exchange of a vertex operator as a sum over bulk CFT states.\footnote{The vertex operator is not special in this regard. Even a free scalar field can be safely expanded in power laws, or bulk CFT states, in the EAdS region. However, in that case it is not very useful to make this identification.}

To calculate the spectral density with the EAdS inversion formula (\ref{eq:Eads_inv_formula}) we can utilize (\ref{eq:vertex_to_bulkcft}) to resolve the vertex propagator into a sum over bulk CFT states. This is legitimate since the EAdS region ($\lambda \in (0,1)$) never sees the essential singularity in $\lambda$. Exchanging the sum with the inversion integral allows us to therefore write,
\begin{equation}\label{eq:vertex_spectral_function}
    \begin{aligned}
        &\rho_{\mathcal{V}}(\Delta) = 2^\beta\sum_{n=0}^{\infty} \frac{\beta^n}{n!}\rho_{\beta+n}(\Delta)\\
        &= \frac{16\pi \Gamma(1-\beta) \sin(\pi \beta)}{\Gamma(\beta-1)} (3-2\Delta)\cos(\pi \Delta) \Gamma(\beta-\Delta)\Gamma(\Delta-3+\beta)\tFt{\beta-\Delta}{\Delta-3+\beta}{\beta-1}{\beta}{\frac{\beta}{2}} 
    \end{aligned}
\end{equation}
where ${}_2 F_{2}\left[a_1,a_2;b_1,b_2|z\right]$ is a generalised hypergeometric function. First, let us demonstrate the positivity of this spectral function. For $\beta>1$, positivity is obvious since all the bulk CFT spectral densities we have summed have scaling dimension $\delta=\beta+n > 1$, and are therefore unitary. For $\beta<1$, we have to do a bit more work since the first term, corresponding to a dimension $\delta=\beta$ CFT state, is not unitary. Specifically in (\ref{eq:vertex_spectral_function}) we can isolate the first two terms which sum up to produce,
\begin{equation}
    \rho_{\beta}(\Delta)+ \beta \rho_{\beta+1}(\Delta) = 4\pi^2 \nu \sinh(\pi \nu) \left[(1-2\beta)^2+4\nu^2\right]\frac{|\Gamma \left(\beta +i \nu -\frac{3}{2}\right)|^2}{\Gamma^2(\beta)}
\end{equation}
which is manifestly positive. Of course all the $\delta=\beta+2,3,\cdots$ contributions are also positive since they correspond to unitary bulk CFT operators. Therefore, we conclude that this spectral density is unitary for any $\beta>0$.

Examining the singularities of $\rho_\mathcal{V}(\Delta)$ we see that it has poles at $\Delta_*=\beta+n$ on the right side of the principal series line.\footnote{There are also poles at $\Delta=3-\beta-n$ but they are irrelevant.} This merely reflects the fact that the vertex operator does not admit an infrared OPE organised with scaling operators and shadows, unlike a free-field $\sigma$ or even composite operators such as $\sigma^2$. This is not worrying since $\mathcal{V} \propto e^{i\varphi}$ is highly composite and we observe a similar singularity structure for another highly composite operator, the bulk CFT operator, as well. Its appearance in perturbative calculations can therefore be interpreted as a sign of strong-coupling. 

Integrating the spectral function along the principal axis reproduces the Wightman function for $\beta>3/2$. However, for $\beta<3/2$ we can analytically continue from larger $\beta$ to find that the pole $\Delta_*=\beta$ crosses the integration contour, so we must account for the residue generated by it and be careful to remove the contribution from the shadow symmetric piece (i.e.\ multiply by a factor of $2$). This removal of the shadow is equivalent to accounting for the residue of the momentum coefficient $[G_\mathcal{V}]_{-\Delta}$, as illustrated in Figure~\ref{fig:watson_to_kallen}. To summarize, for $\beta<3/2$ the spectral representation is,
\begin{equation}
    G_{\mathcal{V}}(\lambda) = -2\,{\rm Res}_{\Delta=\beta}\,\rho_{\mathcal{V}}(\Delta) G(\Delta;\lambda) + \int_{-\infty}^{\infty} \frac{\ud \nu}{2\pi} \rho_{\mathcal{V}}\left(\tfrac{3}{2}+i \nu\right) G\left(\tfrac{3}{2}+i\nu;\lambda\right)
\end{equation}

Another check of the spectral density (\ref{eq:vertex_spectral_function}) is to use it to determine the infrared expansion of the vertex propagator. We can simply use (\ref{eq:ir_expansion_spec}) which, we have checked, reproduces the infrared expansion of the vertex propagator up to $\mathcal{O}(\lambda^{20})$.

Lastly, we show in Appendix~\ref{app:vertex_spec_asymp} that $\rho_\mathcal{V}(\Delta) \sim \Delta^{\tfrac{2}{3}(\beta-2)}\exp(\beta^{1/3}\Delta^{2/3})$ as $\Delta \to \infty$, which is slower than $\exp(\frac{\pi}{2}\Delta)$, as required by the convergence of the K\"all\'en-Lehmann representation. The spectral integral can therefore easily be performed numerically. Moreover, this rapid growth for large $\Delta$ can be understood to be due to the essential short distance singularity of the vertex propagator.

\section{Non-Gaussianities in the EFT of Inflation}\label{sec:eft_of_inf}

In this section, we will move from pure de Sitter correlators to inflationary ones. We start with a review of the EFT of inflation, which is where all of the forthcoming discussion will be situated \cite{Cheung:2007st}. The inflationary state corresponds to a weak breaking of the de Sitter isometries by selecting a ``clock'' or preferred time-slicing $\tilde{t}(t,\mb{x})$. To construct the EFT action we can write all possible terms consistent with the residual symmetries of the theory, i.e.\ spatial diffeomorphisms. At leading order the action can be found to be,
\begin{equation}
    S = \int \ud t \,\ud ^3 \mb{x}\sqrt{-g}\left[\tfrac{1}{2}M_{\rm pl}^2 R - M_{\rm pl}^2(3H^2+\dot{H})+M_{\rm pl}^2 \dot{H}g^{00}\right] + \cdots\,,
\end{equation}
where $H$ is the inflationary Hubble parameter and the metric enters as $g^{00}$ and $R$. We can use the Stueckelberg trick to introduce the Goldstone mode corresponding to the broken spacetime symmetry and make the transformation $t \mapsto t + \pi(t,\mb{x})$ and $\mb{x} \mapsto \mb{x}$. The action is then
\begin{equation}
    S = \int \ud t \,\ud ^3 \mb{x}\sqrt{-g}\left[\tfrac{1}{2}M_{\rm pl}^2 R - M_{\rm pl}^2(3H^2(t+\pi)+\dot{H}(t+\pi))+M_{\rm pl}^2 \dot{H}\left(g^{00}+2\partial_\mu \pi g^{0\mu}+\partial_\mu \pi \partial_\nu \pi g^{\mu \nu}\right)\right] \,.
\end{equation}
Evaluating this on the exact dS background gives $g^{00}\to -1-2\dot{\pi}+(\partial \pi)^2$ and the action is
\begin{equation}
    S_{\pi} = \int \ud t \,\ud ^3 \mb{x}\,\sqrt{-g} M_{\rm pl}^2 \dot{H} (\partial \pi)^2+\cdots\,,
\end{equation}
where we have isolated the leading quadratic term. The comoving curvature perturbation $\zeta=-H \pi$ and the corresponding symmetry breaking scale can be read off this action as $f_\pi^4=2 M_{\rm pl}^2 \dot{H}=-2\epsilon M_{\rm pl}^2 H^2$,\footnote{In the presence of terms quadratic in $g^{00}$, one can admit a nontrivial sound speed $c_\pi$ for the inflaton in which case the symmetry breaking scale changes to $f_\pi^4=2 M_{\rm pl}^2 \dot{H} c_\pi$. We will assume $c_\pi=1$ here.} where the slow roll parameter $\epsilon=-\dot{H}/H^2$. It is also useful to canonically normalise the Goldstone to $\pi_c = \pi f_\pi^2$ so that the action becomes
\begin{equation}
    S_{\pi} = -\int \ud \eta \,\ud ^3 \mb{x}\,  \tfrac{1}{2} a^{2}\left[-(\pi_c')^2 + (\partial_i \pi_c)^2\right]\,,
\end{equation}
where we have also changed to conformal time. The inflaton is a massless free field and its mode functions, i.e.\ the solutions to the free field equations of motion in Fourier space, are therefore
\begin{equation}
    \pi_c^\pm(k,\eta) = \frac{i H}{\sqrt{2k^3}}(1\pm i k\eta)e^{\mp i k\eta}\,,
\end{equation}
where the $\pm$ corresponds to the positive/negative frequency mode function. Using these we can determine the $\zeta$ power spectrum to be
\begin{equation}
    P_\zeta(k) = \frac{2\pi^2 A_\zeta}{k^3}\,, \qquad A_\zeta = \frac{1}{4\pi^2}\left(\frac{H}{f_\pi}\right)^4\,,
\end{equation}
where $A_\zeta \approx 2 \times 10^{-9}$ is the amplitude of scalar fluctuations and is experimentally measured from CMB data \cite{Planck:2018jri}.

We are interested in perturbatively calculating the bispectrum $\langle \pi_c(\mb{k}_1) \pi_c(\mb{k}_2) \pi_c(\mb{k}_3)\rangle$ at equal times $\eta_0 \to 0$. To do so we will implement the standard rules of quantum mechanics and use the in-in formalism \cite{Weinberg:2005vy}
\begin{equation}
    \langle \pi_c(\mb{k}_1) \pi_c(\mb{k}_2) \pi_c(\mb{k}_3) \rangle = \left \langle \left(\bar{T}e^{i\int_{-\infty(1+i\epsilon)}^{\eta_0}\ud \eta H_{I}(\eta)}\right) \pi_c(\mb{k}_1) \pi_c(\mb{k}_2) \pi_c(\mb{k}_3) \left(Te^{-i\int_{-\infty(1-i\epsilon)}^{\eta_0}\ud \eta H_{I}(\eta)}\right)\right \rangle\,,
\end{equation}
where $H_{I}(\eta)$ is the interaction Hamiltonian. Upon expanding the time-evolution operator, one can write Feynman rules to compute these correlators diagrammatically \cite{Chen:2017ryl}. To do so we will require the momentum space propagators of all the fields involved. We need both the bulk-bulk propagators for any fields propagating in the inflationary bulk spacetime and bulk-boundary propagators for our external fields, which for us will just be $\pi_c$. The bulk-bulk propagators can be determined using the $\pi_c$ mode functions to be
\begin{equation}
    \begin{aligned}
        G^{\pm,\pm}_{\pi}(k;\eta,\eta') &= \pi_c^{\pm}(k,\eta) \pi_c^{\mp}(k,\eta')\theta(\eta-\eta') + \pi_c^{\mp}(k,\eta) \pi_c^{\pm}(k,\eta')\theta(\eta'-\eta) \, ,\\
        G^{+-}_{\pi}(k;\eta,\eta') &= \pi_c^{+}(k,\eta')\pi_c^{-}(k,\eta)\, , \\
        G^{-+}_{\pi}(k;\eta,\eta') &= \pi_c^{-}(k,\eta')\pi_c^{+}(k,\eta) \,,
    \end{aligned}
\end{equation}
where these are the (anti)time-ordered propagators and the Wightman functions respectively. Likewise the bulk-boundary propagators are
\begin{equation}
    \begin{aligned}
        K^{\pm}_{\pi}(k;\eta) &= \pi_c^{\mp}(k,\eta) \pi_c^{\pm}(k,\eta_0)\,.
    \end{aligned}
\end{equation}
Since we are only interested in the $\pi_c$ bispectrum, $\pi_c$ will only appear as an external state and we will only need to use the bulk-boundary propagator. We will represent it diagrammatically as
\begin{equation}
    \begin{tikzpicture}[thick, baseline=-28pt]
        \coordinate (c1) at (-1.0, -1.75);
        \coordinate (c2) at (1.0, -1.75);
        \coordinate (c3) at (-1.75, 0);
        \coordinate (c4) at (-0.25, 0);
        \coordinate (c5) at (1.75, 0);
        \coordinate (c6) at (0.0, -1.75);
        \draw[inflSty] (c2) -- (c5) node[midway, shift={(0.4, 0)}] {$\mb{k}$};
        \draw[line width=0.6mm, gray] (0, 0) -- (2.75, 0);
        \end{tikzpicture} = K_{\pi}^{\tt a}(k;\eta,\eta_0)\,,
\end{equation}
where ${\tt a}=\pm$ represents the two operator orderings and the grey bar represents the late-time boundary of inflation, which is also called the reheating surface.

\subsection{Goldstone Mixing with Composite Operators}
Now we will outline the leading interactions which mix the inflaton with a spectator local scalar operator $\mathcal{O}(x)$. We will consider $\mathcal{O}=\tfrac{1}{2}\sigma^2$ and $\tfrac{1}{2}\left[\mathcal{V}+\mathcal{V}^\dagger\right]$ for the non-compact and compact cases respectively. Our discussion here will hold for any scalar local operator $\mathcal{O}(x)$ and we will only specialise at the end. For this discussion, we will primarily follow the strategy of \cite{Noumi:2012vr,Jazayeri:2022kjy,Werth:2023pfl,Pinol:2023oux} and assume the local operator $\mathcal{O}(x)$ is drawn from a de Sitter invariant sector. In other words, we will ignore slow-roll corrections to the propagation of this operator since we expect such corrections to be slow-roll suppressed.

Before discussing the mixing with $\pi_c$, let us situate the discussion from Section~\ref{sec:pure_ds} in the inflationary context. For perturbative calculations we will need the momentum space Wightman function as well, i.e.\ the two-point correlation function for the spatial Fourier transformed operator
\begin{equation}
    \mathcal{O}_{\mb{k}}(\eta) \equiv \int \ud ^3 x \, e^{i \mb{k}\cdot \mb{x}} \mathcal{O}(\mb{x},\eta)\,.
\end{equation}
The momentum space propagator can be obtained from the position space one using an $i\epsilon$ prescription
\begin{equation}
    \langle \mathcal{O}_{\mb{k}}(\eta) \mathcal{O}_{-\mb{k}}(\eta') \rangle'\equiv G_{\mathcal{O}}^{+-}(k;\eta,\eta') = \int \ud ^3 x \, e^{i \mb{k}\cdot \mb{x}} G_\mathcal{O}(\lambda - i \epsilon)\,,
\end{equation}
where we assume $\eta>\eta'$, taking the limit $\epsilon \to 0^+$ after integrating. Note that the notation $\langle \cdots\rangle'$ indicates that the momentum conserving Dirac delta function is stripped off the correlator. The Wightman function for the opposite operator ordering $G^{-+}(k;\eta,\eta')$ can be obtained by complex conjugation as usual. Using these, both the time-ordered and anti time-ordered two-point correlators can be built,
\begin{equation}
    G^{\pm \pm}_{\mathcal{O}}(k;\eta,\eta') \equiv G^{\pm \mp}_{\mathcal{O}}(k;\eta,\eta') \theta(\eta-\eta') + G^{\mp \pm}_{\mathcal{O}}(k;\eta,\eta') \theta(\eta'-\eta)\,.
\end{equation}
For composite operators such as $\sigma^2$ and $\mathcal{V}$, it is difficult to carry out these Fourier integrals analytically as they amount to carrying out loop-integrals in momentum space. Instead we will note that the spectral representation also can be trivially applied to momentum space propagators
\begin{equation}
    G_{\mathcal{O}}^{\pm \pm}(k;\eta,\eta') = \int_{\tfrac{3}{2}-i\infty}^{\tfrac{3}{2}+i\infty} \frac{\ud \Delta}{2\pi i}\rho_{\mathcal{O}}(\Delta) G^{\pm \pm}(\Delta;k;\eta,\eta') + (\text{complementary states})\,,
\end{equation}
where $G^{\pm \pm}(\Delta;k;\eta,\eta')$ is the momentum space free-field propagator with scaling dimension $\Delta$. Diagrammatically we will refer to this propagator as
\begin{equation}
    \begin{tikzpicture}[thick, baseline=-28pt]
        \coordinate (c1) at (-1.0,-0.875);
        \coordinate (c2) at (1.0, -0.875);
        \begin{scope}[shift={(0, 0)}]
            \draw[double] (c1) -- (c2) node[midway, below] {$\mathcal{O}$};
        \end{scope}
        \end{tikzpicture} = G_\mathcal{O}^{\tt ab}(k; \eta, \eta')\,,
\end{equation}
where ${\tt a,b}=\pm$ correspond to any four of the possible propagators involved in the bulk propagation of $\mathcal{O}$. Note that although $\mathcal{O}$ is in principle highly composite, since we are working at tree-level we will only ever have to deal with a single propagator and we will not have to worry about potential complications with Wick factorization that appear while dealing with loops of the compact scalar \cite{Chakraborty:2023eoq}. Therefore the standard Feynman rules can be implemented without any change.

The leading coupling between the Goldstone $\pi_c$ and $\mathcal{O}$ is determined by
\begin{equation}
    S_{\mathcal{O}\pi} = \int \ud t \ud ^3 \mb{x}\sqrt{-g} \left[\tfrac{1}{2}g \Lambda^3(1+g^{00})\mathcal{O} \right]\,,
\end{equation}
where we normalise $\mathcal{O}$ to keep mass dimension $[\mathcal{O}]=1$ and the coupling dimensionless, leaving the cutoff $\Lambda$ explicit. In doing so we are hiding a factor of $(H/\Lambda)^{\Delta_{\mathcal{O}}-1}$ in the coupling $g$, which restores the canonical mass dimensions of $\mathcal{O}$ to $[\mathcal{O}]=\Delta_\mathcal{O}$. However we can always reintroduce this factor at a later stage. Introducing the Goldstone mode leaves us with
\begin{equation}
    S_{\mathcal{O} \pi} = \int \ud \eta \ud ^3 \mb{x}\,a^{2}(\eta)\left[g \Lambda^3 f_\pi^{-2} a(\eta) \pi_c' \mathcal{O} - \tfrac{1}{2}g \Lambda^3 f_\pi^{-4} (\pi_c')^2 \mathcal{O} +  \tfrac{1}{2}g \Lambda^3 f_\pi^{-4} (\partial_i\pi_c)^2 \mathcal{O}\right]\,.
\end{equation}
We will only pay attention to the couplings with time-derivatives and discard the $(\partial_i \pi_c)^2$ term. This is because we are primarily interested in the squeezed bispectrum and additional spatial derivatives produce a suppressed contribution in this limit (see e.g.~\cite{Jazayeri:2022kjy}). Therefore for our purposes the interactions we need to keep track of are
\begin{equation}\label{eq:Sint}
    S_{\mathcal{O} \pi} = \int \ud \eta \ud ^3 \mb{x}\,a^{2}\left[g \Lambda^3 f_\pi^{-2} a \pi_c' \mathcal{O} - \tfrac{1}{2}g \Lambda^3 f_\pi^{-4} (\pi_c')^2 \mathcal{O}\right]\,.
\end{equation}
These interaction terms produce a quadratic vertex and a cubic vertex
\begin{equation}
    \begin{tikzpicture}[thick, baseline=-28pt]
        \coordinate (c1) at (-1.0, -1.75);
        \coordinate (c2) at (1.0, -1.75);
        \coordinate (c3) at (-1.75, 0);
        \coordinate (c4) at (-0.25, 0);
        \coordinate (c5) at (1.75, 0);

        \coordinate (c6) at (0.0, -1.75);

        \begin{scope}[shift={(0, -1.75)}]
            
            \draw[double] (c6) -- (c2) node[midway, below] {$\mathcal{O}$};
        \end{scope}

        \draw[inflSty] (c2) -- (c5) node[midway, shift={(0.4, 0)}] {$\mb{k}_3$};
        \fill[intSty] (c2) circle (0.07) ;

        \end{tikzpicture} = -i g \Lambda^3 f_\pi^{-2} a^3(\eta) \partial_\eta\, \quad {\rm and}\quad \begin{tikzpicture}[thick, baseline=-28pt]
        \coordinate (c1) at (-1.0, -1.75);
        \coordinate (c2) at (1.0, -1.75);
        \coordinate (c3) at (-1.75, 0);
        \coordinate (c4) at (-0.25, 0);
        \coordinate (c5) at (1.75, 0);
        \coordinate (c6) at (0.0, -1.75);

        \begin{scope}[shift={(0, -1.75)}]
            
            \draw[double] (c1) -- (c6) node[midway, below] {$\mathcal{O}$};
        \end{scope}
        \draw[inflSty] (c1) -- (c3) node[midway, shift={(-0.3, 0)}] {$\mb{k}_1$};
        \draw[inflSty] (c4) -- (c1) node[midway, shift={(0.4, 0)}] {$\mb{k}_2$};
        \fill[intSty] (c1) circle (0.07) ;
        \end{tikzpicture} = -i g \Lambda^3 f_\pi^{-4} a^2(\eta) \partial_\eta \partial_\eta\,,
\end{equation}
where it is understood that the time-derivatives act on the $\pi_c$ mode functions. Note that momentum is conserved at the vertices due to spatial translation invariance of the background.

\subsection{Primordial Bispectrum}
There are many available tools to perturbatively compute inflationary correlation functions. One can implement the standard in-in formalism \cite{Weinberg:2005vy,Chen:2016uwp,Chen:2017ryl,Wang:2021qez,Xianyu:2022jwk,Qin:2023bjk,Chowdhury:2023arc,Werth:2024mjg}, analytically continue Euclidean AdS Witten diagrams \cite{Sleight:2019mgd,Sleight:2019hfp,DiPietro:2021sjt,Loparco:2023rug}, or analytically continue from the Euclidean sphere \cite{Marolf:2010zp,Lu:2021wxu,Chakraborty:2023qbp,Chakraborty:2023eoq}. The bootstrap approach, on the other hand, recasts the standard in-in diagrams into second order differential equations and we will particularly be relying on insights and results from this approach \cite{Arkani-Hamed:2015bza,Arkani-Hamed:2018kmz,Baumann:2019oyu,Baumann:2020dch,Pajer:2020wnj,Pajer:2020wxk,Jazayeri:2022kjy,Liu:2024xyi,Liu:2024str}. 

Let us begin with the simplest case of tree-level exchange of a scalar with scaling dimension $\Delta$. In the bootstrap approach, the primary object of interest is the seed function, called $b(\Delta;k_1,k_2,k_3)$, which is the three-point correlation function of conformally coupled scalars, in place of $\pi_c$, due to a massive scalar exchange. Since the massless propagator is related to the propagator of a conformally coupled scalar via certain weight-shifting operations, the respective bispectra can also be related similarly \cite{Arkani-Hamed:2015bza}. Here we will recap some essential features of the tree-level exchange bispectrum and defer the details to Appendix~\ref{app:inin}.

The seed function $b(\Delta; k_1,k_2,k_3)$ is highly constrained by the de Sitter isometries on the late-time boundary, where they become the three dimensional Euclidean conformal group. Conformal symmetry forces this function to be of the form $b(\Delta; k_1, k_2, k_3)=k_3^{-1} \hat{b}(\Delta; u)$, where $u\equiv \frac{k_3}{k_1 + k_2}$ is called the momentum cross-ratio, which for physical kinematics (set by momentum conservation $\mb{k}_1+\mb{k}_2+\mb{k}_3=0$) takes values $u\in [0,1]$. The limit $u \to 0$ corresponds to \textit{squeezed} triangles ($k_3 \ll k_1, k_2$) and the limit $u \to 1$ corresponds to \textit{flattened} triangles ($k_3 \sim k_1+k_2$). The squeezed limit will be the primary focus of this paper. The reason for these simplifications is essentially because we are working in the slow-roll approximation, where all the propagators are taken to be de Sitter invariant, and all knowledge of slow-roll is hidden inside the symmetry breaking scale $f_\pi$. 

In terms of the seed-function the $\pi_c$ bispectrum takes the form
\begin{equation}
    B_{\pi}(k_1, k_2, k_3)= -\frac{g^2 \Lambda^6 H^3}{16 f_\pi^6}\frac{1}{k_1 k_2 k_3 k_{12}^3}(u\partial_u^2 + 2 \partial_u)\hat{b}(\Delta;u) +\text{cyc.},
\end{equation}
where ``cyc.'' denotes cyclic permutations in the momenta $k_1$, $k_2$ and $k_3$. Therefore the form of the primordial bispectrum is entirely dictated by the seed function $\hat{b}(\Delta; u)$. This function is conventionally organised into two parts,
\begin{equation}
    \hat{b}(\Delta; u) = \hat{b}_{\rm EFT}(\Delta; u) + \frac{1}{2}\left[\hat{b}_{\rm NA}(\Delta;u)+\hat{b}_{\rm NA}(\bar{\Delta};u)\right]\,.
\end{equation}
The EFT piece $\hat{b}_{\rm EFT}(u)$, as the name suggests, is an analytic function of $u$ and is entirely mimicable by local self-interactions. It can be expressed as a series expansion
\begin{equation}\label{eq:tree_seed_eft}
    \hat{b}_{\rm EFT}(u) = \sum_{n=0}^\infty \left(-\frac{1}{2}\right)^{n}\frac{\sqrt{\pi} \,n!}{\Gamma(n+\tfrac{1}{2})} c_n(\Delta)u^{n+1} \tFo{\frac{n+1}{2}}{\frac{n+2}{2}}{n+\tfrac{3}{2}}{u^2}
\end{equation}
where the series coefficients are
\begin{equation}
    c_n(\Delta) =  \frac{1}{(\Delta-n-2)(n-1-\Delta)}\,,
\end{equation}
which carry all the mass dependence in the tree-level exchange process. Importantly, this series converges in the physical domain, except for $u \to 1$, and we review this limit in more detail in Appendix~\ref{app:eft_seed}. The non-analytic piece is
\begin{equation}
    \hat{b}_{\rm NA}(\Delta; u)  
    = \mathcal{F}(\Delta) 
    u^{\Delta-1} \tFo{\tfrac{1}{2}(\Delta-1)}{\tfrac{\Delta}{2}}{\Delta-\tfrac{1}{2}}{u^2}\,, \quad \mathcal{F}(\Delta) \equiv 
    \frac{\Gamma(\tfrac{3}{2}-\Delta)
    \Gamma(1-\tfrac{\Delta}{2})
    \Gamma^{2}(\tfrac{\Delta}{2})}
    {2\Gamma(\tfrac{3}{2}-\tfrac{\Delta}{2})}.
\end{equation}
In the squeezed limit this scales as $\hat{b}_{\rm NA}(\Delta; u) \sim \mathcal{F}(\Delta)u^{\Delta-1}$. For heavy masses $\nu \gg 1$ (where $\Delta = \tfrac{3}{2}+i \nu$) it is useful to recall that
\begin{equation}
    \hat{b}_{\rm NA}(\tfrac{3}{2}+i \nu;u)+\hat{b}_{\rm NA}(\tfrac{3}{2}-i \nu;u) \sim \nu^{-1/2}e^{-\pi \nu} u^{\tfrac{1}{2}+ i \nu} + \nu^{-1/2}e^{-\pi \nu} u^{\tfrac{1}{2} - i \nu},
\end{equation}
where the exponential decay is the characteristic Boltzmann suppression associated with particle production in de Sitter. A crucial point is that $\hat{b}_{\rm NA}(\Delta; u)$ admits simple poles in the complex $\Delta$-plane, specifically when $ {\rm Re}(\Delta)\geq \tfrac{3}{2}$. The form of the amplitude $\mathcal{F}(\Delta)$ makes it apparent that these poles are at $\Delta=\tfrac{3}{2}+n$ and $\Delta = 2(n+1)$, for $n \in \mathbb{Z}_{\geq 0}$.

Now let us move on to the case where we exchange the composite operator $\mathcal{O}$. In the exchange diagram for the inflationary bispectrum, we can resolve $\mathcal{O}$ into its spectral representation in order to express it as a sum over free field exchange diagrams, weighted by the spectral density. Schematically
\begin{equation}
    \def\circSize{0.6}
    \begin{tikzpicture}[thick, baseline=-28pt]
        \coordinate (c1) at (-1.0, -1.75);
        \coordinate (c2) at (1.0, -1.75);
        \coordinate (c3) at (-1.75, 0);
        \coordinate (c4) at (-0.25, 0);
        \coordinate (c5) at (1.75, 0);

        \begin{scope}[shift={(0, -1.75)}]
            
            \draw[double] (c1) -- (c2) node[midway, below] {$\mathcal{O}$};
        \end{scope}

        \draw[inflSty] (c1) -- (c3) node[midway, shift={(-0.3, 0)}] {$\mb{k}_1$};
        \draw[inflSty] (c4) -- (c1) node[midway, shift={(0.4, 0)}] {$\mb{k}_2$};
        \draw[inflSty] (c2) -- (c5) node[midway, shift={(0.4, 0)}] {$\mb{k}_3$};
        \draw[line width=0.6mm, gray] (-2.75, 0) -- (2.75, 0);
        \fill[intSty] (c1) circle (0.07) ;
        \fill[intSty] (c2) circle (0.07) ;
        \fill[cornellRed, intSty] (c3) circle (0.07);
        \fill[cornellRed, intSty] (c4) circle (0.07);
        \fill[cornellRed, intSty] (c5) circle (0.07);

        \end{tikzpicture}= \int_{\tfrac{3}{2}-i\infty}^{\tfrac{3}{2}+i \infty}  \frac{\ud \Delta}{2\pi i}\, \rho_{\mathcal{O}}(\Delta)\times \left[\begin{tikzpicture}[thick, baseline=-28pt]
        \coordinate (c1) at (-1.0, -1.75);
        \coordinate (c2) at (1.0, -1.75);
        \coordinate (c3) at (-1.75, 0);
        \coordinate (c4) at (-0.25, 0);
        \coordinate (c5) at (1.75, 0);

        \begin{scope}[shift={(0, -1.75)}]
            
            \draw (c1) -- (c2) node[midway, below] {$\Delta$};
        \end{scope}

        \draw[inflSty] (c1) -- (c3) node[midway, shift={(-0.3, 0)}] {$\mb{k}_1$};
        \draw[inflSty] (c4) -- (c1) node[midway, shift={(0.4, 0)}] {$\mb{k}_2$};
        \draw[inflSty] (c2) -- (c5) node[midway, shift={(0.4, 0)}] {$\mb{k}_3$};
        \draw[line width=0.6mm, gray] (-2.75, 0) -- (2.75, 0);
        \fill[intSty] (c1) circle (0.07) ;
        \fill[intSty] (c2) circle (0.07);
        \fill[cornellRed, intSty] (c3) circle (0.07);
        \fill[cornellRed, intSty] (c4) circle (0.07);
        \fill[cornellRed, intSty] (c5) circle (0.07);

        \end{tikzpicture}\right]\,.
\end{equation}
As with the tree-level bispectrum, we are constrained by the de Sitter invariance of the composite operator. Therefore the form of the composite-exchange bispectrum is fixed by a single function
\begin{equation}\label{eq:composite_seed_spectral_rep}
    \hat{\mathcal{B}}_{\mathcal{O}}(u) \equiv \int_{\tfrac{3}{2}-i\infty}^{\tfrac{3}{2}+i \infty} \frac{\ud \Delta}{2\pi i}\rho_{\mathcal{O}}(\Delta) \hat{b}(\Delta;u)\,,
\end{equation}
where $\hat{b}(\Delta;u)$ is the standard tree-level seed function, and we have assumed no complementary states contribute. The bispectrum generated by $S_{\mathcal{O}\pi}$ can be readily expressed as
\begin{equation}
    B_{\pi}(k_1,k_2,k_3) = -\frac{g^2 \Lambda^6 H^3}{16 f_\pi^6}\frac{1}{k_1 k_2 k_3 k_{12}^3}(u\partial_u^2 + 2 \partial_u)\hat{\mathcal{B}}_{\mathcal{O}}(u) +\text{cyc.}
\end{equation}
Recalling that $\zeta = -(H/f_\pi^2)\pi_c$, the $\zeta$ bispectrum can then be expressed as
\begin{equation}\label{eq:comp_zeta_bisp}
    B_{\zeta}(k_1,k_2,k_3) = \tfrac{1}{2}g^2 (2\pi^2 A_\zeta)^3 \left(\frac{\Lambda}{H}\right)^6\frac{1}{k_1 k_2 k_3 k_{12}^3}(u\partial_u^2 + 2 \partial_u) \hat{\mathcal{B}_{\mathcal{O}}(u)} +\text{cyc.}
\end{equation}
Therefore in order to obtain the bispectrum, we must solve for the seed function $\hat{\mathcal{B}}_{\mathcal{O}}(u)$. We will proceed by separately performing the spectral integral for $\hat{b}_{\rm EFT}(\Delta;u)$ and $\hat{b}_{\rm NA}(\Delta; u)$. In other words we can express the complete answer as
\begin{equation}
    \hat{\mathcal{B}}_{\mathcal{O}}(u) = \hat{\mathcal{B}}_{\mathcal{O}}^{\rm EFT}(u) + \hat{\mathcal{B}}_{\mathcal{O}}^{\rm NA}(u)\,.
\end{equation}
Let us discuss the EFT piece first. Note that in (\ref{eq:tree_seed_eft}) since all the mass dependence is captured by the series coefficients $c_n(\Delta)$, the series expansion will remain analytic in $u$ and only the coefficients will change. Therefore the EFT piece can be expressed as
\begin{equation}\label{eq:comp_seed_eft}
    \hat{\mathcal{B}}^{\rm EFT}_{\mathcal{O}}(u) = \sum_{n=0}^\infty \left(\minus \frac{1}{2}\right)^{n}\frac{\sqrt{\pi} \,n!}{\Gamma(n+\tfrac{1}{2})}C_n^{\mathcal{O}}u^{n+1} \tFo{\frac{n+1}{2}}{\frac{n+2}{2}}{n+\tfrac{3}{2}}{u^2}\,,
\end{equation}
where the series coefficients are
\begin{equation}
    C_n^{\mathcal{O}} = \int_{\tfrac{3}{2}-i\infty}^{\tfrac{3}{2}+i\infty}\frac{\ud \Delta}{2\pi i}\, \frac{\rho_{\mathcal{O}}(\Delta)}{(\Delta-n-2)(n+1-\Delta)}\,.
\end{equation}
Note that these coefficients can very easily be UV divergent, provided the spectral density grows like $\rho(\tfrac{3}{2}+i\nu) \sim \nu$ as $\nu \to \infty$. However, our focus here is on the squeezed limit of the bispectrum, in which this piece scales at most as $\hat{\mathcal{B}}^{\rm EFT}_{\mathcal{O}}(u) \sim u$. As we will see explicitly below, this is typically subdominant compared to the non-analytic piece. Therefore we will move on to the non-analytic piece and defer a more complete discussion of the EFT piece to Appendix~\ref{app:eft_seed}.\footnote{The EFT piece is nevertheless important in order to capture the full shape of the bispectrum. As we discuss more thoroughly in Appendix~\ref{app:eft_seed}, the EFT piece is crucial in cancelling an unphysical singularity when $u\to1$, or for flattened triangular configurations.}

At tree-level, the non-analytic part is written as a sum of shadow symmetric terms. Since the spectral density is invariant under shadow symmetry on the principal series line, we only need to integrate over one of the terms. The result is then,
\begin{equation}\label{eq:comp_seed_na}
    \hat{\mathcal{B}}_{\mathcal{O}}^{\rm NA}(u) = \int_{\tfrac{3}{2}-i\infty}^{\tfrac{3}{2}+i\infty}\frac{\ud \Delta}{2\pi i}\, \rho_{\mathcal{O}}(\Delta) \hat{b}_{\rm NA}(\Delta; u)\,.
\end{equation} 
Recall that by virtue of Boltzmann suppression, $\hat{b}_{\rm NA}(\tfrac{3}{2}\pm i \nu;u) \sim e^{-\pi |\nu|}$ when $\nu \to \pm \infty$. Since a healthy spectral function must grow slower than that for the K\"all\'en-Lehmann integral of the two-point function to converge, the non-analytic piece must be free from UV divergences. Of course, this makes sense since the non-analytic piece is fixed by infrared physics, i.e.\ the long-distance propagation of the exchanged operator $\mathcal{O}$. Note that this integral can also readily be done numerically. In case any of singularities of the spectral density $\Delta_*$ cross the integration contour, we must be more careful in also accounting for the residues accumulated along the way. We will comment on those cases separately.
\begin{figure}
    \centering
    \includegraphics[width=0.5\textwidth]{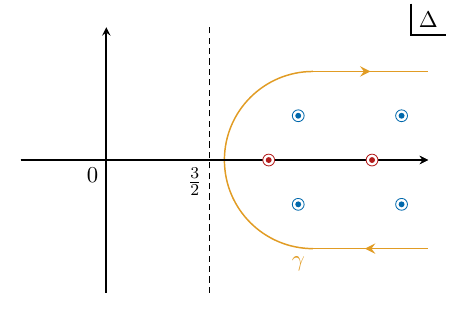}
    \caption{The contour deformation to resolve the spectral integral of $\hat{b}_{\rm NA}(\Delta;u)$ in terms of a sum over residues. The blue points schematically indicate the poles of the spectral density $\Delta_*$ and the red points the poles of the tree-level seed function $\hat{b}_{\rm NA}(\Delta;u)$ at $\Delta=2,4,\cdots$.}
    \label{fig:NA_contour_deform}
\end{figure}

Now we wish to deform the contour to the right of the principal series line and resolve our integral into a sum over residues, as shown in Figure~\ref{fig:NA_contour_deform}. We note that for large and positive $\Delta$ the function $\hat{b}_{\rm NA}(\Delta;u)$ scales as
\begin{equation}
    \hat{b}_{\rm NA}(\Delta;u) \sim \Delta^{-\tfrac{1}{2}} \exp\left[\frac{1}{4}\Delta \log\left(\frac{u^4+4 \left(\sqrt{1-u^2}-2\right) u^2-8 \sqrt{1-u^2}+8}{u^4}\right)\right]\,,
\end{equation}
where the coefficient of $\Delta$ in the exponent is negative for all physical kinematics $u \in [0,1)$ (see Appendix~\ref{app:btree_largeDlim} for details). Therefore as long as the spectral density $\rho_{\mathcal{O}}(\Delta)$ grows sub-exponentially this deformation can always be done. We will find that this is the case for both $\sigma^2$ and the vertex operator.

The potential poles we encounter are either those of $\hat{b}_{\rm NA}(\Delta; u)$ or of $\rho_{\mathcal{O}}(u)$. Let us discuss the former first. We encounter one set of poles at $\Delta_*=\tfrac{3}{2}+n$, for $n \in \mathbb{Z}_{\geq 0}$. Fortunately as discussed in Section~\ref{sec:pure_ds}, all known spectral densities have zeros at these locations so these poles are spurious. We also have singularities at $\Delta_*=2n+2$, for $n \in \mathbb{Z}_{\geq 0}$, which do not have any reason to be eliminated. These poles will contribute terms which are analytic in $u$, i.e.\ they will resemble EFT type terms.\footnote{At first glance it may seem odd that summing the non-analytic bispectrum, which is what the spectral integral accomplishes, produces EFT-type terms at all. However, the function $\hat{b}_{\rm NA}(u)$ as we call it receives a partial contribution from the non-factorizable in-in diagrams. In simple terms, the separation between the EFT and non-analytic piece at tree-level is not sharp. A better representation of the tree-level bispectrum therefore may not see the singularities at $\Delta=2,4,\cdots$. We thank Hayden Lee for explaining this to us.}

We will also receive contributions from the singularities of the spectral density. These have a clear interpretation as the infrared states we observe in the squeezed limit of the inflationary bispectrum. Putting everything together we have
\begin{equation}\label{eq:comp_seed_na_res}
    \hat{\mathcal{B}}_{\mathcal{O}}^{\rm NA}(u) = -\sum_{\Delta_*}\left[{\rm Res}_{\Delta_*} \rho_{\mathcal{O}}(\Delta)\right] \hat{b}_{\rm NA}(\Delta_*; u) - \sum_{n=0}^{\infty} \left[{\rm Res}_{2n+2} \hat{b}_{\rm NA}(\Delta;u)\right] \rho_{\mathcal{O}}(2n+2) \,.
\end{equation}
An implicit assumption here is that the singularities of the spectral density do not overlap with those of $\hat{b}_{\rm NA}(\Delta;u)$. The singularities can overlap e.g. for the compact scalar when $\beta\in \mathbb{Z}_{\geq 0}$, but we will ignore these cases here. Let us now evaluate the seed function for the non-compact and compact theories. 

\subsection{Non-Compact Scalar}
For $\mathcal{O}=\sigma^2$ one fairly straightforward way to calculate the non-analytic seed function is to simply numerically integrate the spectral integral. When $2\Delta_\sigma>\tfrac{3}{2}$, i.e.\ when the poles of the spectral density lie to the right side of the principal series line we have
\begin{equation}
    \hat{\mathcal{B}}_{\sigma^2}^{\rm NA}(u) = \frac{1}{2}\int_{-\infty}^{\infty}\frac{\ud \nu}{2\pi} \rho_{\sigma^2}(u)\left[\hat{b}_{\rm NA}(\tfrac{3}{2}+i \nu;u) + \hat{b}_{\rm NA}(\tfrac{3}{2}-i \nu;u)\right]\,.
\end{equation}
Along the principal series line when $\nu\to \infty$ the spectral density scales as
\begin{equation}
    \rho_{\sigma^2}(\alpha+i \nu) = \frac{1}{4(16\pi)^{\alpha-1} \Gamma(\alpha)}\nu^{2\alpha-2} + \cdots \xrightarrow{\alpha \to \tfrac{3}{2}}\frac{\nu}{8\pi}+\cdots\,,
\end{equation}
so the numerical integral converges for all $u$. When the poles cross over to the left of the principal series line, the spectral representation of $\sigma^2$ does not only include principal series states, but also some finite number of complementary states. These can be interpreted as pole crossings through the principal series contour as we continue from $2\Delta_\sigma>\tfrac{3}{2}$ to a smaller $\Delta_\sigma$. When the pole $\Delta_*=2\Delta_\sigma<\tfrac{3}{2}$ the numerical integral is amended to
\begin{equation}
    \begin{aligned}
        \hat{\mathcal{B}}_{\sigma^2}^{\rm NA}(u) = &\,-\left[\hat{b}_{\rm NA}(2\Delta_\sigma;u) + (\Delta_\sigma \leftrightarrow \bar{\Delta_\sigma})\right]{\rm Res}_{2\Delta_\sigma}\rho_{\sigma^2}(\Delta) \\
        &\,+\frac{1}{2}\int_{-\infty}^{\infty}\frac{\ud \nu}{2\pi} \rho_{\sigma^2}(u)\left[\hat{b}_{\rm NA}(\tfrac{3}{2}+i \nu;u) + \hat{b}_{\rm NA}(\tfrac{3}{2}-i \nu;u)\right]\,,
    \end{aligned}
\end{equation} 
where the residue can equivalently be seen as the contribution from the momentum coefficient $[G_{\sigma^2}]_{-\Delta}$ which now has a singularity on the complementary series line (see Figure~\ref{fig:watson_to_kallen}).

When we work with the sum over residues instead we have three sets of spectral function singularities to keep track of. We will start with the assumption that $\sigma$ is a principal series scalar and analytically continue our result to the complementary series. This prescription is the same as requiring that the contour of integration in (\ref{eq:comp_seed_na}) must separate the ``left'' and ``right'' poles of the spectral density, as is commonly implemented in Mellin space \cite{Sleight:2019mgd,Sleight:2019hfp}. Lastly, note that as $\Delta \to +\infty$ the spectral density scales as $\rho_{\sigma^2}(\Delta)\sim \Delta$. Following the discussion in the previous section, the exponential decay of $\hat{b}_{\rm NA}(\Delta;u)$ ensures we can apply the residue theorem for all physical $u$.

In (\ref{eq:comp_seed_na_res}), in principle we have contributions from $\Delta_* = 3,5,7,\cdots$, however we find that $\hat{b}_{\rm NA}(3+2n;u)=0$ so this series of poles does note contribute. Then, the set of poles of $\hat{b}_{\rm NA}(\Delta;u)$ at $\Delta=4,\cdots$ also don't contribute---only the one at $\Delta=2$. This is because $\rho_{\sigma^2}(4+2n)=0$ for $n \in \mathbb{Z}_{\geq 0}$. The only remaining poles which contribute are the ones at $2\Delta_\sigma+2n$ and $2\bar{\Delta}_\sigma + 2n$. Putting everything together we have
\begin{equation}
    \begin{aligned}
        \hat{\mathcal{B}}_{\sigma^2}^{\rm NA}(u) = &\frac{(3-2\Delta_\sigma)}{8\pi}\csc{2\pi \Delta_\sigma} \arctan(u) \\
        &\,- \sum_{n=0}^{\infty} \left[\left[{\rm Res}_{2\Delta_\sigma+2n}\rho_\sigma^2(\Delta)\right] \hat{b}_{\rm NA}(2\Delta_\sigma + 2n;u) + (\Delta_\sigma \leftrightarrow \bar{\Delta}_\sigma)\right]\,.
    \end{aligned}
\end{equation}
\begin{figure}
    \centering
    \includegraphics[width=0.47\textwidth]{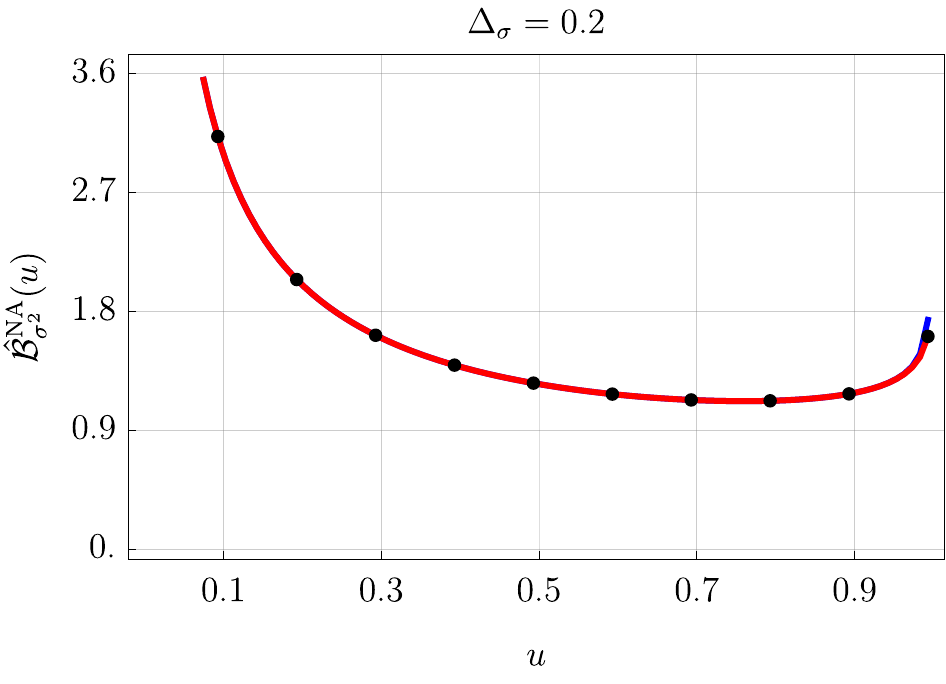}
    \includegraphics[width=0.5\textwidth]{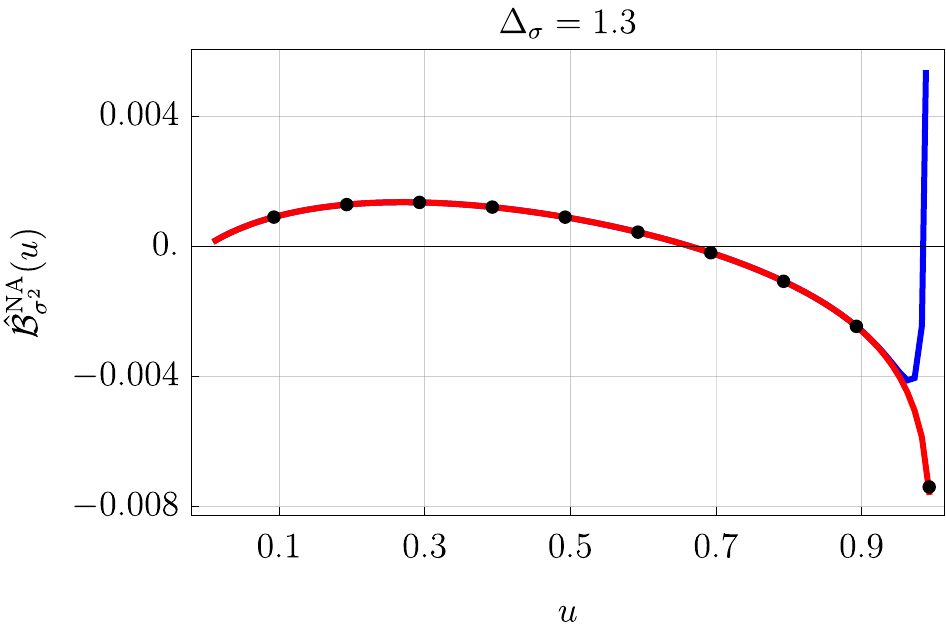}
    \caption{We compare the result of the numerical integral (\ref{eq:comp_seed_na}) for the operator $\sigma^2$ with the corresponding analytical result for scaling dimensions $\Delta_\sigma=0.2$ and $\Delta_\sigma=1.3$. We sum up to $\textcolor{blue}{N=10}$ and $\textcolor{Red}{N=40}$ residues for the solid lines and show the numerical integral in black dots.}
    \label{fig:sig2_num_analyt_comparison}
\end{figure}
We show a comparison of the analytical result with the numerical one in Figure~\ref{fig:sig2_num_analyt_comparison} for scaling dimensions $\Delta_\sigma=0.2$ and $\Delta_\sigma=1.3$. We observe that as we approach $u \to 1$, we need to sum over more terms in the series, especially for larger scaling dimensions. This is because the convergence of the series worsens as we approach the flattened configuration, especially for larger $\Delta_\sigma$. Nevertheless, most triangle configurations are captured with great accuracy using $N \sim 10$ residues. We also see that for larger mass, the small $u$ behavior is regular, whereas for small $\Delta_\sigma$ we observe a singularity. This is the usual singularity in the bispectrum $\hat{\mathcal{B}}_{\sigma^2}(u)\sim u^{2\Delta_\sigma-1}$ if we exchange a light scalar, and the threshold for seeing this singularity is $\Delta_\sigma = 1$. Lastly, the growth as $u\to 1$ is the standard $\log(1-u)$ growth associated with the non-analytic piece. As observed for the tree-level seed function, we expect that it should be cancelled by the EFT piece, which we partially discuss in Appendix~\ref{app:eft_seed}, but leave a complete investigation for future work.

Now we are primarily interested in the seed function when $\Delta_\sigma \to 0$. It is instructive to note that the leading residue in this limit is at $\Delta_*=2\Delta_\sigma$, where it scales as ${\rm Res}_{2\Delta_\sigma}\rho_{\sigma^2}(\Delta)\approx -1/(8\pi^2 \Delta_\sigma)+\mathcal{O}(\Delta_\sigma^0)$. This is of course the usual infrared divergence observed for light scalars in de Sitter. The key point is that all other residues are regular in this limit. Therefore the seed function in this light limit is given by
\begin{equation}
   \hat{\mathcal{B}}_{\sigma^2}^{\rm NA}(u) \approx \frac{1}{8\pi^2 \Delta_\sigma} \hat{b}_{\rm NA}(2\Delta_\sigma;u) = \frac{\Gamma \left(\frac{3}{2}-2 \Delta_\sigma \right) \Gamma (1-\Delta_\sigma ) \Gamma (\Delta_\sigma )^2}{16\pi^2 \Delta_\sigma\Gamma \left(\frac{3}{2}-\Delta_\sigma \right)} u^{2\Delta_\sigma-1}+ \cdots\,,
\end{equation}
where in the second equality we have expanded the seed function in the squeezed limit. We can translate this result onto the inflationary bispectrum using (\ref{eq:comp_zeta_bisp}). This produces the well known local shape in the inflationary bispectrum
\begin{equation}
    B_\zeta(k_1,k_2,k_3) \xrightarrow{k_3 \to 0}-g^2\frac{A_\zeta}{16 \Delta_\sigma^2}\left(\frac{\Lambda}{H}\right)^6 P_{\zeta}(k_1) P_{\zeta}(k_3) \left(\frac{k_3}{k_1}\right)^{0}\,,
\end{equation}
where the amplitude $f_{\rm NL} \propto \Delta_\sigma^{-2}$ is infrared enhanced by the mass. 

As an aside, let us briefly comment on the role played by $\Delta_\sigma$ in the expression above. The mass functions as an infrared regulator by causing the $\sigma$ propagator to decay at super-horizon distances. The mass here should be understood to be small but nonzero, since exactly massless non-compact scalars do not admit a dS invariant vacuum \cite{PhysRevD.32.3136}. The same does not hold in inflation, since it occurs for a finite number of e-folds, and therefore one can set the mass exactly to zero, typically finding a logarithmic enhancement in the soft-momentum. For example, the tree-level exchange of a massless $\sigma$ is well known to yield (see e.g.~\cite{Wang:2022eop})
\begin{equation}
    B_\zeta(k_1,k_2,k_3) \xrightarrow{k_3 \to 0} g^2 \frac{\log(-k_3\eta_0)}{k_1^3 k_3^3},
\end{equation}
where the enhancement can be understood, by setting $k_3 = -1/\eta_3$, as the number of e-folds spent by the soft mode random walking outside the horizon until the end of inflation at $\eta_0$ \cite{Starobinsky:1986fx,Starobinsky:1994bd}. Denoting $\eta_0/\eta_3 \equiv \exp[N(k_3)]$ we have
\begin{equation}
    B_\zeta(k_1,k_2,k_3) \sim g^2 \frac{N(k_3)}{k_1^3 k_3^3}.
\end{equation}
We have instead considered above the exchange of $\sigma^2$, which exhibits a similar but more severe infrared enhancement. To compare, note that upon accounting for a small but finite mass, the exchange of $\sigma$ is enhanced as $B_{\zeta} \sim \Delta_\sigma^{-1}$ compared to $\Delta_\sigma^{-2}$ for the exchange of $\sigma^2$ obtained above.\footnote{The tree-level seed function in the $\Delta \to 0 $ limit scales as $\hat{b}(\Delta;u)\sim \Delta^{-1}$.}  Therefore we expect that, upon regulating the infrared for an exactly massless $\sigma$ with $\eta_0$, one would find $B_\zeta \sim N^2(k_3)$. In other words, we expect a hierarchy of large IR logarithms, which is well known from self-interacting theories of massless scalars in inflation \cite{Starobinsky:1986fx,Starobinsky:1994bd}. Let us verify this expectation. To do so, we will need to refer more directly to the corresponding in-in integral, since the seed function we have worked with so far assumes $\eta_0=0$. We refer the reader to appendix~\ref{app:inin} for details regarding the in-in integrals.

In order to obtain the leading late-time singularity we should first extract the leading infrared scaling of the momentum space propagator. This can be done straightforwardly by relying on the dominant contribution from the K\"all\'en-Lehmann representation of $\sigma^2$. In four-dimensions the spectral density for $\sigma^2$, taking $\sigma$ to be massless, is
\begin{equation}
    \rho_{\sigma^2}(\Delta) =\frac{1}{16\pi} \frac{(\Delta -4) (\Delta +1) (2 \Delta -3) \cot (\pi  \Delta )}{(\Delta -3) \Delta }.
\end{equation}
The leading contribution from the spectral representation is therefore a $\Delta=0$ complementary state at which the spectral density admits a \textit{double} pole $\rho_{\sigma^2}(\Delta) \to -1/(4\pi^2 \Delta^2)$. As a consequence the infrared, or late-time, behavior of the momentum space propagator is logarithmically enhanced
\begin{equation}
    G_{\sigma^2}(k;\eta,\eta') = -2 {\rm Res}_{\Delta=0} \rho_{\sigma^2}(\Delta) G(\Delta;k;\eta,\eta') \xrightarrow{\eta,\eta'\to 0}\frac{1}{8\pi^2} \frac{\log(-k \eta)+ \log(-k \eta')}{k^3}.
\end{equation}
Therefore the massless $\sigma^2$ propagator diverges logarithmically in the late-time limit whereas the massless $\sigma$ propagator does not. Since we are only interested in isolating the leading contribution in the limit $k_3 \to 0$ we do not need to solve the corresponding in-in integrals exactly, and instead will only extract the leading late-time divergence. It is straightforward to then obtain the four-point seed function by inserting this propagator in (\ref{eq:4pt_seed_inin}). Upon simplifying the sum over the various time-orderings we obtain
\begin{equation}
    F(k_1,k_2,k_3,k_4;s)=-\frac{1}{2\pi^2 s^3}\left[J(k_{12},k_{34};s)+J(k_{34},k_{12};s) \right]+\cdots,
\end{equation}
where for convenience we set $k_{ij}\equiv k_i + k_j$ and define
\begin{equation}
    J(k_{12},k_{34};s) \equiv \left[\int_{-\infty}^{\eta_0} \frac{\ud \eta}{\eta^2}\log(-s \eta)\sin(k_{12}\eta)\right]\left[\int_{-\infty}^{\eta_0} \frac{\ud \eta'}{\eta'^2}\sin(k_{34}\eta')\right].
\end{equation}
Note that we have only kept track of the leading late-time terms of the propagator and omitted the $i \epsilon$ prescription for brevity. We can further approximate the integrand to leading order in $\eta_0$ and therefore also the integral to obtain
\begin{equation}
    J(k_{12},k_{34};s) = \frac{1}{2}k_{12}k_{34} \log^2(-s \eta_0)\log(-k_{34} \eta_0)+\cdots.
\end{equation}
Now, as with the tree-level exchange  (\ref{eq:bisp_seed_from_4pt_seed}), acting on $F$ with $\partial^2/\partial k_{12}^2$ and taking the limit $k_{34} \to k_3$ and $s\to k_3$, we can obtain the scalar bispectrum in the $k_3 \to 0$ limit
\begin{equation}
    B_\zeta(k_1,k_2,k_3) \sim g^2 \frac{\log^2(-k_3 \eta_0)}{4\pi^2 k_1^3 k_3^3} \sim g^2 \frac{N^2(k_3)}{4\pi^2 k_1^3 k_3^3},
\end{equation}
which, as expected, is stronger than the infrared divergence at tree-level. Indeed for sufficiently large number of e-folds one encounters the same loss of perturbativity as from demanding $\Delta_\sigma$ to be too small.\footnote{In passing, we highlight a perturbativity bound on the number of e-folds for a massless noncompact scalar coupled to the inflaton via the operators $\sigma$ and $\sigma^2$, where $\sigma$ is massless. Assuming order one couplings for both operators and comparing with the tree-level bispectrum and assuming order one couplings, we find that perturbation theory breaks down for $N(k_3)\gtrsim 4\pi^2$. This could be problematic for our universe depending on the precise physics of reheating. For example, under the standard assumption of efficient reheating $N(k_3)\approx 60$ for large-scale modes of the CMB, which falls outside this bound.}

To summarize, the infrared enhancement for a light non-compact scalar is signalling that the leading order answer is incomplete for sufficiently small $\Delta_\sigma$ or for sufficiently long inflation. In other words, it is a sign of the theory losing perturbativity. Higher order operators such as $\sigma^4, \sigma^6,\cdots$ need to enter the picture to potentially rescue this divergence. We will see shortly that for a compact scalar, this does indeed happen, and keeping track of the full gauge-invariant operator will change the scaling of the bispectrum in the squeezed limit.

\subsection{Compact Scalar}
For the compact scalar the composite operator being exchanged is $\tfrac{1}{2}[\mathcal{V}+ \mathcal{V}^\dagger]$\footnote{We could have instead worked with the operator $\cos(\varphi + \varphi_0)$, where $\varphi_0$ is an arbitrary phase and arrived at $\mathcal{O}=\tfrac{1}{2}e^{i\varphi_0} \mathcal{V}+{\rm c.c.}$ instead. At the level of the bispectrum, this phase can be completely absorbed by the coupling $g$, so we find it convenient to set it to zero. This also enables us to match with the non-compact theory since $\cos(\varphi)=1 + \tfrac{1}{2}\varphi^2 + \cdots$.} so we rely on our computation of the spectral density $\rho_{\mathcal{V}}(\Delta)$. With the spectral density, we can calculate the non-analytic seed function by numerically performing the spectral integral. For $\beta>3/2$ it is
\begin{equation}
    \hat{\mathcal{B}}_{\mathcal{V}}^{\rm NA}(u) = \frac{1}{2}\int_{-\infty}^{\infty}\frac{\ud \nu}{2\pi} \rho_{\mathcal{V}}(u)\left[\hat{b}_{\rm NA}(\tfrac{3}{2}+i \nu;u) + \hat{b}_{\rm NA}(\tfrac{3}{2}-i \nu;u)\right]\,.
\end{equation}
Recall that the spectral density of the vertex operator grows as $\Delta ^{\frac{2}{3}(\beta -2)} \exp\left(\frac{3 \beta^{1/3}  \Delta ^{2/3}}{2^{4/3}}\right)$ which ensures that this integral converges (see Appendix~\ref{app:vertex_spec_asymp} for details).

This is sufficient when $\beta>\tfrac{3}{2}$, i.e.\ in the absence of pole crossings. For $\beta<3/2$ the spectral representation of the vertex operator also includes a complementary series state. In that case the spectral representation of the non-analytic piece becomes
\begin{equation}
    \begin{aligned}
        \hat{\mathcal{B}}_{\mathcal{V}}^{\rm NA}(u) = &\,-\left[\hat{b}_{\rm NA}(\beta;u) + \hat{b}_{\rm NA}(2\alpha-\beta;u)\right]{\rm Res}_{2\beta}\rho_{\mathcal{V}}(\Delta) \\
        &\,+\frac{1}{2}\int_{-\infty}^{\infty}\frac{\ud \nu}{2\pi} \rho_{\mathcal{V}}(u)\left[\hat{b}_{\rm NA}(\tfrac{3}{2}+i \nu;u) + \hat{b}_{\rm NA}(\tfrac{3}{2}-i \nu;u)\right]\,.
    \end{aligned}
\end{equation} 
\begin{figure}
    \centering
    \includegraphics[width=0.49\textwidth]{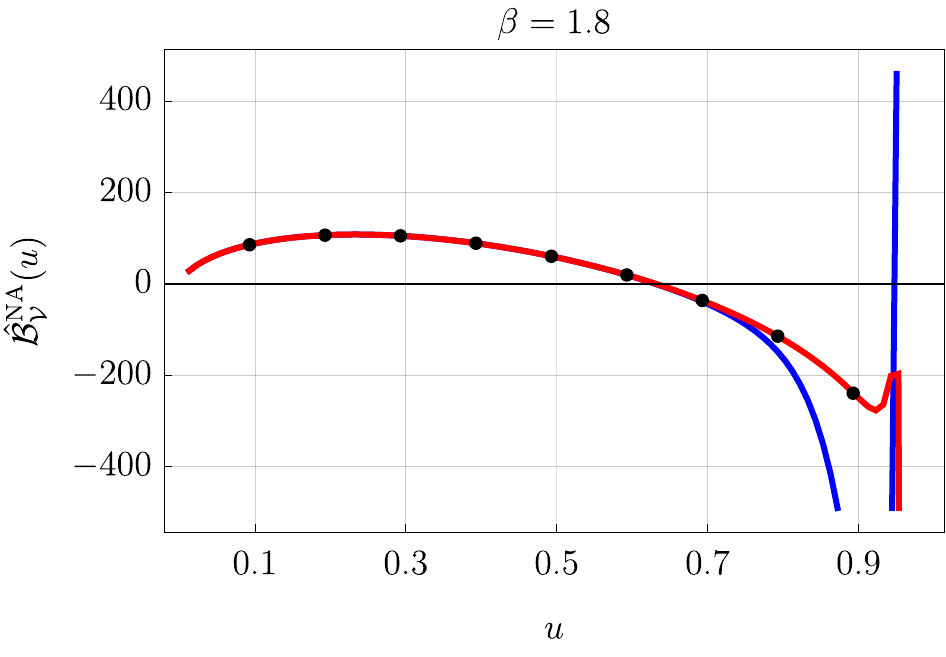}
    \includegraphics[width=0.49\textwidth]{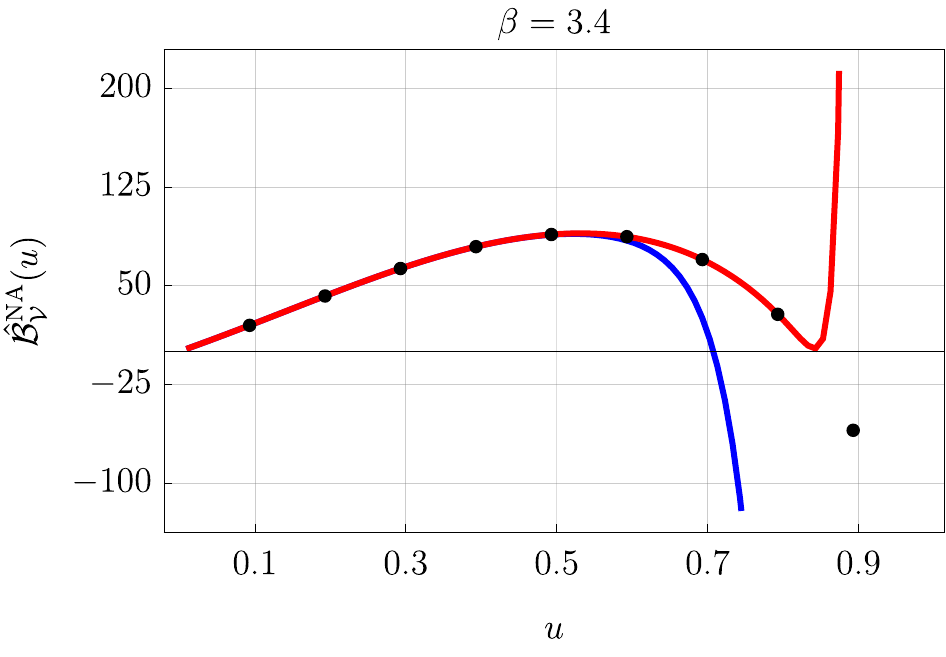}
    \caption{The same as Figure~\ref{fig:sig2_num_analyt_comparison} but for the compact scalar. We sum up to $\textcolor{blue}{N=10}$ and $\textcolor{Red}{N=40}$ residues for the solid lines and show the numerical integral in black dots. On the left we show the comparison for $\beta=1.8$ and on the right for $\beta=3.4$. We see that as we increase $\beta$, we need to sum over more residues to match the exact numerical answer.}
    \label{fig:vertex_num_analyt_comparison}
\end{figure}
As discussed previously, we can also simply apply the residue theorem to the spectral integral, assuming $\beta>3/2$, where the resulting expression applies for all $\beta$ by analytic continuation. In order to do so we note that the relevant poles of the spectral density lie at $\Delta_* = \beta + n$ where $n \in \mathbb{Z}_{\geq 0}$. In addition to this, all of the poles of $\hat{b}_{\rm NA}(\Delta; u)$ at $\Delta=2,4,\cdots$, will also contribute. The complete answer then becomes
\begin{equation}
    \hat{\mathcal{B}}_{\mathcal{V}}(u) = -\sum_{n=0}^{\infty}\left[{\rm Res}_{\beta+n} \rho_{\mathcal{V}}(\Delta)\right] \hat{b}_{\rm NA}(\beta+n; u) - \sum_{n=0}^{\infty} \left[{\rm Res}_{2n+2} \hat{b}_{\rm NA}(\Delta;u)\right] \rho_{\mathcal{V}}(2n+2) \,.
\end{equation}
The sub-exponential growth of the spectral density ensures that the sum over residues converges. Nevertheless, the spectral density grows rapidly compared to $\rho_{\sigma^2}(\Delta)$ which makes numerical convergence extremely slow as $u \to 1$. For these points it is more efficient to instead numerically perform the spectral integral. We compare the numerical evaluation of the spectral representation with the analytical result in Figure~\ref{fig:vertex_num_analyt_comparison}. Indeed we observe that the convergence of the series becomes worse as we approach $u \to 1$. Nevertheless since we are only interested in the leading asymptotic behavior as $u \to 0$, we only need to keep track of the first term of the sum corresponding to $n=0$. When $\beta<2$ the leading scaling is determined by the residue at $\Delta_*=\beta$ to be
\begin{equation}
    \hat{\mathcal{B}}_{\mathcal{V}}^{\rm NA}(u) \approx 16\pi^2(3-2\beta)\cos(\pi \beta) \frac{\Gamma(2\beta-3)}{\Gamma(\beta)\Gamma(\beta-1)} \hat{b}_{\rm NA}(\beta;u) = \frac{4\pi^3 2^{\beta}\cot(\tfrac{\pi}{2}\beta)}{1-\beta}u^{\beta-1} + \cdots\,,
\end{equation}
whereas for $\beta>2$ the leading scaling is analytic in $u$
\begin{equation}
    \hat{\mathcal{B}}_{\mathcal{V}}^{\rm NA}(u) \approx -2 \rho_{\mathcal{V}}(2) \arctan(u) = -2 \rho_{\mathcal{V}}(2) u + \cdots.
\end{equation}
The interpretation of this result is also clear--- for sufficiently large $\beta$, the vertex propagator decays rapidly which means that it becomes difficult to produce an on-shell compact scalar state. In this sense, the compact scalar becomes very ``heavy''. Its effect on the bispectrum becomes degenerate with that of a series of local self interactions of the inflaton. This is consistent with the observations of \cite{Chakraborty:2023eoq} where it was found that loop effects of the compact scalar exponentially diminish in magnitude as we send $\beta \to \infty$.

We therefore see that working with the appropriate gauge-invariant vertex operators for a massless compact scalar leads to a qualitatively different prediction for the squeezed bispectrum. Instead of the local shape, the squeezed limit is determined by another important dimensionless ratio, namely $\beta\propto (H/f)^2$. For the primordial $\zeta$ bispectrum we therefore find that for $\beta<2$
\begin{equation}
    B_\zeta(k_1, k_2, k_3) \xrightarrow{k_3 \to 0} f_{\rm NL}(\beta)\left(\frac{k_3}{k_1}\right)^{\beta}P_\zeta(k_1) P_\zeta(k_3) + \cdots\,,
\end{equation}
where the amplitude is
\begin{equation}
    f_{\rm NL}(\beta) = - 2\pi^5 g^2 A_\zeta \left(\frac{\Lambda}{H}\right)^{6}\,\beta \cot\left(\frac{\pi \beta}{2}\right)\,.
\end{equation}
Note that $f_{\rm NL}(\beta)$ in the limit $\beta \to 0$, i.e.~the limit in which the scalar `decompactifies', is finite. This is due to our definition of the coupling. Upon restoring mass dimensions of $\varphi$ and matching the coupling of the non-compact scalar, i.e.~sending $g \to g (f/H)^2$, $f_{\rm NL}(\beta)$ would also diverge. Additionally, we caution that for $\beta \sim 1$, one expects a suppression in the coupling $g$, as discussed in more detail in Appendix~\ref{app:uv_cutoff}. 
We also note that this amplitude diverges at $\beta=2,4,6,\cdots$. This is a manifestation of the fact that for positive integer $\beta$, the poles of $\rho_\mathcal{V}(\Delta)$ overlap with those of $\hat{b}_{\rm NA}(\Delta;u)$. In a previous study \cite{Chakraborty:2023eoq}, it was found that it is necessary to be careful around integer $\beta$ values in order to extract physical effects. Moreover, in Appendix~\ref{app:bulk_cft} we analyse the bulk CFT exchange case which exhibits a qualitatively similar feature when the CFT scaling dimension $\delta$ takes integer values. Specifically we analyse the bispectrum when $\delta=2$ and show that it is no longer built up solely of power-laws, but also contains $\log(u)$ terms. We expect a very similar result for the compact scalar with positive integer $\beta$ as well. Nevertheless, we will leave a careful treatment of these values of $\beta$ for future work. Finally, when $\beta>2$ the analytic scaling dominates and the bispectrum in the squeezed limit is
\begin{equation}
    B_\zeta(k_1, k_2, k_3) \xrightarrow{k_3 \to 0} \left(\frac{k_3}{k_1}\right)^{2}P_\zeta(k_1) P_\zeta(k_3) + \cdots\,,
\end{equation}
which is degenerate with local self-interactions of $\zeta$.

For our final point, note that in general we could have considered a gauge-invariant operator with any frequency $\cos(p \varphi)$, for positive integer $p$. The two-point function for such a vertex operator $\mathcal{V}_p \propto e^{i p \varphi}$ is instead \cite{Chakraborty:2023eoq}
\begin{equation}
    \langle \mathcal{V}_p^{\dagger} \mathcal{V}_p \rangle \propto \exp\left(\frac{p^2 \beta}{2} \lambda\right) \lambda^{p^2 \beta}\,,
\end{equation}
i.e.\ its effect is the same as working with our $\mathcal{V}$ and a rescaled $\beta \to p^2 \beta$. As we have just observed, vertex operators with large $\beta$ behave like heavy operators in that they decay very rapidly as they propagate through spacetime, which implies that they are difficult to produce in the expanding inflationary spacetime. Therefore their effects in the squeezed limit of the primordial bispectrum will be subdominant. In this way, we can see that the predictions for the compact theory are dominated by a single vertex operator.

\section{Discussion and Conclusions}\label{sec:conclusions}
In this paper, we have studied the phenomenology of light compact scalar fields, or axions, which acted as spectators during inflation. Our study is a step in fully characterising the impact of such scalars on the non-Gaussianity of the comoving curvature fluctuations $\zeta$. Specifically, we have argued that to capture the effects of the compact scalar $\varphi$, it is necessary to properly account for its gauge symmetry, and in doing so demonstrated a qualitatively different prediction than that of the non-compact scalar $\sigma$, as was previously found in \cite{Chakraborty:2023eoq}. We targeted the squeezed limit of the bispectrum as a particular observable, and subsequently found that working with the gauge invariant vertex operators $\mathcal{V}$ produced the novel scaling
\begin{equation}
    B_\zeta(k_1, k_2, k_3) \sim \left(\frac{k_3}{k_1}\right)^{\beta} + \left(\frac{k_3}{k_1}\right)^{2}
\end{equation}
in the squeezed $k_3 \to 0$ limit. The bispectrum decays with an exponent set by the ratio $H/(2\pi f)$, i.e. the ratio of the inflationary Hubble scale to the field-space circumference of $\varphi$. Since we are required to work with a highly composite operator in this theory, the light compact scalar can be understood as a strongly-coupled sector. Indeed, the qualitative features of the spectral density $\rho_\mathcal{V}(\Delta)$ in Section~\ref{sec:pure_ds} revealed similarities with another example of a strongly-coupled sector--- the bulk CFT operator. We additionally discuss the bispectrum generated by the CFT operator in Appendix~\ref{app:bulk_cft} and find a result which qualitatively displays a similar scaling in the squeezed limit as the compact theory.

One observable in which the squeezed bispectrum explicitly manifests is the power spectrum of a late-time cosmological tracer, e.g.~galaxy clustering catalogues or line intensity maps. Let us call the fluctuations of this tracer $\delta_t(z,\mb{k})$ at redshift $z$ with power spectrum,
\begin{equation}
    \langle \delta_t(z,\mb{k}) \delta_t(z,\mb{k}')\rangle = (2\pi)^3 \delta_{\rm D}(\mb{k}+\mb{k}') D^2(z) P_t(k)
\end{equation}
where $D(z)$ is called the growth factor since it tracks the growth of fluctuations over time. Recall that for $\beta<2$ the squeezed $\zeta$ bispectrum is given by,
\begin{equation}
    B_{\zeta}(k_1, k_2, k_3) \xrightarrow{k_3 \to 0} f_{\rm NL}(\beta)\left(\frac{k_3}{k_1}\right)^\beta\,.
\end{equation}
In general, primordial non-Gaussianity impacts the clustering of tracers in the late universe \cite{Dalal:2007cu,Matarrese:2008nc}. In the tracer's power spectrum this scaling is reflected as \cite{Baumann:2012bc,Assassi:2015fma,MoradinezhadDizgah:2017szk}
\begin{equation}
    P_t(k) = \left[b_1^2 + b_\zeta f_{\rm NL}(\beta)k^{\beta-2}\right] P_{m}(k)\,,
\end{equation}
where $P_m(k)$ is the dark matter power spectrum, $b_1$ is the linear bias and $b_\zeta$ is known as the scale-dependent bias which is generated by non-Gaussianities in $\zeta$.\footnote{For the reader who is unfamiliar with these terms---the linear and scale-dependent biases can be understood as free parameters which track our ignorance of astrophysical processes involved with the production of the specific tracer being considered. For example, if our tracers are galaxies, these parameters capture any unknown physics involved with galaxy formation.} The scale-dependent bias is a useful probe of primordial non-Gaussianity as it causes the power spectrum $P_t(k)$ to grow in magnitude for longer wavelengths, i.e. in the $k\to 0$ limit. When $\beta>2$ the situation is equivalent to setting $f_{\rm NL}=0$ as the power spectrum is regular in the limit $k \to 0$, and moreover the overall amplitude is suppressed, since the coupling $g$ is exponentially suppressed in $\beta$ in this regime. Therefore we have a range of exponents $\beta \in (0,2)$ which have the potential to be measured by cosmological surveys. 

Unfortunately, the compact scalar bispectrum is not unique in this regard. We have also calculated the bispectrum generated by the operator $\sigma^2$. Assume we keep the mass $m_\sigma$ non-zero, but still in the complementary series so that the scaling dimension $\Delta_\sigma \in (0, \tfrac{3}{2})$. Recall that $\hat{\mathcal{B}}_{\sigma^2}(u) \sim u^{2\Delta_\sigma-1}$ for small but non-zero $\Delta_\sigma$. This produces a scale-dependent bias $\delta P_t(k) \sim k^{2\Delta_\sigma-2}$, where $2\Delta_\sigma \in (0,3)$. Therefore this theory is also capable of generating the same numerical exponent in the scale-dependent bias term. As we discuss in Appendix~\ref{app:bulk_cft}, the exchange of a bulk CFT operator is also capable of generating the same squeezed bispectrum. Seen from the opposite perspective, this also demonstrates that a measurement of the exponent in the squeezed bispectrum can correspond to vastly different physical mechanisms. For the non-compact scalar we measure $\Delta_\sigma$, or the mass, and for the compact scalar the exponent measures the ratio of the inflationary Hubble scale to the circumference of the compact scalar's field space $2\pi f$. Therefore one must be careful while interpreting a measurement of this exponent with future cosmological surveys.\footnote{We thank John Stout for highlighting this point.}

This does not of course preclude the possibility of distinguishing between these scenarios using a complete bispectrum template which is valid for other triangular configurations. In other words, measuring subleading terms in the series expansion of $\hat{\mathcal{B}}_{\mathcal{O}}^{\rm NA}(u)$, which as we have seen are fixed by the spectral density $\rho_\mathcal{O}(\Delta)$, provides sufficient information to specify the operator $\mathcal{O}$. In Appendix~\ref{app:eft_seed} we find preliminary signs that for the compact scalar case, the non-analytic piece $\hat{\mathcal{B}}^{\rm NA}_{\mathcal{V}}$ is of comparable size to the EFT piece $\hat{\mathcal{B}}^{\rm EFT}_{\mathcal{V}}$ even for $\beta>2$. This suggests that utilizing all available triangles can allow us to identify the vertex operator exchange. Of course, this is only practical insofar as systematics permit so we leave a careful investigation for future work.

This work has set the foundations for studying the effects of light compact scalars, and many of the results can trivially be extended to theories with multiple compact scalars as well. In more realistic quantum gravity models, we typically encounter not one but many such compact scalars with decay constants $\beta_i$ ranging across many orders of magnitude \cite{Svrcek:2006yi,Arvanitaki:2009fg,Mehta:2021pwf,Demirtas:2018akl,Demirtas:2021gsq,Gendler:2023kjt}. These scalars typically mix very weakly and we can therefore expect them to combine linearly at the level of the $\zeta$-bispectrum 
\begin{equation}
    B_\zeta(k_1, k_2, k_3) \sim \sum_{i|\beta_i<2} f_{\rm NL}(\beta_i) \left(\frac{k_3}{k_1}\right)^{\beta_i}\,,
\end{equation}
and therefore also at the level of the tracer's power spectrum
\begin{equation}
    P_t(k) = \left[b_1^2 + \sum_{i| \beta_i<2}b_\zeta f_{\rm NL}(\beta_i)k^{\beta_i-2}\right] P_{m}(k)\,.
\end{equation}
Depending on the density of decay constants in a given window of mass scales, this may change the scaling with $k$ from a power law entirely. We leave a more careful examination for future work. 

This paper is a step in characterising the inflationary phenomenology of spectator axions during inflation. More work needs to be done in order to be in a position to fully exploit the available cosmological data to constrain such models. As discussed above, there are many interesting directions to explore. We summarize them below:
\begin{itemize}
    \item Our focus here has been on the squeezed primordial bispectrum, using which we identified an interesting range of axion decay constants to be $\beta<2$ or $f> \frac{1}{4\pi}H$. It would be interesting to calculate the bispectrum for all triangular configurations which may allow us to identify other configurations in order to detect a light compact scalar. We have taken the first steps towards this goal in Appendix~\ref{app:eft_seed}. In this vein, the techniques used in this paper can also be trivially extended to calculate the primordial trispectrum as well. We leave a complete exploration of these targets for future work.
    \item A related aspect which deserves further investigation is the flattened configuration, i.e. when $u \to 1$. For the compact scalar, it is unclear if the bispectrum exhibits a singularity in this configuration or not. As discussed more thoroughly in Appendix~\ref{app:eft_seed}, this singularity is typically expected to be absent for any bispectrum in the Bunch-Davies state. However, the compact scalar is a very composite operator and may not satisfy this expectation. In order to study this, it would be necessary to carefully renormalise the EFT piece of the seed function. We leave this for future work.
    \item In this paper, we have worked with a toy model for the compact scalar field. In more realistic scenarios, there may be several compact scalars active during inflation. It would be interesting to apply the results developed here to different UV models in order to develop a more realistic phenomenological picture, and to particularly assess the detectability of these signals when $\beta \sim 1$. Moreover, it would be interesting to place constraints on such compact scalars using \textit{Planck} CMB data and galaxy clustering data, e.g.~the \textit{SDSS-BOSS} catalogue. We leave this for future work.
\end{itemize}

\textbf{Acknowledgements}\newline We are especially indebted to Matt Reece and John Stout for encouragement and many helpful conversations. We are also grateful to Timothy Cohen, Daniel Green and Yiwen Huang for many useful conversations and collaboration on related work. We also thank Aur\'elien Dersy, Hayden Lee, Rashmish Mishra, Joshua Sandor and Ahmed Sheta for helpful discussions. We are grateful to Hayden Lee, Matt Reece and John Stout for feedback on a draft. We also thank an anonymous referee for useful feedback. This work is supported by the DOE Grant {\tt DE-SC0013607}.

\appendix
\section{In-In Diagrams}\label{app:inin}
In this section we will provide a detailed derivation of the bispectrum exchange diagram calculated in the main text. We will first review the derivation of the tree-level seed function for the \textit{trispectrum}, thereby reviewing the work of \cite{Arkani-Hamed:2018kmz}, and using that obtain the same for the bispectrum $\hat{b}(\Delta;u)$. We will then demonstrate the details of the in-in calculation for the exchange of the composite operator $\mathcal{O}$ which were omitted in the main text.

\subsection{Tree-Level Seed Function}
We will begin our discussion with the tree-level seed function for the trispectrum. We will recount the $s$-channel trispectrum of conformally coupled scalars $\confsc$, generated by the exchange of a free-field scalar with scaling dimension $\Delta$, which we call $F(k_1, k_2, k_3, k_4;s)$. This is convenient for us since several different computations for this function exist in the literature \cite{Arkani-Hamed:2018kmz,Werth:2024mjg}. Setting the coupling to unity for convenience, the corresponding in-in integral for the $s$-channel is
\begin{equation}\label{eq:4pt_seed_inin}
    \begin{aligned}
        F(k_1 ,k_2 ,k_3, k_4; s)&= \sum_{{\tt a,b }=\pm}({\tt ab})\int_{-\infty(1\mp i \epsilon)}^{0}\frac{\ud \eta}{\eta^2}\, \int_{-\infty(1\mp i \epsilon)}^{0}\frac{\ud \eta'}{\eta'^2} \, e^{-{\tt a} i (k_1+k_2) \eta}e^{-{\tt b} i (k_3+k_4) \eta'} G^{\tt ab}(\Delta;s;\eta,\eta')\,,
    \end{aligned}
\end{equation}
where $s \equiv |\mb{k}_1+\mb{k}_2|$ is the momentum of the exchanged particle. Note that the free propagators $G^{\tt ab}(\Delta;s;\eta,\eta')$ are factorizable into functions of $s \eta$ and $s \eta'$. Therefore rescaling $\eta \to s^{-1}\eta$ and $\eta' \to s^{-1} \eta'$ shows that the kinematic variables fixing the form of $F$ are the so-called momentum cross-ratios $u=s/(k_1+k_2)$ and $v=s/(k_3+k_4)$. Specifically, it can be shown that $F(\{k_i\};s)=s^{-1} \hat{F}(u,v)$ where $\hat{F}$ is called the trispectrum seed function. 

In \cite{Arkani-Hamed:2018kmz} it was shown that any correlator on the late-time boundary of de Sitter is constrained by conformal invariance, which is the more fundamental way of understanding why $F(k_1,k_2,k_3,k_4;s)$ only depends on the cross-ratios $u$ and $v$. The conformal Ward identities on the boundary lead to a differential equation for the seed function
\begin{equation}
    (\Delta_u-\Delta_v)\hat{F}(u,v) = 0\,,
\end{equation}
where the differential operator $\Delta_u=u^2(1-u^2)\partial_u^2 -2 u^3 \partial_u$, and likewise for $v$. For the tree-level exchange of a massive scalar, one can derive two second order differential equations. These can be straightforwardly obtained by manipulating the equations of motion for the massive scalar. The first, in terms of $u$, is
\begin{equation}\label{eq:tree_seed_eq}
    (\Delta_u+m^2-2)\hat{F}(u,v) = \frac{u v}{u+v}\,.
\end{equation}
There is also an identical second equation in $v$. The space of solutions must be restricted by implementing several boundary conditions. First, the choice of the Bunch-Davies vacuum is enforced by requiring that $\lim_{u\to 1}\hat{F}(u,v)$ must be finite \cite{Green:2020whw,Holman:2007na}. This corresponds to the so-called folded configuration which must be regular for correlators in the Bunch-Davies vacuum \cite{Flauger:2013hra,Green:2020whw}. Another condition we must impose on the solution is symmetry under the exchange $u\leftrightarrow v$, which can be seen to be true from the bulk in-in integrals. Lastly, we need to demand that the correlator factorizes into a product of three-particle amplitudes as we send $u,v \to -1$, which is known as the total energy singularity.

Imposing these conditions with the help of the homogenous solutions of (\ref{eq:tree_seed_eq}) provides the complete solution
\begin{equation}
    \hat{F}(u,v) = \frac{\pi}{2 \cosh(\pi \nu)}\hat{F}_{\rm NA}(u,v) + \hat{F}_{\rm EFT}(u,v)\,.
\end{equation}
The solution is conveniently separated into a term which is non-analytic in $u$ and $v$ and therefore knows about particle production in the bulk, and an EFT part which is analytic in $u$ and $v$. The EFT part is organised into a power series
\begin{equation}
    \hat{F}_{\rm EFT}(u,v) = \sum_{m,n=0}^{\infty}c_{mn} u^{2m+1}\left(\frac{u}{v}\right)^{n}\,,\qquad \text{when }u\leq v\,,
\end{equation}
and the coefficients are
\begin{equation}
    c_{mn}(\nu) \equiv \frac{(-1)^n (n+1)(n+2)\cdots (n+2m)}{\left[(n+\tfrac{1}{2})^2+\nu^2\right]\left[(n+\tfrac{5}{2})^2+\nu^2\right]\cdots \left[(n+\tfrac{1}{2}+2m)^2+\nu^2\right]}\,.
\end{equation}
An alternative representation of the EFT piece will be especially useful for our purposes. We will invoke the form calculated in \cite{Werth:2024mjg} which evaluated the bulk integrals instead of solving the boundary differential equations to obtain
\begin{equation}\label{eq:werth_eft}
    \hat{F}_{\rm EFT}(u,v) = \sum_{n\geq 0}\frac{(-1)^{n}}{(n+\tfrac{1}{2})^2 + \nu^2} u \left(\frac{u}{v}\right)^{n} \tFo{\tfrac{n+1}{2}}{1+\tfrac{n}{2}}{\tfrac{3}{2}+n}{u^2} \, \tFo{\tfrac{1-n}{2}}{-\tfrac{n}{2}}{\tfrac{1}{2}-n}{v^2}\,.
\end{equation}
It can be checked numerically that this expression agrees with the double-sum from \cite{Arkani-Hamed:2018kmz}. The non-analytic piece is
\begin{equation}\label{eq:Fhat_na}
    \begin{aligned}
        \hat{F}_{\rm NA}(u,v) &= \hat{F}_{+}(u) \hat{F}_{-}(v)-\hat{F}_{-}(u) \hat{F}_{+}(v) - \frac{\alpha_{-}}{\alpha_{+}}(\beta_0+1)\hat{F}_{+}(u) \hat{F}_{+}(v) \\
        & - \frac{\alpha_{+}}{\alpha_{-}}(\beta_0-1)\hat{F}_{-}(u) \hat{F}_{-}(v) + \beta_0\left[\hat{F}_{-}(u) \hat{F}_{+}(v)+\hat{F}_{+}(u) \hat{F}_{-}(v)\right]\,,
    \end{aligned}
\end{equation}
where
\begin{equation}
    \begin{aligned}
        \hat{F}_{\pm}(z) \equiv &\, \left(\frac{i z}{2\nu}\right)^{\frac{1}{2}\pm i \nu}  \tFo{\tfrac{1}{4}\pm i \tfrac{\nu}{2}}{\tfrac{3}{4}\pm i \tfrac{\nu}{2}}{1 \pm i \nu}{z^2}\,, \\
        \alpha_{\pm} \equiv &\, -\left(\frac{i}{2\nu}\right)^{\frac{1}{2}\pm i \nu}\frac{\Gamma(1\pm i \nu)}{\Gamma(\tfrac{1}{4}\pm i \tfrac{\nu}{2})\Gamma(\tfrac{3}{4}\pm i \tfrac{\nu}{2})}\,,\quad \beta_0 \equiv \frac{1}{i \sinh(\pi \nu)}\,.
    \end{aligned}
\end{equation}
Note that the non-analytic piece can be reorganised into a manifestly shadow symmetric form
\begin{equation}
    \hat{F}_{\rm NA}(u,v) = \hat{F}_{+}(u) \hat{F}_{-}(v) - \frac{\alpha_{-}}{\alpha_{+}}(\beta_0+1)\hat{F}_{+}(u) \hat{F}_{+}(v)  + \beta_0\hat{F}_{+}(u) \hat{F}_{-}(v) + (\nu \leftrightarrow -\nu)\,.
\end{equation}
Taking the $v \to 1$ limit of (\ref{eq:Fhat_na}) and (\ref{eq:werth_eft}) leads to the expressions for $\hat{b}_{\rm NA}(\Delta; u)$ and $\hat{b}_{\rm EFT}(\Delta; u)$ provided in the main text.

\subsection{In-In Diagrams for Composite Operator Exchange}
The leading perturbative contribution to the $\pi_c$ bispectrum due to the interactions in (\ref{eq:Sint}) can be determined using the standard diagrammatic rules to be
\begin{equation}\label{eq:treeLevelBS}
    \def\circSize{0.6}
    \begin{aligned}
        &B_\pi(k_1, k_2, k_3) = \begin{tikzpicture}[thick, baseline=-28pt]
        \coordinate (c1) at (-1.0, -1.75);
        \coordinate (c2) at (1.0, -1.75);
        \coordinate (c3) at (-1.75, 0);
        \coordinate (c4) at (-0.25, 0);
        \coordinate (c5) at (1.75, 0);

        \begin{scope}[shift={(0, -1.75)}]
            
            \draw (c1) -- (c2) node[midway, below] {$\mathcal{O}$};
        \end{scope}

        \draw[inflSty] (c1) -- (c3) node[midway, shift={(-0.3, 0)}] {$\mb{k}_1$};
        \draw[inflSty] (c4) -- (c1) node[midway, shift={(0.4, 0)}] {$\mb{k}_2$};
        \draw[inflSty] (c2) -- (c5) node[midway, shift={(0.4, 0)}] {$\mb{k}_3$};
        \draw[line width=0.6mm, gray] (-2.75, 0) -- (2.75, 0);
        \fill[intSty] (c1) circle (0.07) ;
        \fill[intSty] (c2) circle (0.07) ;
        \fill[cornellRed, intSty] (c3) circle (0.07);
        \fill[cornellRed, intSty] (c5) circle (0.07);

        \end{tikzpicture}  + \text{cyc.} \\
        &= \sum_{{\tt a,b}=\pm}({\tt ab})\left(-\frac{g^2 f_\pi^{-6}}{2}\right) \int_{-\infty}^{\eta_0}\ud \eta\, a^{2}(\eta) \int_{-\infty}^{\eta_0}\ud \eta'\, a^{3}(\eta') K_{\tt a}'(k_1;\eta) K_{\tt a}'(k_2;\eta) K_{\tt b}'(k_3;\eta') G_{\mathcal{O}}^{\tt ab}(k_{3};\eta,\eta') +\text{cyc.}\\
        &\equiv \frac{g^2 f_\pi^{-6} H}{16 k_1 k_2 k_3} \mathcal{B}_{\mathcal{O}}(k_1,k_2,k_3) + \text{cyc.}\,,
    \end{aligned} 
\end{equation}
where, anticipating no late-time infrared divergences, in the last line we have taken $\eta_0 \to 0$. Here 'cyc.' denotes permutations of the momenta $(k_1,k_2,k_3)$. We define
\begin{equation}
    \mathcal{B}_{\mathcal{O}}(k_1,k_2,k_3) \equiv  \sum_{{\tt a,b }=\pm}({\tt ab})\int_{-\infty(1\mp i \epsilon)}^{0}\ud \eta\, \int_{-\infty(1\mp i \epsilon)}^{0}\frac{\ud \eta'}{\eta'^2} \, e^{-{\tt a} i (k_1+k_2) \eta}e^{-{\tt a} i k_3 \eta'} G_{\mathcal{O}}^{\tt ab}(k_{3};\eta,\eta')\,.
\end{equation}
Our goal now is to evaluate this in-in integral. We can resolve the propagators $G_{\mathcal{O}}^{\tt ab}(k;\eta,\eta')$ into their spectral representation and exchange the time integrals with the spectral integral to produce
\begin{equation}
    \mathcal{B}_{\mathcal{O}}(k_1, k_2, k_3;\eta_0) = \int_{\tfrac{3}{2}-i\infty}^{\tfrac{3}{2}+i\infty}\frac{\ud \Delta}{2\pi i} \rho_{\mathcal{O}}(\Delta) b(\Delta; k_1, k_2, k_3)\,,
\end{equation}
where the tree-level bispectrum is
\begin{equation}
    b(\Delta;k_1,k_2,k_3) = \sum_{{\tt a,b }=\pm}({\tt ab})\int_{-\infty(1\mp i \epsilon)}^{0}\ud \eta\, \int_{-\infty(1\mp i \epsilon)}^{0}\frac{\ud \eta'}{\eta'^2} \, e^{-{\tt a} i (k_1+k_2) \eta}e^{-{\tt b} i k_3 \eta'} G^{\tt ab}(\Delta;k_{3};\eta,\eta')\,.
\end{equation}
The tree-level bispectrum is highly constrained by the de Sitter isometries. Its functional form only depends on the momentum cross-ratio $u\equiv k_3/(k_1 + k_2)$. It is related to the four-point function discussed previously by
\begin{equation}\label{eq:bisp_seed_from_4pt_seed}
    b(\Delta; k_1, k_2, k_3) = -\lim_{k_4 \to 0}\lim_{s \to k_3}\frac{\partial^2}{\partial(k_1+k_2)^2}F(k_1, k_2, k_3, k_4;s)\,.
\end{equation}
Therefore we have
\begin{equation}
    b(\Delta; k_1, k_2, k_3) = -\frac{k_3}{k_{12}^3}(u\partial_u^2 + 2 \partial_u)\hat{b}(\Delta;u)\,,
\end{equation}
where $\hat{b}(\Delta;u)$ is the so-called tree-level seed function and corresponds to the equivalent exchange bispectrum for conformally coupled external fields $\confsc$. As a result $\mathcal{B}_{\mathcal{O}}(k_1,k_2,k_3)$ can be expressed as
\begin{equation}
    \mathcal{B}_{\mathcal{O}}(k_1, k_2, k_3) = -\frac{1}{k_{12}^3}(u\partial_u^2 + 2 \partial_u)\hat{\mathcal{B}}_{\mathcal{O}}(u)\,,
\end{equation}
where
\begin{equation}
    \hat{\mathcal{B}}_{\mathcal{O}}(u) \equiv \int_{\tfrac{3}{2}-i\infty}^{\tfrac{3}{2}+i\infty}\frac{\ud \Delta}{2\pi i} \rho_{\mathcal{O}}(\Delta) \hat{b}(u)\,.
\end{equation}
With this we can trivially determine the $\pi_c$ bispectrum to be
\begin{equation}
    B_{\pi}(k_1, k_2, k_3) = -\frac{g^2 f_\pi^{-6} H}{16 k_1 k_2 k_3}\frac{1}{k_{12}^3}(u\partial_u^2 + 2 \partial_u)\hat{\mathcal{B}}_{\mathcal{O}}(u)\,,
\end{equation}
which is the expression used in the main text.

\section{EFT Piece of the Seed Function}\label{app:eft_seed}
\begin{figure}
    \centering
    \includegraphics[width=0.4\textwidth]{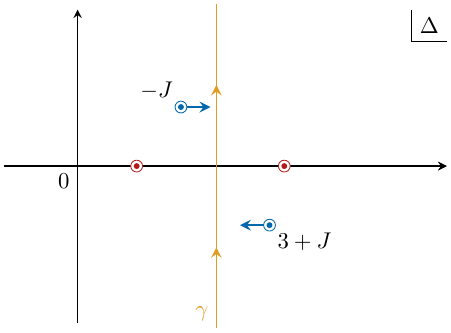}
    \includegraphics[width=0.4\textwidth]{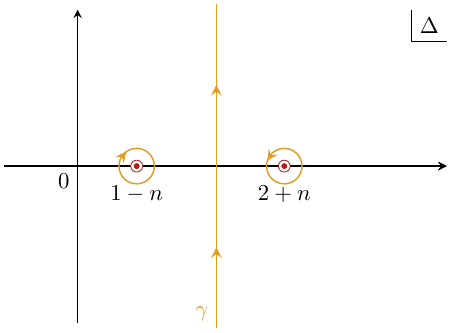}
    \caption{Contour deformation to relate the momentum coefficients with the EFT coefficients $\mathcal{C}_n^{\mathcal{O}}$. Sending $-J \to n+2$ pinches the contour twice, picking up two residues along the way.}
    \label{fig:mom_to_eft_cn}
\end{figure}

In this section we will work out the details of the EFT contribution which were omitted in the main text. There, we showed that the EFT expansion was fixed by the coefficients
\begin{equation}\label{eq:eft_cn_integ}
    C_n^{\mathcal{O}} = \int_{\tfrac{3}{2}-i\infty}^{\tfrac{3}{2}+i\infty}\frac{\ud \Delta}{2\pi i}\, \frac{\rho_{\mathcal{O}}(\Delta)}{(\Delta-n-2)(1-n-\Delta)}\,.
\end{equation}
It is helpful to recall (\ref{eq:spec_density_to_mom_coeff}) to connect this integral with the momentum coefficients
\begin{equation}
    [G_\mathcal{O}]_J = \int_{\tfrac{3}{2}-i\infty}^{\tfrac{3}{2}+i\infty}\frac{\ud \Delta}{2\pi i}\, \frac{\rho_{\mathcal{O}}(\Delta)}{(J+\Delta)(J+\bar{\Delta})}\,,
\end{equation}
where we are assuming $-J$ lies to the left of the principal series line and $3+J$ to the right. To connect the EFT coefficients to the momentum coefficients we need to send $J \to -2,-3, \cdots$, which requires passing through the contour as shown in Figure~\ref{fig:mom_to_eft_cn}. In doing so we will pick up residues at $\Delta=-J$ and $\Delta=3+J$ to produce
\begin{equation}\label{eq:eft_cn_mom}
    C_n^{\mathcal{O}} = [G_{\mathcal{O}}]_{\minus(n+2)} - \left[{\rm Res}_{\Delta=2+n}-{\rm Res}_{\Delta=-n-1}\right]\frac{\rho_\mathcal{O}(\Delta)}{(\Delta-n-2)(\bar{\Delta}-n-2)}\,,
\end{equation}
where we have accounted for the residues accumulated by the $\Delta=-J$ and $\Delta=J+3$ poles crossing the contour as we send $J \to -2,-3,\cdots$.
This is an intriguing result and a convenient tool to organise our calculation. The first thing we learn from this is that since the momentum coefficients are generically UV divergent, the EFT coefficients must be as well. This is good since the EFT terms can be mimicked by local self-interactions of the inflaton, and therefore can be easily renormalised by counterterms. However, these terms will be subdominant in the infrared, or the squeezed limit, so the complete result is not particularly interesting for our purposes. 

An important role played by the EFT piece at tree-level is to cancel an unphysical singularity as $u\to 1$, which is known to correspond to our choice of the Bunch-Davies vacuum. To see this we note that as $u \to 1$ the non-analytic piece produces a singularity
\begin{equation}
    \frac{1}{2}\hat{b}_{\rm NA}(\Delta;u) + (\Delta \leftrightarrow \bar{\Delta}) \xrightarrow{u \to 1^-} \frac{\pi}{2}\sech(\pi \nu) \log(1-u)\,.
\end{equation}
The EFT piece exhibits an equal and opposite singularity
\begin{equation}
    \begin{aligned}
        \hat{b}_{\rm EFT}(\Delta;u) &\xrightarrow{u \to 1^-} \log (1-u)\sum_{n=0}^{\infty} \frac{1}{2}(-1)^{n+1} (2 n+1) \frac{1}{(\Delta+n-1)(n+2-\Delta)} \\
        &= -\frac{\pi}{2}\sech(\pi \nu) \log(1-u)\,,
    \end{aligned}
\end{equation}
where we have isolated the leading divergence in the summand in the $u\to 1$ limit. Thus we can see that the relative normalisation of the EFT and non-analytic pieces is important in order to cancel this flattened singularity. 

Already this aspect of the EFT piece presents an interesting point. As we have seen with our general discussion concerning the exchange of a generic composite operator $\mathcal{O}$, the coefficients which determine the EFT expansion are generically UV divergent and must be renormalised appropriately. Therefore the cancellation of this flattened singularity for a generic case is non-trivial.\footnote{We are grateful to Xingang Chen and Anson Hook for related discussions.} 

Now we will determine the EFT coefficients using these general expressions for the non-compact and compact scalars.

\subsection{Non-Compact Scalar}
Here we will not worry about determining the EFT expansion exactly, but instead focus on making some qualitative remarks. Particularly we will only make a statement about the renormalisation of the bispectrum and therefore are only interested in isolating the UV divergence.

As outlined, there are two ways of calculating the EFT coefficients--- we can either perform the integral (\ref{eq:eft_cn_integ}) or calculate the momentum coefficients. Fortunately, the momentum coefficients for a quadratic operator $\sigma \varphi$ operator in general spacetime dimension were calculated in \cite{Marolf:2010zp} in terms of a ${}_7 F_6[\cdots]$ hypergeometric function, and it is therefore straightforward to regulate these coefficients in dimensional regularisation. Unfortunately, the equivalent form of $\hat{b}_{\rm EFT}(\Delta; u)$ is not known to us in general spacetime dimensions, so we leave a complete calculation of these terms for the future. 

The UV divergence for $\alpha = \frac{3-\epsilon}{2}$ dimensions is
\begin{equation}
    [G_{\sigma^2}]_J \xrightarrow{\epsilon \to 0} \frac{1}{8\pi^2 \epsilon} + \mathcal{O}(\epsilon)\,,
\end{equation}
which leads to a UV divergent contribution to the bispectrum
\begin{equation}
    \hat{\mathcal{B}}^{\rm EFT}_{\sigma^2}(u) = \frac{1}{8\pi^2 \epsilon}\sum_{n=0}^\infty \left(\minus \frac{1}{2}\right)^{n}\frac{\sqrt{\pi} \,n!}{\Gamma(n+\tfrac{1}{2})} u^{n+1} \tFo{\frac{n+1}{2}}{\frac{n+2}{2}}{n+\tfrac{3}{2}}{u^2} + \mathcal{O}(\epsilon)\,.
\end{equation}
While this expression looks complicated it can be checked by numerically performing the sum that it is merely a rewriting of
\begin{equation}
    \hat{\mathcal{B}}^{\rm EFT}_{\sigma^2}(u) = \frac{1}{8\pi^2 \epsilon} \frac{u}{1+u} + \mathcal{O}(\epsilon)\,.
\end{equation}
We could also have guessed this from previous calculations of the trispectrum generated by the exchange of $\sigma^2$, e.g. \cite{Xianyu:2022jwk}. There it was determined that the UV divergence is,
\begin{equation}
    \hat{F}_{\sigma^2}(u,v) = \frac{1}{8\pi^2 \epsilon} \frac{uv}{u+v} + \cdots
\end{equation}
Because setting $v \to 1$ produces the bispectrum seed function, we see that our calculation produces the same result. 

As a cross check, let us try to recover this divergence with a hard cutoff. We will therefore compute (\ref{eq:eft_cn_integ}) up to $\nu \leq \bar{\nu}$, i.e.\ we will kill all states heavier than $\bar{\nu}$. In four dimensions, the spectral density grows as $\rho_{\sigma^2}(\nu) \sim \frac{\nu}{8\pi}$ using which we can isolate the leading logarithmic divergence
\begin{equation}
    \mathcal{C}^{\sigma^2}_{n} = \int_{-\bar{\nu}}^{\bar{\nu}} \frac{\ud \nu}{2\pi} \frac{1}{(n+\tfrac{1}{2})^2+\nu^2} \rho_{\sigma^2}(\nu) \xrightarrow{\bar{\nu} \to \infty} \frac{1}{8\pi^2}\log{\bar{\nu}} + \mathcal{O}(\bar{\nu}^{-1})\,,
\end{equation}
which is consistent with the $\epsilon$-divergence we obtained using the momentum coefficients.

\begin{figure}
    \centering
    \includegraphics[width=0.8\textwidth]{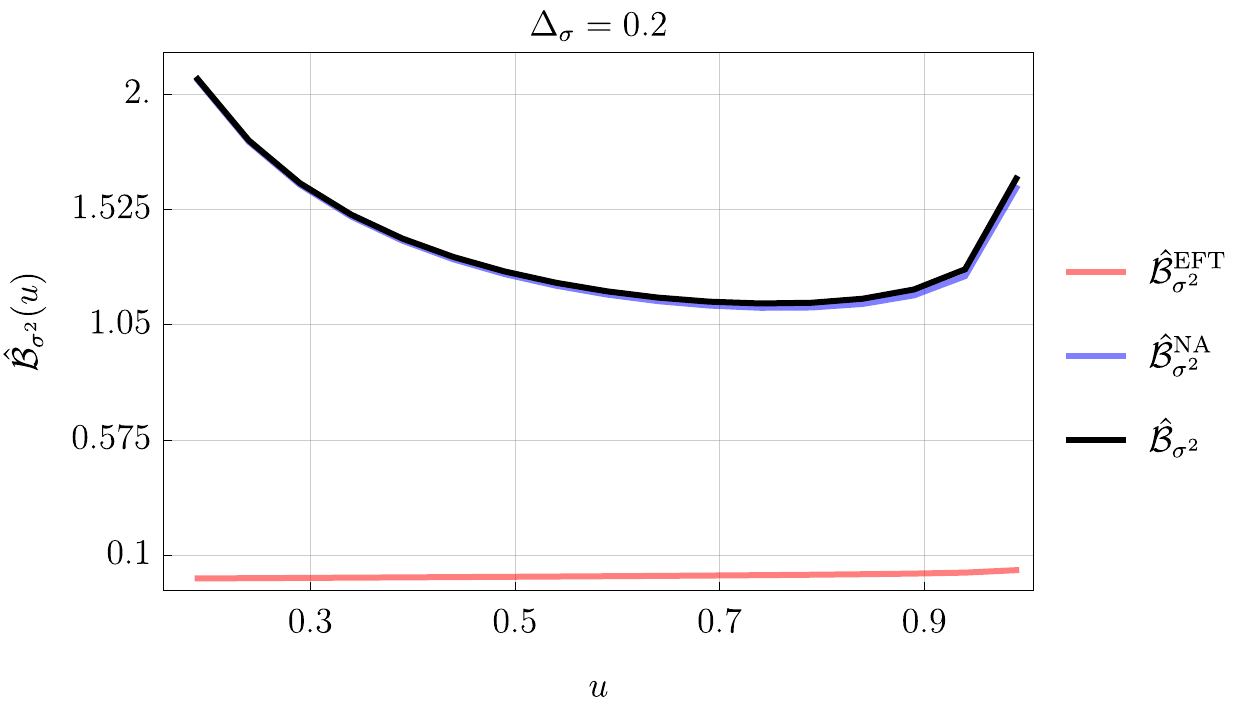}
    \caption{We plot the renormalised seed function for the non-compact scalar $\hat{\mathcal{B}}_{\sigma^2}(u)$ for scaling dimension $\Delta_\sigma=0.2$. We show both the EFT and non-analytic parts separately from the full seed function, i.e.\ the sum of the two.}
    \label{fig:sig2_full_seed}
\end{figure}

Using the hard-cutoff it is possible, albeit inefficient, to numerically regulate the EFT coefficients. Specifically we can define the renormalised coefficients as
\begin{equation}
    \bar{\mathcal{C}}^{\sigma^2}_{n} = \left[\int_{-\bar{\nu}}^{\bar{\nu}} \frac{\ud \nu}{2\pi} \frac{\rho_{\sigma^2}(\nu)}{(n+\tfrac{1}{2})^2+\nu^2} \right] - \frac{1}{8\pi^2}\log{\bar{\nu}}
\end{equation}
for sufficiently large $\bar{\nu}$ so that the renormalised EFT piece of the seed function is
\begin{equation}
    \hat{\mathcal{B}}^{\rm EFT,ren}_{\sigma^2}(u) = \sum_{n=0}^\infty \bar{\mathcal{C}}^{\sigma^2}_{n}\left(\minus \frac{1}{2}\right)^{n}\frac{\sqrt{\pi} \,n!}{\Gamma(n+\tfrac{1}{2})} u^{n+1} \tFo{\frac{n+1}{2}}{\frac{n+2}{2}}{n+\tfrac{3}{2}}{u^2} \,.
\end{equation}
We plot the complete $\hat{\mathcal{B}}_{\sigma^2}(u)$ in Figure~\ref{fig:sig2_full_seed}. For small masses the seed function is dominated by the non-analytic piece. We seem to observe a singularity when $u \to 1$ although it is unclear if this is a numerical artefact or a genuine feature of bispectrum at the loop level.

\subsection{Compact Scalar}
For the compact scalar, it is more convenient to use (\ref{eq:eft_cn_mom}) to determine the EFT coefficients. We will therefore calculate the momentum coefficients for the vertex operator. This was done in \cite{Chakraborty:2023eoq} using dimensional regularization, where it was found that these coefficients are (shockingly) UV finite. There, these coefficients were determined using the Lorentzian inversion formula (\ref{eq:mom_coeff_lorentz}) and the UV finiteness was due to ${\rm disc}\,G_{\mathcal{V}}(\xi)$ being very well behaved on the time-like branch cut ($\xi>1$). Of course, it must be noted that this may be scheme dependent. Dimensional regularization is famously only sensitive to scale-less or logarithmic divergences \cite{Peskin:1995ev}, and may be blind to the divergences we would have encountered using another scheme choice. Indeed, the spectral density $\rho_\mathcal{V}(\Delta)$ grows faster than $\Delta$ so, for instance, applying a hard-cutoff to (\ref{eq:spec_density_to_mom_coeff}) may introduce a UV divergence. We will not follow this approach and instead choose to regulate our momentum coefficients another way.

\begin{figure}
    \centering
    \includegraphics[width=0.49\textwidth]{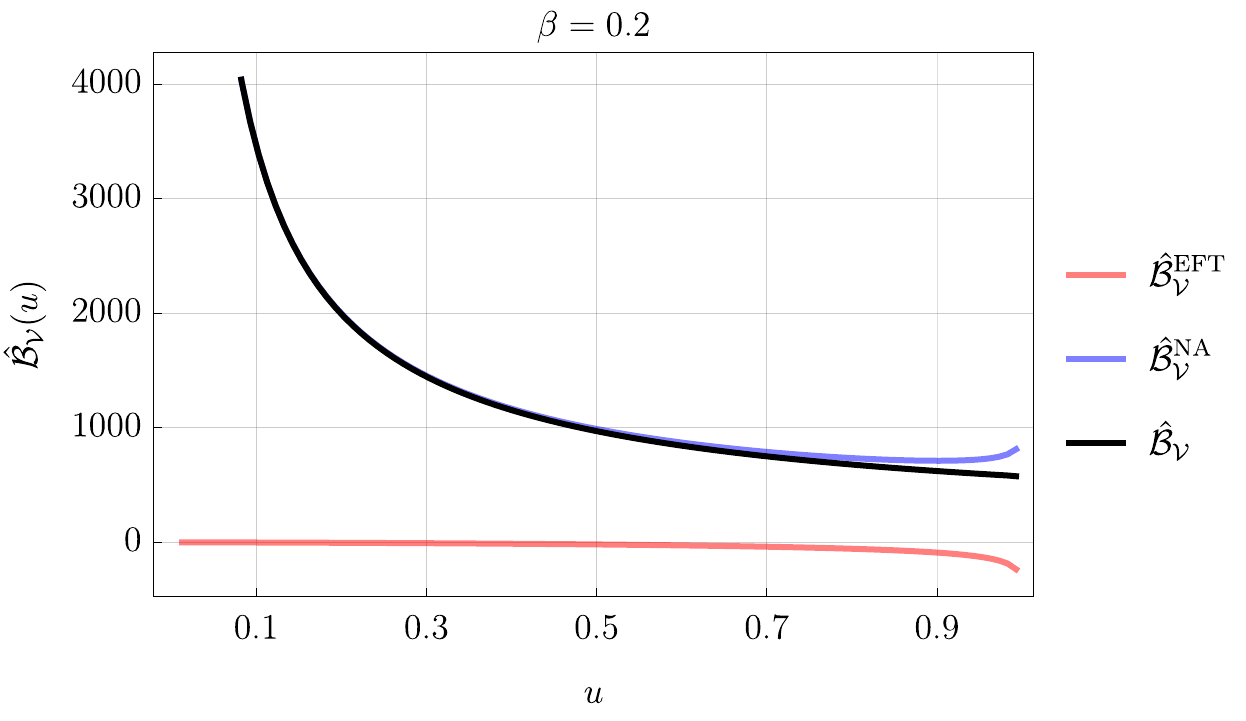}
    \includegraphics[width=0.49\textwidth]{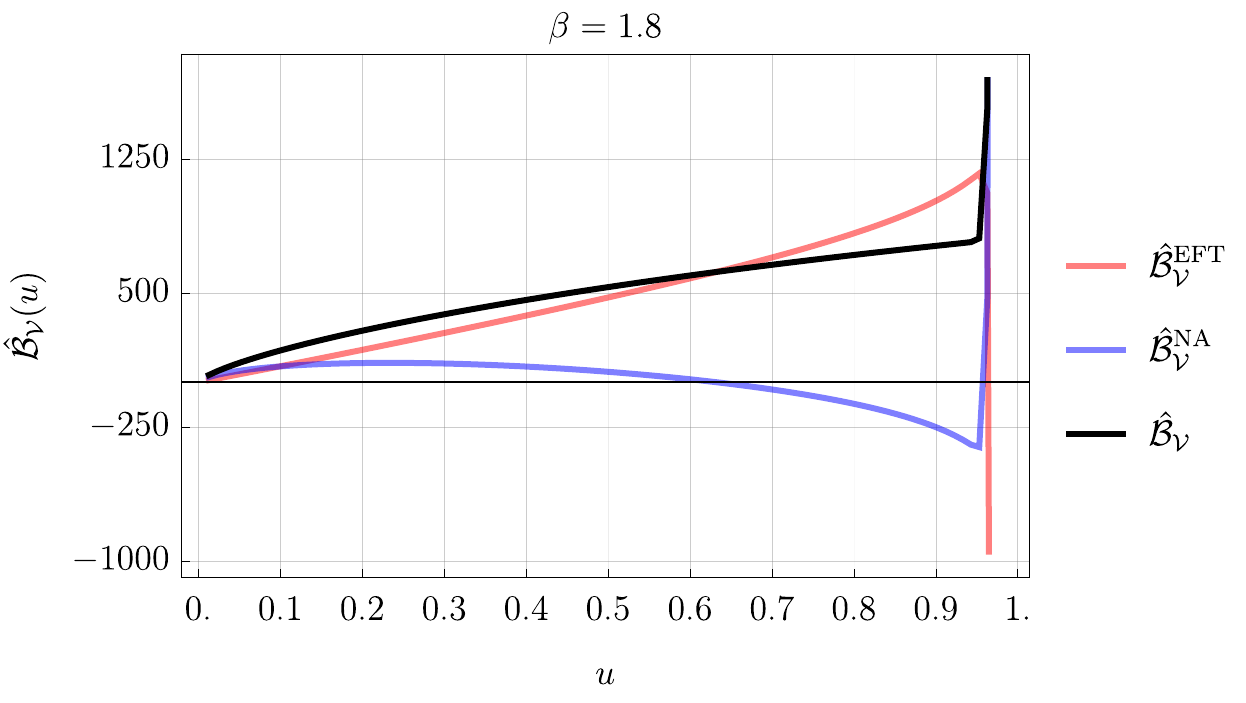}
    \caption{We plot the full seed function for the compact scalar $\hat{\mathcal{B}}_{\mathcal{V}}(u)$ for scaling dimension $\beta=0.2$ and $\beta=1.8$. We show both the EFT and non-analytic parts separately from the full seed function, i.e.\ the sum of the two. We have summed up to $N=100$ residues for the non-analytic piece and likewise in the EFT series.}
    \label{fig:vertex_full_seed}
\end{figure}

We will instead choose to define our momentum coefficients by the same way we calculated the spectral density, i.e.\ as a sum over bulk CFT states. Specifically,
\begin{equation}\label{eq:vert_mom_coeffs}
    \begin{aligned}
        [G_\mathcal{V}]_J &\equiv \, 2^\beta \sum_{n=0}^{\infty} \frac{\beta^n}{n!}[G_{\beta+n}]_J \\
        &\,=-\frac{16\pi^2 \Gamma(1-\beta)}{\Gamma(\beta-1)}\frac{\Gamma(J+\beta)}{\Gamma(4+J-\beta)} \,\tFt{J+\beta}{\beta-3-J}{\beta}{\beta-1}{\frac{\beta}{2}}
    \end{aligned}
\end{equation}
This expression exhibits the correct properties. It has simple poles at the locations one would expect from the spectral density, namely at $J=-\beta,-\beta-2,-\beta-3, \cdots$, and we have checked that the residues at these poles are consistent with the ones determined in \cite{Chakraborty:2023eoq}. Moreover $[G_\mathcal{V}]_J$ also symmetrizes to produce the spectral density calculated in the main text
\begin{equation}
    \rho_\mathcal{V}(\Delta) = \frac{1}{2}(3-2\Delta)[G_{\mathcal{V}}]_{-\Delta} + (\Delta \leftrightarrow \bar{\Delta})\,,
\end{equation}
which implies that the residues at these poles match with those of the spectral density as well. Of course, this does not preclude the possibility that this regularization scheme is insensitive to divergences we would have encountered in another scheme. In other words, these UV divergences are typically shadow symmetric in $J$ and therefore get killed in the spectral density. Supposing, for example, that we were able to evaluate (\ref{eq:eft_cn_integ}) by implementing a hard-cutoff $\bar{\nu}$ we expect to find
\begin{equation}
    [G_\mathcal{V}]_J = -\frac{16\pi^2 \Gamma(1-\beta)}{\Gamma(\beta-1)}\frac{\Gamma(J+\beta)}{\Gamma(4+J-\beta)} \,\tFt{J+\beta}{\beta-3-J}{\beta}{\beta-1}{\frac{\beta}{2}} + f_{J}(\bar{\nu},\beta)\,,
\end{equation}
where $f_{J}(\bar{\nu},\beta)$ is a function such that $f_{-\Delta}(\bar{\nu},\beta)$ is regular on the ${\rm Re}\,(\Delta)>0$ part of the complex plane. It must also satisfy $f_{-\Delta}(\bar{\nu},\beta) = f_{\Delta-3}(\bar{\nu},\beta)$ in order to be killed upon shadow symmetrization to produce the correct spectral density $\rho_\mathcal{V}(\Delta)$. We do not presently have a rigorous proof of this claim and leave a more careful investigation for future work. It is nonetheless helpful to keep in mind as we progress on to the EFT coefficients. Using the momentum coefficients (\ref{eq:vert_mom_coeffs}) in (\ref{eq:eft_cn_mom}) produces
\begin{equation}
    \mathcal{C}_n^{\mathcal{V}} = -[G_{\mathcal{V}}]_{\minus(n + 2)}\,.
\end{equation}
We plot the resulting seed function in Figure~\ref{fig:vertex_full_seed}. While we are missing potential contributions from local counterterms, this solution should nevertheless capture some qualitative features of the bispectrum. For $\beta=0.2$ we can see that the $u\to 0$ limit is singular and the singularity scales as $u^{\beta-1}$ as determined in our analysis of the non-analytic piece in the main text. Therefore $\beta=1.8$ sees no singularity and instead a simple power law decay. Nevertheless, we can see that $\hat{\mathcal{B}}_{\mathcal{V}}(u)$ is not completely determined by the non-analytic piece but in fact receives an appreciable contribution from the EFT piece as well. We also observe that the non-analytic piece is comparable in magnitude to the EFT piece, which holds for $\beta>2$ as well. This is unlike the standard behavior seen for heavy free fields, for which the non-analytic piece is exponentially suppressed as we increase the mass $\nu$.

As with the non-compact case, we may observe a singularity as $u \to 1$ but again it is unclear if that is a numerical artefact or not. We leave a more careful investigation for future work.

\section{Bulk CFT Exchange}\label{app:bulk_cft}
In this section we will apply our main methodology to the exchange of a bulk CFT operator with scaling dimension $\delta>1$. This is useful to do because it serves as a non-trivial example where we can compare with known results at least in some regimes \cite{Green:2013rd}. Additionally, since the CFT operator is extremely constrained (it is simply a power law), it is especially amenable to an analytical treatment. Therefore it will be helpful for us to analyse as a test case as well.

First let us discuss the non-analytic piece. Using the spectral density (\ref{eq:bulk_cft_spect}), specialised to four dimensions ($\alpha=3/2$) we can immediately obtain
\begin{equation}
    \hat{\mathcal{B}}_\delta^{\rm NA} (u)= -\sum_{n \geq 0} \left([{\rm Res}_{\delta + n} \rho_\delta(\Delta)] \hat{b}_{\rm NA}(\delta+n;u)+ [{\rm Res}_{2n+2}\hat{b}_{\rm NA}(\Delta;u)] \rho_\delta(2n+2) \right)\,.
\end{equation}
The complete analytical expression is not particularly illuminating but nevertheless we can put it together by noting that
\begin{equation}
    \begin{aligned}
        {\rm Res}_{\delta + n}\rho_\delta(\Delta) &= 2^{3-\delta}\pi \sin(2\pi \delta)\Gamma(1-\delta)\frac{(2\delta+2n-3)\Gamma (2 \delta +n-3)}{\Gamma(n+1) \Gamma(\delta-1)} \\
        {\rm Res}_{2n+n}\hat{b}_{\rm NA}(\Delta;u) &= 2\pi^{1/2} \frac{\Gamma(2n+1)}{\Gamma(2n+\tfrac{3}{2})}\left(\frac{u}{2}\right)^{1+2n} \tFo{n+1}{n+\tfrac{1}{2}}{2n+ \tfrac{3}{2}}{u^2}\,.
    \end{aligned} 
\end{equation}
To determine the EFT piece we need to compute the momentum coefficients $[G_\delta]_J$. This can be done using the Lorentzian inversion formula (\ref{eq:mom_coeff_lorentz}) and result is
\begin{equation}
    [G_\delta]_J = \frac{\pi ^2 2^{4-\delta } \Gamma (2-\delta )}{\Gamma (\delta )} \frac{\Gamma(J+\delta)}{\Gamma(4+J-\delta)}\,,
\end{equation}
where we have again specialised to four dimensions. Note that the Lorentzian inversion integral strictly only converges for $\delta<2$. In other words, we do not encounter any UV divergences for sufficiently small $\delta$. Qualitatively this is because the CFT correlator in that case is not sufficiently singular to produce UV divergences. Nevertheless, it is possible to analytically continue this answer to larger $\delta$, and this may be interpreted as a regularization scheme. In general we must be wary of missing potential counterterms, as a similar subtlety arises when flat-space scattering amplitudes are computed using such CFT operators \cite{Grinstein:2008qk}.

For the time being let us keep $\delta<2$. The EFT piece is then determined via (\ref{eq:comp_seed_eft}) and the series coefficients can be determined using (\ref{eq:eft_cn_mom})
\begin{equation}
    \mathcal{C}^{\delta}_n = -[G_\delta]_{-(n+2)}\,.
\end{equation}

The squeezed limit of the seed function takes the form
\begin{equation}
    \hat{\mathcal{B}}(u) = \frac{4\pi^3  \cot(\tfrac{\pi}{2}\delta)}{\delta-1} u^{\delta-1} + \frac{\pi ^2 2^{4-\delta }}{\delta ^2-3 \delta +2}u + \cdots \,,
\end{equation}
so we see that for $\delta<2$, the non-analytic scaling dominates in the squeezed bispectrum whereas for $\delta>2$ the scaling is mimicable by local self-interactions. Up to the normalisation, this is the same behavior as was found in \cite{Green:2013rd}. We therefore see that the qualitative behavior of this correlator is the same as that of the compact scalar. The physical explanation for why $\delta=2$ is the threshold is again clear--- for larger values of $\delta$ the CFT correlator decays too rapidly so of course the exchange of $\mathcal{O}_\delta$ should effectively be a contact interaction. 

Now, as a cross-check, we will take the $\delta \to 1$ limit of this result since this corresponds to the exchange of a free conformal scalar, which has a known analytical answer. Then we will take the $\delta \to 2$ limit, which can be interpreted as a loop-level exchange of two conformal scalars via the interactions $\dot{\pi}_c \confsc^2$ and $\dot{\pi}_c^2 \confsc^2$. 

\subsection{Conformal Scalar Limit}
Now let us examine the $\delta \to 1$ limit of this expression, for which our CFT exchange reduces to the familiar problem of exchanging a free conformally coupled scalar. This case has an extremely simple analytical solution which is \cite{Arkani-Hamed:2018kmz}\footnote{Note that the result quoted in \cite{Arkani-Hamed:2018kmz} is for $\hat{F}(u,v)$. We have firstly set $v=1$ to their result and secondly multiplied it by a factor of $8\pi^2$ to match our normalisation convention for the CFT operator.}
\begin{equation}\label{eq:conf_ex_bisp}
    \hat{\mathcal{B}}_{\delta=1}(u) = \frac{4\pi^4}{3} + 4\pi^2 \text{Li}_2\left(\frac{1-u}{1+u}\right)\,.
\end{equation}
As a check, we would like to match this result by taking the $\delta \to 1$ limit of the result we have obtained for general $\delta$ using the spectral density. 

\begin{figure}
    \centering
    \includegraphics[width=0.85\textwidth]{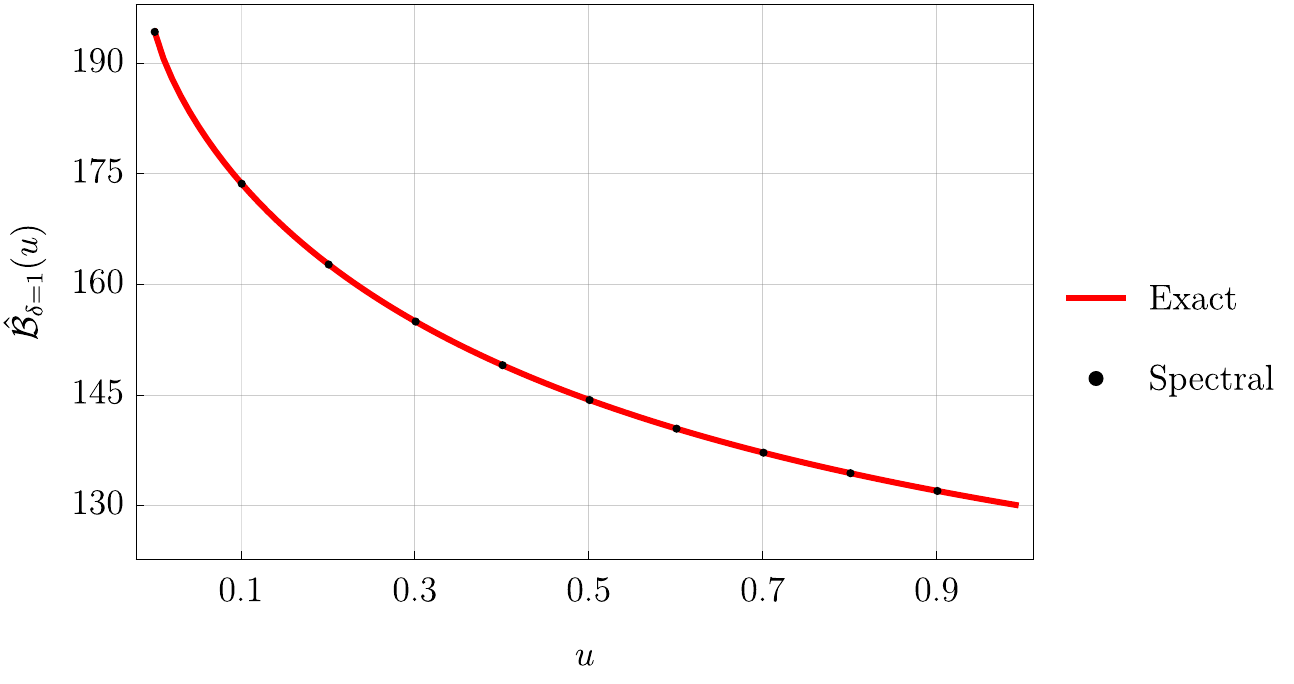}
    \caption{Comparison of the analytical expression for a tree-level conformal scalar exchange bispectrum with the result obtained using the spectral method.}
    \label{fig:cftd1_seedfunc}
\end{figure}

On the non-analytic side the $\Delta_*=\delta$ residue produces a constant contribution of $2\pi^4$. The residue at $\Delta_*=\delta +1$ and the one at $\Delta=2$ are both singular and they sum up to produce
\begin{equation}
    [{\rm Res}_{\delta + 1} \rho_\delta(\Delta)] \hat{b}(\delta+1;u)+ [{\rm Res}_{2}\hat{b}(\Delta;u)] \rho_\delta(2) \to \frac{8\pi^2}{\delta-1}\arctanh(u) + \cdots
\end{equation}
All of the other residues sum up to zero. On the EFT side, the $n=0$ piece is also divergent as $\delta \to 1$, but cancels the divergence from the non-analytic piece exactly. However, all of the $n>0$ terms in the EFT sum are left non-zero. The final result is then
\begin{equation}
    \hat{\mathcal{B}}_{\delta=1}(u) = 2\pi^4+8\pi^2\left[\log \left(\frac{u}{2}\right)+1\right] \arctanh(u) + \sum_{n \geq 0} f_n(u)\,,
\end{equation}
where
\begin{equation}
    \begin{aligned}
        f_n(u) \equiv &-\frac{\pi ^{5/2} 2^{2-n} u^2 (-u)^n \Gamma (n+1)}{(n+2) \Gamma \left(n+\frac{3}{2}\right)} \tFo{\frac{n+2}{2}}{\frac{n+3}{2}}{n+\tfrac{5}{2}}{u^2} \\
        &+\frac{8\pi^2}{2n+1}\left[\psi ^{(0)}(2 n+1)-\psi ^{(0)}(n+\tfrac{3}{2})\right] u^{2n+1}
    \end{aligned}
\end{equation}
and $\psi^{(0)}(z)$ is the Digamma function. The first term here is from the non-vanishing parts of the EFT series and the second from the non-analytic piece. While this expression looks completely different from (\ref{eq:conf_ex_bisp}) it can be checked numerically (summing up to $n \leq 30$ is sufficient) that they are in fact the same. We compare this expression with the known answer in Figure~\ref{fig:cftd1_seedfunc} finding excellent agreement.

\subsection{Double Conformal Scalar Limit}
Now for the sake of completion, we will take the $\delta \to 2$ limit of our result. As we mentioned previously, this case can be interpreted as the exchange of a conformal scalar at one-loop. This intuition will be helpful for us to carry forward especially in interpreting the divergences we will encounter. Before carrying out any calculations, it is also helpful to remind ourselves where the poles of the spectral density occur. They occur at $\Delta_*=2,3,\cdots$, i.e.\ at discrete series states. We should therefore expect a qualitatively similar result, namely terms which scale as $\log(u)$.

\begin{figure}
    \centering
    \includegraphics[width=0.7\textwidth]{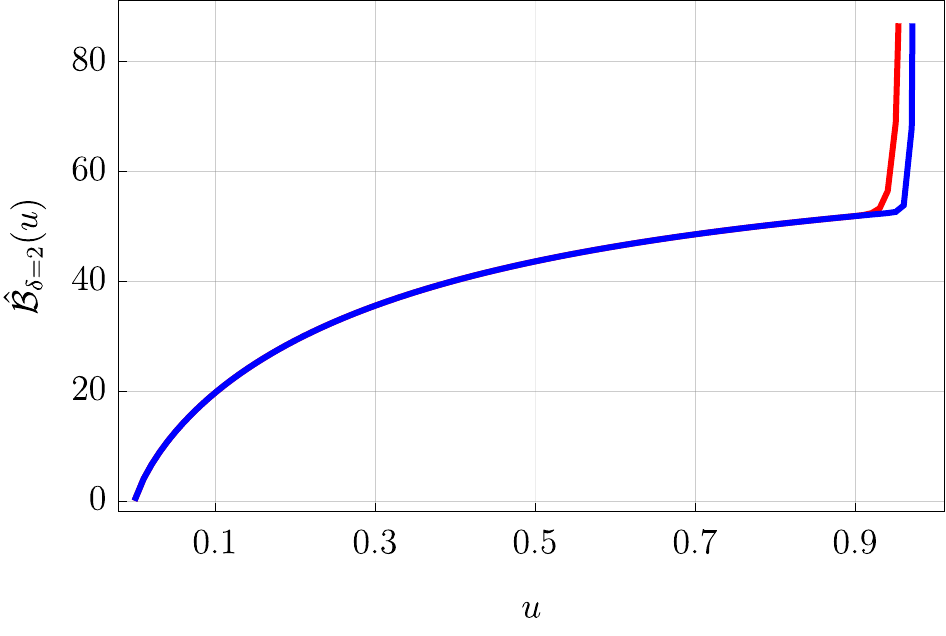}
    \caption{The seed function $\hat{\mathcal{B}}_{\delta=2}(u)$ for a CFT operator with scaling dimension $\delta=2$, where we sum up to $\textcolor{red}{N=50}$ and $\textcolor{blue}{N=100}$ respectively. Summing over more terms appears to soften the apparent $u\to 1$ singularity, albeit slowly.}
    \label{fig:cftd2_seedfunc}
\end{figure}

The residues at $\Delta_*=\delta+ 2n+1$, i.e.\ the odd integer poles, all vanish as $\delta \to 2$. The residues at $\Delta_*=\delta + 2n$ on the other hand exhibit a singularity
\begin{equation}
    \begin{aligned}
        [{\rm Res}_{\delta + 2n} &\rho_\delta(\Delta)] \hat{b}(\delta+2n;u) = \frac{\pi^{5/2}}{\delta-2}\frac{2^{3-2n}\Gamma(2n+1)}{\Gamma(2n+\tfrac{1}{2})}u^{2n+1} \tFo{n+1}{n+\tfrac{1}{2}}{2n+\tfrac{3}{2}}{u^2} \\
        &+ \frac{\pi ^{5/2} 4^{1-n} \Gamma (2 n+1) }{\Gamma \left(2 n+\frac{3}{2}\right)}u^{2 n+1} \left[(4 n+1) \log \left(\frac{u}{4}\right)-4 n+1\right] \\
        &+\sum_{k \geq 0}\frac{\pi ^{5/2} (4 n+1) 2^{-2 (k+n-1)} (2 (k+n))! \left(H_{2 (k+n)}-H_{k+2 n+\frac{1}{2}}+2 H_{2 n}\right)}{k! \Gamma \left(k+2 n+\frac{3}{2}\right)}u^{2 k+2 n+1}\,.
    \end{aligned}
\end{equation} 
Meanwhile the residues at $\Delta=2n+2$ are also divergent as $\delta \to 2$, which exactly cancel out the divergence from the other set of residues. Combining the two produces a result which is finite
\begin{equation}
    \begin{aligned}
        \hat{\mathcal{B}}_{\delta=2}^{\rm NA}(u) &= \sum_{n\geq 0} \frac{\pi ^{5/2} 4^{1-n} \Gamma (2 n+1)}{\Gamma \left(2 n+\frac{3}{2}\right)}u^{2 n+1} \tFo{n+1}{n+\tfrac{1}{2}}{2n+\tfrac{3}{2}}{u^2}\left[(8 n+2) H_{2 n}-2-(4 n+1) \log \left(\frac{u}{2}\right)\right] \\
        &-\sum_{n,k\geq 0}\frac{\pi ^{5/2} (4 n+1) 2^{-2 (k+n-1)} (2 (k+n))! \left(H_{2 (k+n)}-H_{k+2 n+\frac{1}{2}}+2 H_{2 n}\right)}{k! \Gamma \left(k+2 n+\frac{3}{2}\right)}u^{2 k+2 n+1}\,.
    \end{aligned}
\end{equation}
The EFT series on the other hand also exhibits a singularity as $\delta \to 2$. The series becomes
\begin{equation}
    \begin{aligned}
        \hat{\mathcal{B}}_{\delta=2}^{\rm EFT}(u) & = \frac{1}{\delta-2}\sum_{n \geq 0}\frac{\pi ^{5/2} 2^{2-n} n! }{\Gamma \left(n+\frac{1}{2}\right)} (-u)^{n+1} \tFo{\frac{n+1}{2}}{1+\frac{n}{2}}{n+\tfrac{3}{2}}{u^2}\\
        &+\sum_{n \geq 0}\frac{\pi ^{5/2} 2^{2-n} n! }{\Gamma \left(n+\frac{1}{2}\right)}(-u)^{n+1}\tFo{\frac{n+1}{2}}{1+\frac{n}{2}}{n+\tfrac{3}{2}}{u^2}(2 H_n-1-\log (2))\,.
    \end{aligned}
\end{equation}
First let us discuss the $\frac{1}{\delta-2}$ singularity. The parameter $\delta$ can be interpreted as a regulator. Similarly to the role played by $\alpha$ in dimensional regularisation, it controls the strength of the short-distance singularity. Just from the form of this term, it is clear that it is completely analytic in $u$ and therefore we should expect that it can be removed by at least a set of local counterterms. In fact it is the same divergence we have previously encountered in the non-compact theory. Indeed this is just a special case of the general $\Delta_\sigma$ bubble. Indeed we can check numerically that the divergent piece in the EFT expansion is a complicated rewriting of
\begin{equation}
    \frac{1}{\delta-2}\sum_{n \geq 0}\frac{\pi ^{5/2} 2^{2-n} n! }{\Gamma \left(n+\frac{1}{2}\right)} (-u)^{n+1} \tFo{\frac{n+1}{2}}{1+\frac{n}{2}}{n+\tfrac{3}{2}}{u^2} = \frac{4\pi^2}{\delta-2}\frac{u}{1+u}\,.
\end{equation}
Our agreement here makes sense of course--- since the UV divergence is a short-distance effect it should be insensitive to the precise choice of the mass. In the squeezed limit the seed function behaves as
\begin{equation}
    \hat{\mathcal{B}}_{\delta=2}(u) = -4\pi^2 u \left[3+ \log\left(\frac{u^2}{8}\right)\right] + 8 \pi ^2 H_{\frac{1}{2}} u + \mathcal{O}(u^2)\,,
\end{equation}
where the first term is from the non-analytic piece and the second one from the EFT piece. Here we see the characteristic logarithmic OPE scaling associated with discrete series states in de Sitter. In Fig.~\ref{fig:cftd2_seedfunc} we show a plot of this seed function. As we see from the plot, this solution appears to show a singularity as $u \to 1$, similar to our observation with the non-compact scalar bubble. We notice however, that summing over more terms appears to soften this singularity so expect that it is a numerical artefact which would be circumvented with a better organisation of the final answer. We leave a more careful analysis for future work.

\section{Asymptotics of the Tree-Level Seed Function}\label{app:btree_largeDlim}
In this section we will fill in some details about asymptotics of the tree-level seed function. Recall that the non-analytic piece is
\begin{equation}
    \hat{b}_{\rm NA}(\Delta; u)  
    = \mathcal{F}(\Delta) 
    u^{\Delta-1} \tFo{\tfrac{1}{2}(\Delta-1)}{\tfrac{\Delta}{2}}{\Delta-\tfrac{1}{2}}{u^2}\,, \qquad \mathcal{F}(\Delta) \equiv 
    \frac{\Gamma(\tfrac{3}{2}-\Delta)
    \Gamma(1-\tfrac{\Delta}{2})
    \Gamma^{2}(\tfrac{\Delta}{2})}
    {2\Gamma(\tfrac{3}{2}-\tfrac{\Delta}{2})}.
\end{equation}
We are interested in determining the $\Delta \to +\infty$ asymptotics of this expression. Firstly note that
\begin{equation}
    \begin{aligned}
        \mathcal{F}(\Delta) u^{\Delta-1} \xrightarrow{\Delta \to \infty} &\,\pi^{\tfrac{3}{2}}\left(\frac{u}{2}\right)^{\Delta-1}\Delta^{-\tfrac{1}{2}}\sin \left[\pi  \left(\frac{3}{2}-\frac{\Delta }{2}\right)\right] \csc \left[\pi  \left(\frac{3}{2}-\Delta \right)\right] \csc \left[\pi  \left(1-\frac{\Delta }{2}\right)\right] \\
        \sim&\, \pi^{\tfrac{3}{2}}\left(\frac{u}{2}\right)^{\Delta-1}\Delta^{-\tfrac{1}{2}}\,.
    \end{aligned}
\end{equation}
To determine the large $\Delta$ asymptotics of the hypergeometric function we can rely on the integral identity \cite{dmlf}
\begin{equation}
    \tFo{a}{b}{c}{z} = \frac{\Gamma(c)}{\Gamma(b)\Gamma(c-b)}\int_{0}^\infty \ud t\,t^{c-b-1}(1+t)^{a-c}(1-z+t)^{-a}\,,
\end{equation}
which holds whenever ${\rm Re}\,(c)>{\rm Re}\,(b)>0$ and $|{\rm arg}(1-z)|<\pi$. For our choice of $a$, $b$ and $c$ this integral is particularly amenable to a saddle point approximation. The location of the saddle can be determined to be
\begin{equation}
    t_* = \frac{3 u^2+\sqrt{(1-2 \Delta )^2+9 u^4+(-4 (\Delta -1) \Delta -6) u^2}-5}{2 (\Delta +2)} \approx \sqrt{1-u^2}\,,
\end{equation}
where at the end we have taken the limit $\Delta \to \infty$. We can evaluate this integrand at the saddle and account for the fluctuations to produce a rather lengthy expression. Further simplifying it in the limit $\Delta \to \infty$ produces
\begin{equation}
    \begin{aligned}
        \tFo{\tfrac{1}{2}(\Delta-1)}{\tfrac{\Delta}{2}}{\Delta-\tfrac{1}{2}}{u^2} \sim &\sqrt{\left(u^2-1\right) \left[u^2 \left(\sqrt{1-u^2}+ 3\right)-4 \left(\sqrt{1-u^2}+1\right)\right]} \\ 
        &\times  2^{\Delta -\frac{3}{2}} \left(1-u^2\right)^{\frac{\Delta -3}{4}} \left[2 \left(\sqrt{1-u^2}+1\right)-u^2 \left(\sqrt{1-u^2}+2\right)\right]^{-\frac{\Delta }{2}}\,.
    \end{aligned}
\end{equation}
Putting everything together we find that
\begin{equation}
    \hat{b}_{\rm NA}(\Delta;u) \sim \Delta^{-\tfrac{1}{2}} \exp\left[\frac{1}{4}\Delta \log\left(\frac{u^4+4 \left(\sqrt{1-u^2}-2\right) u^2-8 \sqrt{1-u^2}+8}{u^4}\right)\right]\,.
\end{equation}
Therefore for $u \in (0,1)$, i.e.\ in the physical domain, the non-analytic seed function decays exponentially with $\Delta$.

\section{Asymptotics of the Spectral Density of the Vertex Operator}\label{app:vertex_spec_asymp}
In this section we will derive the large $\Delta$ scaling of the spectral density
\begin{equation}
    \rho_{\mathcal{V}}(\Delta) = \frac{16\pi \Gamma(1-\beta) \sin(\pi \beta)}{\Gamma(\beta-1)} (3-2\Delta)\cos(\pi \Delta) \Gamma(\beta-\Delta)\Gamma(\Delta-3+\beta)\tFt{\beta-\Delta}{\Delta-3+\beta}{\beta-1}{\beta}{\frac{\beta}{2}} \,.
\end{equation}
The large-$\Delta$ behavior of the Gamma function prefactors can be straightforwardly obtained
\begin{equation}
    \frac{16\pi \Gamma(1-\beta) \sin(\pi \beta)}{\Gamma(\beta-1)} (3-2\Delta)\cos(\pi \Delta) \Gamma(\beta-\Delta)\Gamma(\Delta-3+\beta) \xrightarrow{\Delta \to +\infty} -\frac{32 \pi ^3 \Delta ^{2 \beta -3} \cos (\pi  \Delta ) \csc (\pi  (\beta -\Delta ))}{\Gamma (\beta -1) \Gamma (\beta )}\,,
\end{equation}
whereas the hypergeometric function is more challenging. For that piece we will start with the assumption that $\beta<2$ so that we can utilize the series expansion of the hypergeometric function
\begin{equation}
    \begin{aligned}
        \tFt{\beta-\Delta}{\Delta-3+\beta}{\beta-1}{\beta}{\frac{\beta}{2}} = &\,\sum_{n=0}^{\infty} a_n(\Delta,\beta)\left(\frac{\beta}{2}\right)^{n} \\
        a_n(\Delta,\beta) = &\, \frac{\Gamma (\beta -1) \Gamma (\beta ) \Gamma (n+\beta -\Delta ) \Gamma (n+\beta +\Delta -3)}{\Gamma (n+1) \Gamma (\beta -\Delta ) \Gamma (\beta +\Delta -3) \Gamma (n+\beta -1) \Gamma (n+\beta )}\,.
    \end{aligned}
\end{equation}
Now we can extract the large $\Delta$ asymptotics of the series coefficients
\begin{equation}
    a_n(\Delta,\beta) \xrightarrow{\Delta \to +\infty} \frac{\Gamma (\beta -1) \Gamma (\beta ) \Delta ^{2 n-1} \sin (\pi  (\beta -\Delta )) (2 \Delta +n (2 \beta +n-7)) \csc (\pi  (\beta -\Delta +n))}{2 \Gamma (n+1) \Gamma (n+\beta -1) \Gamma (n+\beta )}
\end{equation}
and sum over $n$ finding
\begin{equation}
    \begin{aligned}
        &\,\tFt{\beta-\Delta}{\Delta-3+\beta}{\beta-1}{\beta}{\frac{\beta}{2}} \sim {}_0F_{2}\left[\beta,\beta-1;-\frac{1}{2}\beta  \Delta ^2\right] \\
        &\,- \frac{\Delta}{4(\beta-1)}\left[(2\beta-7){}_0F_{2}\left[\beta,\beta+1;-\frac{1}{2}\beta  \Delta ^2\right] +{}_1F_2\left[2;1,\beta,\beta+1;-\frac{1}{2} \beta  \Delta ^2\right]\right]\,.
    \end{aligned}
\end{equation}
Simplifying this expression by taking the limit $\Delta \to +\infty$ produces
\begin{equation}
    \rho_{\mathcal{V}}(\Delta) \sim  \Delta ^{\frac{2}{3}(\beta -2)} \exp\left(\frac{3 \beta^{1/3}  \Delta ^{2/3}}{2^{4/3}}\right)\,.
\end{equation}
While we assumed that $\beta<2$ in order to derive this scaling, it can be verified numerically that it holds for $\beta>2$ as well.

The $\Delta \to \infty$ limit is probing heavier states so we can understand it as a UV limit. We should therefore expect to recover the flat-space spectral density as $\Delta \to \infty$ with the understanding that $\Delta \approx i m/H$ for heavy masses. Let us briefly calculate the spectral density for the vertex operator in $(d+1)$-dimensional flat-space and determine its behavior in the UV. We will start our calculation in Euclidean signature and subsequently analytically continue to Lorentzian signature. In flat-space the action for the free massless compact scalar is
\begin{equation}
    S_{\rm E} = \int \ud^{d+1}\mb{x}\, \left[\tfrac{1}{2}f^2 \mu^{d-3}(\partial \varphi)^2\right]\,,
\end{equation}
where $f$ is the decay constant and we have introduced the mass scale $\mu$ to ensure $[f]=1$ for all $d$ (note that as in de Sitter, we keep $\varphi$ dimensionless). As in de Sitter, the position space vertex propagator is the exponential of the massless flat-space propagator $G(x)$ and can be evaluated to be
\begin{equation}
    G_{\mathcal{V}}(x) = \exp\left[\frac{1}{\mu^{d-3}f^2} G(x)\right]=\exp\left[\tfrac{1}{2}\tilde{\beta} (\mu x)^{1-d}\right]\,,
\end{equation}
where we define $\tilde{\beta}\equiv (4\pi)^{\frac{d+1}{2}} \Gamma(\tfrac{d-1}{2})\frac{\mu^{2}}{2f^2}$ for convenience, and can be interpreted as a flat-space version of the $\beta$ encountered in de Sitter. The momentum space propagator can be obtained by taking the Fourier transform
\begin{equation}
    \begin{aligned}
        G_{\mathcal{V}}(k) = \,& \int \ud^{d+1}\mb{x}\, e^{i \mb{k}\cdot \mb{x}} G_{\mathcal{V}}(x) \\
        =\, & \sum_{n=0}^{\infty} \frac{1}{n!}\left(\frac{1}{\mu^{d-3}f^2}\right)^{n} \left[\int \ud^{d+1}\mb{x}\, e^{i \mb{k}\cdot \mb{x}} G^n(x)\right]\,,
    \end{aligned}
\end{equation} 
where in the last line we have expanded the exponential into its series and exchanged the sum with integral. This is equivalent to calculating the momentum space propagator as a sum over massless $n$-loop integrals. As is standard, each of these loop integrals can be done analytically
\begin{equation}
    \begin{aligned}
        \frac{1}{n!}\left(\frac{1}{\mu^{d-3}f^2}\right)^{n}\int \ud^{d+1}\mb{x}\, e^{i \mb{k}\cdot \mb{x}} G^n(x) = \, & \frac{(2\pi)^{\frac{d+1}{2}}}{n!}\left(\frac{1}{\mu^{d-3}f^2}\right)^{n}\int_{0}^\infty \ud x\, x^{d} (k x)^{\frac{1-d}{2}} J_{\frac{d}{2}-\frac{1}{2}}(k x) G^n(x) \\
        = \,&  (4\pi)^{\frac{d+1}{2}} \tilde{\beta}^n \frac{2^{-d n} \mu ^{(1-d) n} \Gamma \left(\frac{1}{2} (-n d+d+n+1)\right)}{n! \Gamma \left(\frac{1}{2} (d-1) n\right)} k^{d (n-1)-n-1}\,,
    \end{aligned}
\end{equation}
where we have regulated the loops corresponding to $n \geq 2$ using dimensional regularisation. In order to extract the spectral density, we analytically continue $k \to (-s)^{\frac{1}{2}}$ and extract the discontinuity across the cut along physical energies $s>0$
\begin{equation}
    \begin{aligned}
        \rho_{\mathcal{V}}(s) = -i\,&{\rm Disc}_s G_\mathcal{V}(s) \\
        =\, & -(2\pi) (4 \pi )^{\frac{d+1}{2}} \sum_{n=0}^{\infty} \tilde{\beta}^n\frac{2^{-d n} \mu ^{(1-d) n}}{n! \Gamma \left(\frac{1}{2} (d-1) (n-1)\right) \Gamma \left(\frac{1}{2} (d-1) n\right)} s^{\frac{1}{2}d (n-1)-\tfrac{1}{2}(n+1)}\,.
    \end{aligned}
\end{equation}
Note that the spectral density is UV finite and we can readily specialise to our choice of $d$. Unfortunately we are unable to evaluate this sum for general $d$. However, it can be performed using {\tt Mathematica} for $d=2,3,4,\cdots$. We quote the results for $d=2$ and $d=3$
\begin{equation}
    \begin{aligned}
        \rho_{\mathcal{V}}(s) = \,& 2\pi \tilde{\beta}  \left(\frac{\pi}{\mu s}\right) I_2\left[(2 \tilde{\beta} \mu^{-1})^{\frac{1}{2}}s^{\frac{1}{4}}\right] \sim \exp\left(\sqrt{2 \mu^{-1} \tilde{\beta}}s^{\frac{1}{4}}\right) \,, \qquad (d=2) \\
         = \,& \frac{\pi^3 \tilde{\beta}^2 \mu^{-4}}{4} {}_0 F_2\left[2,3; \frac{1}{8}\tilde{\beta} \mu^{-2} s\right] \sim \exp\left[\frac{3}{2}\left(\tilde{\beta}\mu^{-2}s\right)^{\frac{1}{3}}\right]\,,\qquad (d=3)\,.
    \end{aligned}
\end{equation}
It is straightforward to obtain the same for $d=4,5,\cdots$. Doing so allows us to conjecture a general form for the factor appearing in the exponent to be $s^{\frac{d-1}{2d}}$. Written in terms of $m^2=s$ we have
\begin{equation}
    \log\rho_{\mathcal{V}}(m^2) \sim m^{\frac{d-1}{d}}\,,
\end{equation}
and we have checked that this scaling holds for up to $d=10$. In $d=3$ we therefore see that the flat-space calculation agrees with the scaling of the de Sitter spectral density in the limit $\Delta \to \infty$. Specifically in flat-space we have
\begin{equation}
    \rho_{\mathcal{V}}(m^2) \sim \exp\left[\left(\frac{\tilde{\beta} m^2}{\mu^2}\right)^{1/3}\right] \sim \exp\left[\left(\frac{m}{f}\right)^{2/3}\right] \,.
\end{equation}
Recalling that $\beta^{1/3}\Delta^{2/3} \sim (m/f)^{\frac{2}{3}}$ we see that this is the same scaling (up to prefactors) with respect to the decay constant $f$ and $m$ as in the dS result. This constitutes a check of our spectral density, but also explains the factor of $\Delta^{\tfrac{2}{3}}$ appearing in the exponential in the de Sitter result.\footnote{We are grateful to Hayden Lee for raising a question about this.} We see that it is fixed by the number of spacetime dimensions. Indeed, upon increasing $d$, the short distance singularity of the massless propagator $G(x)$ worsens, in turn making the vertex operator more singular, which is consistent with the spectral density becoming more badly divergent.

Notably, the flat-space result is not able to reproduce the power-law prefactor $\Delta^{\frac{2}{3}(\beta-2)}$. This is a consequence of the fact that upon taking a flat-space limit, a massless field in de Sitter \textit{does not} map onto a massless field in flat-space. One way to see this is to recall that the massless de Sitter propagator contains a logarithmic piece which leads to a divergence in the IR but also contributes to the UV divergence. This extra logarithmic UV divergence does not occur for the corresponding flat-space propagator. It is precisely this logarithmic piece in the massless de Sitter propagator which led to the de Sitter vertex propagator to come dressed with a power law prefactor. The de Sitter propagator is
\begin{equation}
    G_{\mathcal{V}}(\lambda) = \lambda^{\beta} e^{\tfrac{1}{2}\beta \lambda} \sim \left[\frac{4\eta \eta'}{x^2-(\eta-\eta')^2}\right]^{\beta} \exp\left[\beta\frac{2\eta \eta'}{x^2-(\eta-\eta')^2}\right]\,.
\end{equation}
So the short-distance singularity is not merely the essential singularity due to the exponential, but it also comes dressed with a power-law. Thus we may expect to recover the exponential scaling in the flat-space calculation, but can miss power laws.

\section{UV cutoff of the compact scalar}\label{app:uv_cutoff}
In this section we will provide an example for a specific UV completion of the compact scalar effective theory considered in the main text. Our goal is to determine the UV cutoff, particularly aiming to justify the claim that one may generally expect $f$ to span a wider range of values than what may be expected. Specifically, we will justify our assumption that our EFT can be sensible for $\beta= H^2/(2\pi f)^2>1$ as well, at the cost of an exponentially suppressed coupling to the inflaton. While this statement can be established using EFT arguments (see \cite{Hook:2023pba}), we will rely on an explicit extra-dimensional UV construction.  

Certainly, the cutoff associated with field theory (or Peccei-Quinn) axions is $f$, since it is the Peccei-Quinn symmetry breaking scale, and therefore we require $H \ll f$ or $\beta \ll 1$ for such scalars. However for an extra-dimensional compact scalar, there is no such symmetry breaking scale. Instead the cutoff could take one of two values---the Kaluza-Klein (KK) scale or the scale of quantum gravity. Both of these scales depend on the volume of the compact extra-dimensional manifold, or cycles therein, and their specific values are fairly model dependent. As such, such extra-dimensional scenarios are more flexible than field theory ones, and by relying on this point we will argue that it is possible for both of these scales to be parametrically larger than $f$ such that the ratio $\beta$ can safely be large. We will provide an explicit example in which this is the case.\footnote{We thank Matt Reece for useful discussions on this topic.}

As argued in \cite{Reece:2025thc}, the scale of quantum gravity in the context of extra-dimensional axions is determined by the axion string tension $\mathcal{T}$. In string theory UV completions, these are fundamental strings and their tension is observed to be related to axion parameters as
\begin{equation}
    \mathcal{T}\approx 2\pi S_{\rm inst} f^2
\end{equation}
where $S_{\rm inst}=8\pi^2/g^2$ is the Yang-Mills instanton action and $g$ the gauge coupling of the gluons coupled to the axions. Therefore, the string tension can be parametrically larger than $f$ if we assume $g \ll 1$. This is of course problematic if we are referring to the QCD axion, in which case $f$ is typically two orders of magnitude lower than the string tension.

On the other hand the Kaluza-Klein scale is determined by the volume of the compact dimensions as $M_{\rm KK}\sim M_s/\mathcal{V}_p^{1/p}$, where $\mathcal{V}_p$ is the larger of two possibilities: the volume of the largest $p$-cycle, or the volume of the full six dimensional manifold, in which case $p=6$. We express the volume in units of the string length $\ell_s$ and the string scale $M_s \equiv 2\pi/\ell_s$. Meanwhile, the decay constant also scales with the volume of its associated cycle in a model dependent fashion. Broadly speaking, there is sufficient flexibility among these models such that the volumes of various cycles can admit hierarchies which allow $f \ll H \ll M_{\rm KK}$.

Let us discuss a concrete example. We will consider a Type IIB string theory compactified on a 6d Calabi-Yau orientifold $Y$ which contains three four-cycles $\tau_i$, which yields three axions $\theta_i$ upon dimensionally reducing on the three different cycles (the fibered ``Swiss-Cheese'' example of \cite{Reece:2025thc}). The volume of the Calabi-Yau expressed in terms of the volume of the cycles is then
\begin{equation}
    \mathcal{V} = \frac{1}{6}\sqrt{\tau_1}\tau_2 - \frac{\sqrt{2}}{3}\tau_3^{3/2},
\end{equation}
and we will assume the hierarchy $\tau_3 \ll \tau_1,\tau_2$. This choice yields three decoupled axions. As is standard, their decay constants $f_i$ are determined by the K\"ahler potential $k= -2\log{\mathcal{V}}$ via the kinetic matrix
\begin{equation}
    \kappa_{ij} = \frac{g_s^2}{8\pi^2}M_{\rm pl}^2 \frac{\partial^2 k}{\partial \tau_i \tau_j}.
\end{equation}
Let us focus on the first axion which has the decay constant
\begin{equation}
    f_1^2 = \frac{g_s^2}{8\pi^2}M_{\rm pl}^2 \frac{1}{(\tau_1)^2},
\end{equation}
where we have neglected terms suppressed in $\tau_3$. The gauge field associated with this scalar has the gauge coupling $g_1^2 = 4\pi g_s/\tau_1$. Now let us assume a hierarchy in the volumes of the three four-cycles $\tau_1 \gg \tau_2 , \tau_3$ such that $\tau_1^{3/2} \gg \mathcal{V}$. Upon recalling that the four dimensional Planck scale is $M_{\rm pl}^2 \sim (\mathcal{V}/g_s^2)M_s^2$ and $M_{\rm KK}\sim M_s/(\tau_1)^{1/4}$ we note
\begin{equation}
    f_1^2 \sim \frac{\mathcal{V}}{(\tau_1)^{3/2}} M_{\rm KK}^2.
\end{equation}
Thus we observe that the KK scale can be parametrically larger than the axion decay constant, and therefore even the Hubble scale. The requirement that our EFT is valid, i.e.~$H \ll M_{\rm KK}$, can thus be re-expressed as
\begin{equation}
    \beta \ll \frac{(\tau_1)^{3/2}}{\mathcal{V}}.
\end{equation}
Thus the upper bound on $\beta$ can also be parametrically large. Another condition, which we mention in passing, is that SUSY must be broken in our EFT. This requires that the vacuum energy $H^2 M_{\rm pl}^2 \lesssim M_s^4$, which is safely satisfied in this example since
\begin{equation}
    \frac{H^2 M_{\rm pl}^2}{M_s^4} \sim \frac{H^2}{M_{\rm KK}^2} \tau_2 \ll 1,
\end{equation}
where we have approximated $\mathcal{V}\sim \sqrt{\tau_3}\tau_2$ and $\tau_2 \sim 1$.

An important caveat in this example is that the condition $\tau_3 \gg 1$ makes the QCD coupling $g_1$ too small, so this axion cannot be the QCD axion. Relatedly, $\tau_3$ cannot be arbitrarily large since we expect vertex operators to come dressed with an instanton factor $\exp(-2\pi S_{\rm inst})$, and $S_{\rm inst}\sim \tau_3$. Such an exponential suppression is consistent with known EFT arguments \cite{Hook:2023pba}, and can be understood from the low-energy perspective to arise due to the fact that taking the coupling to zero achieves a free theory, valid to any scale. Therefore the true cutoff of the theory is (in fact, logarithmically) sensitive to the couplings.

To put some numbers, we can take $M_s\approx 10^{17}\,{\rm GeV}$ and $g_s \approx 1$ which requires $\mathcal{V}\approx 10$, which can be accomplished by setting $\tau_1 \approx 10^{2}$ and $\tau_2, \tau_1 \sim 1$. Therefore the parameter $\beta \ll 10^{2}$ and can therefore easily be order ten. Unfortunately, this tunes the coefficient of the vertex operator to be exponentially small since $2\pi S_{\rm inst}\sim 4\pi^2 \tau_1\sim 10^{3}$.

\phantomsection
\addcontentsline{toc}{section}{References}
\bibliographystyle{utphys}
\bibliography{refs.bib}

\end{document}